\documentclass[traditabstract]{aa}
\usepackage{longtable}
\usepackage{supertabular}
\usepackage{rotating}
\usepackage{graphicx}
\usepackage[usenames,dvipsnames]{color}

\usepackage{txfonts}
\begin{document}

\newcommand{\mic}{$\mu$m$\,$}
\newcommand{\mica}{$\mu$m}
\newcommand{\lir}{$\rm{L}_{\rm{IR}} \,$}
\newcommand{\lira}{$\rm{L}_{\rm{IR}}$}
\newcommand{\tsfr}{$\Sigma$(SFR)$\,$}
\newcommand{\tsfra}{$\Sigma$(SFR)}
\newcommand{\tsfrm}{$\Sigma$(SFR)/M$\,$}
\newcommand{\tsfrma}{$\Sigma$(SFR)/M}

    \renewcommand{\topfraction}{0.9}	
    \renewcommand{\bottomfraction}{0.8}	
    \setcounter{topnumber}{2}
    \setcounter{bottomnumber}{2}
    \setcounter{totalnumber}{4}     
    \setcounter{dbltopnumber}{2}    
    \renewcommand{\dbltopfraction}{0.9}	
    \renewcommand{\textfraction}{0.07}	
    \renewcommand{\floatpagefraction}{0.7}	
    \renewcommand{\dblfloatpagefraction}{0.7}	


\title{The effect of the high-pass filter data reduction technique on the {\it{Herschel}} PACS Photometer PSF and noise.}
\author{P. Popesso\inst{1}, B. Magnelli\inst{1}, S. Buttiglione\inst{2}, D. Lutz\inst{1}, A. Poglitsch\inst{1}, S. Berta\inst{1}, R. Nordon\inst{1}, B. Altieri\inst{3},  H. Aussel\inst{4}, N. Billot\inst{5}, R. Gastaud\inst{4} , B. Ali\inst{5}, Z. Balog\inst{6}, A. Cava\inst{7}, H. Feuchtgruber\inst{1}, B. Gonzalez Garcia\inst{3}, N. Geis\inst{1}, C. Kiss\inst{8}, U. Klaas\inst{6}, H. Linz\inst{6},  X.C. Liu\inst{5}, A. Moor\inst{8}, B. Morin\inst{4}, T. M\"uller\inst{1}, M. Nielbock\inst{6}, K. Okumura\inst{4},  S. Osterhage\inst{1}, R. Ottensamer\inst{9}, R. Paladini\inst{5}, S. Pezzuto\inst{10}, V. Dublier Pritchard\inst{1}, S. Regibo\inst{11},G. Rodighiero\inst{2}, P. Royer\inst{11}, M. Sauvage\inst{4}, E. Sturm\inst{1}, M. Wetzstein\inst{1}, E. Wieprecht\inst{1}, E.Wiezorrek\inst{1}
\fnmsep\thanks{ Herschel is an ESA space observatory with science instruments provided by 
European-led Principal Investigator consortia and with important 
participation from NASA.}}
\offprints{Paola Popesso, popesso@mpe.mpg.de}

\institute{Max-Planck-Institut f\"{u}r Extraterrestrische Physik (MPE), Postfach 1312, 85741 Garching, Germany.
\and Dipartimento di Fisica e Astronomia "G. Galilei", vicolo dell'Osservatorio 3, 35122 Padova, Italy
\and Herschel Science Centre, European Space Astronomy Centre, ESA, Villanueva de la Ca\~nada, 28691 Madrid, Spain
\and Laboratoire AIM, CEA/DSM-CNRS-Universit{\'e} Paris Diderot, IRFU/Service d'Astrophysique,  B\^at.709, CEA-Saclay, 91191 Gif-sur-Yvette Cedex, France.
\and IPAC, California Institute of Technology, Pasadena, CA 91125, USA 
\and Max-Planck-Institut für Astronomie, Königstuhl 17, D-69117 Heidelberg, Germany
\and Departamento de Astrof´ısica, Universidad Complutense de Madrid, E28040-Madrid, Spain
\and Konkoly Observatory of the Hungarian Academy of Sciences, PO Box 67, 1525 Budapest, Hungary 
\and Department of Astronomy, University of Vienna, Wien, Austria
\and Istituto di Astrofisica e Planetologia Spaziali - INAF, Via del Fosso del Cavaliere, 100, I-00133 Roma, Italy
\and Institut voor Sterrenkunde, KU Leuven, Leuven, Belgium.  }

\date{Received / Accepted}

\abstract{We investigate the effect of the ``high-pass filter'' data reduction technique on the {\it{Herschel}} PACS PSF and noise of the PACS maps at the 70, 100 and 160 $\mu$m bands and in medium and fast scan speeds. This branch of the PACS Photometer pipeline is the most used for cosmological observations and for point-source observations.The calibration of the flux loss due to the median removal applied by the PACS pipeline (high-pass filter) is done via dedicated simulations obtained by ``polluting'' real PACS timelines with fake sources at different flux levels. The effect of the data reduction parameter settings on the final map noise is done by using selected observations of blank fields with high data redundancy. We show that the running median removal can cause significant flux losses at any flux level. We analyse the advantages and disadvantages of several masking strategies and suggest that a mask based on putting circular patches on prior positions is the best solution to reduce the amount of flux loss. We provide a calibration of the point-source flux loss for several masking strategies in a large range of data reduction parameters, and as a function of the source flux. We also show that, for stacking analysis, the impact of the high-pass filtering effect is to reduce significantly the clustering effect. The analysis of the global noise and noise components of the PACS maps shows that the dominant parameter in determining the final noise is the high-pass filter width. We also provide simple fitting functions to build the error map from the coverage map and to estimate the cross-correlation correction factor in a representative portion of the data reduction parameter space.}


\titlerunning{HP-filtering effect on the PACS PSF}
\authorrunning{Popesso et al.}

\maketitle

\section{Introduction}

The {\it{Herschel}} satellite (Pilbratt et al. 2010) is designed to explore the "cool Universe" during its expected 3.5 year mission lifetime. To achieve its scientific goals, HERSCHEL is equipped with a 3.5 m main mirror and marks the beginning of a new generation of "space giants". Bigger than any of its predecessors at approximately 7.5 m height and 4 m width, its science payload consists of three instruments: PACS and SPIRE, both cameras and spectrometers that allow {\it{Herschel}} to perform imaging in six different "colors" in the far-infrared, and HIFI, a spectrometer with extremely high spectral resolution. In particular, the Photoconducting Array Camera and Spectrometer PACS (Poglitsch et al. 2010) provides HERSCHEL with far-infrared imaging and spectroscopic capabilities from 60 to 210 $\mu$m.  

The largest fraction of the {\it{Herschel}} requested observing time has been spent on deep and/or large scale photometric surveys performed in scan map mode with the PACS photometer. Indeed, the opening of the 60-210 $\mu$m window by PACS photometer will address a wide range of key questions of current astrophysics concerning the origins of stars, planetary systems, galaxies, and the evolution of the Universe. 

PACS Photometer data are heavily affected by the so-called $1/f$ noise. The PACS pipeline in the HIPE environment consists of two branches able to deal in different ways with the 1/f noise. The recommended method for observations of extended sources is the MadMap branch, that removes the $1/f$ noise through an ``inversion'' algorithm (Cantalupo et al. 2010) able to preserve structures on large scales. The recommended method for point-source observations, such as miniscap map of point-sources or cosmological surveys of large blank fields, is the high-pass-filtering $+$ ``naive'' projection route. This method can strongly affect the shape of the PSF and, consequently, it can cause significant flux loss but it is not affected by artifacts around point-sources as in the MadMap case. The source flux loss can be limited by the use of a source mask during the high-pass-filtering in the so-called ``masked'' high-pass-filtering. However, there are no calibrations yet available of the effect of the high-pass-filtering on the shape of the PSF and the percentage of flux loss. In this paper we present a full calibration of this method of data reduction based on semi-simulated observations, based on real PACS timelines ``polluted'' with artificial sources.

The PACS pipeline does also not provide yet a reliable method to estimate the final error map after projection and there are no indications about the amount of cross-correlated noise due to the projection itself and the remaining $1/f$ noise in the final map. We present here two methods for estimating the two error components, error per pixel and cross correlated noise, for the high-pass-filtering $+$ ``naive'' projection branch of the PACS Photometer pipeline.

The paper is organized as follows. We present a brief description of the PACS data reduction steps for scan mode in Section 2. The simulations are described in Section 3 and 4. The analysis of the data reduction effect on the PACS PSF at 70, 100 and 160 $\mu$m is analyzed in Section 5 and 6. Section 7 describes how to correct the final extracted flux for the effect of the data reduction. A full characterization of the noise of PACS maps is presented in Section 8. The conclusions are given in section 9. 

\section{PACS data and PACS data reduction}

\subsection{PACS Photometer}

The PACS photometer detectors are bolometer arrays. Each pixel of the array can be considered as a little cavity in which sits an absorbing grid. The incident infrared radiation is registered by each bolometer pixel by causing a tiny temperature difference, which is measured by a thermometer implanted on the grid. The blue channel offers two filters, 60$–$85 $\mu$m and 85$–$130 $\mu$m and has a 32$\times$64 pixel array. The red channel has a 130$–$210 $\mu$m filter has a 16$\times$32 pixel array. Both channels cover a field-of-view of $\sim$1.75'$\times$3.5', with full beam-sampling in each band. The two short wavelength bands are selected by two filters via a filter wheel. The field-of-view is nearly filled by the square pixels, however the arrays are made of sub-arrays which have a gap of $\sim$1 pixel in between. For the long wavelength channel 2 matrices of 16$\times$16 pixels are tiled together. During a PACS scan map observation, the telescope moves back and forth in a pattern of parallel scan lines that are connected by short turnaround loops.

\subsection{Scan map technique}
The scan$-$map technique is the most frequently used {\it{Herschel}} observing mode. Scan maps are the default to map large areas of the sky, for galactic as well as extragalactic surveys, but they are also recommended for small fields and even for point-sources. Scan maps are performed by slewing the spacecraft at a constant speed along parallel lines. Available satellite speeds are 10, 20, 60  arcsec/s in PACS prime mode and 20, 60  arcsec/s (medium, fast) in PACS/SPIRE parallel mode. The number of satellite scans, the scan leg length, the scan leg separation, and the orientation angles (in array and sky reference) are freely selectable by the observer. Via a repetition parameter the specified map configuration can be repeated $n$ times. The PACS/SPIRE parallel mode sky coverage maps are driven by the fixed 21 arcmin separation between the PACS and SPIRE footprints. 

During the full scan-map duration the bolometers are constantly read-out with 40 Hz. However, due to satellite data-rate limitations there are on-board reduction and compression steps needed before the data is down-linked. In PACS prime modes 4 subsequent frames are averaged in both bands for a final sampling rate of 10 Hz. In PACS/SPIRE parallel mode 8 consecutive frames are averaged in the blue/green bands and 4 in the red band. In addition to the averaging process there is a supplementary compression stage ``bit rounding'' for high gain observations required, where the last n bits of the signal values are rounded off. The default value for n is 1 or 2, depending on the period (quantization step of $n\times 10^{-5}$ V) for all high gain PACS/SPIRE parallel mode observations, 1 for all high gain PACS prime mode observations, and 0 for all low gain observations. A combination of two different scan directions is recommended for a better field and PSF reconstruction. The data are organized in user friendly structures called Frames or Frames class, which are basically data cubes. The main information contained in the Frames are the Signal data cube, which is the temporal sequence of PACS detector images, and the Status table which provides a collection of information about each individual image, including info about the World Coordinate System (WCS, coordinates of the reference pixel and position angle, changing from image to image in the temporal sequence). We call timeline of the temporal sequence of Signal detected by an individual pixel. 

Most of the PACS prime observations are performed with a 20  arcsec/s scan speed where the bolometer performance is best and the pre-flight sensitivity estimates are met. This mode is particularly suited for large field, but it shows an excellent performance for very small fields and even for point-sources (see Poglitsch et al. 2010 for more details). The fast speed is used in the PACS/SPIRE parallel mode observations while low speed was never requested by the observers.

\subsection{$1/f$ noise}
\begin{figure}
\centering
\includegraphics[width=\columnwidth]{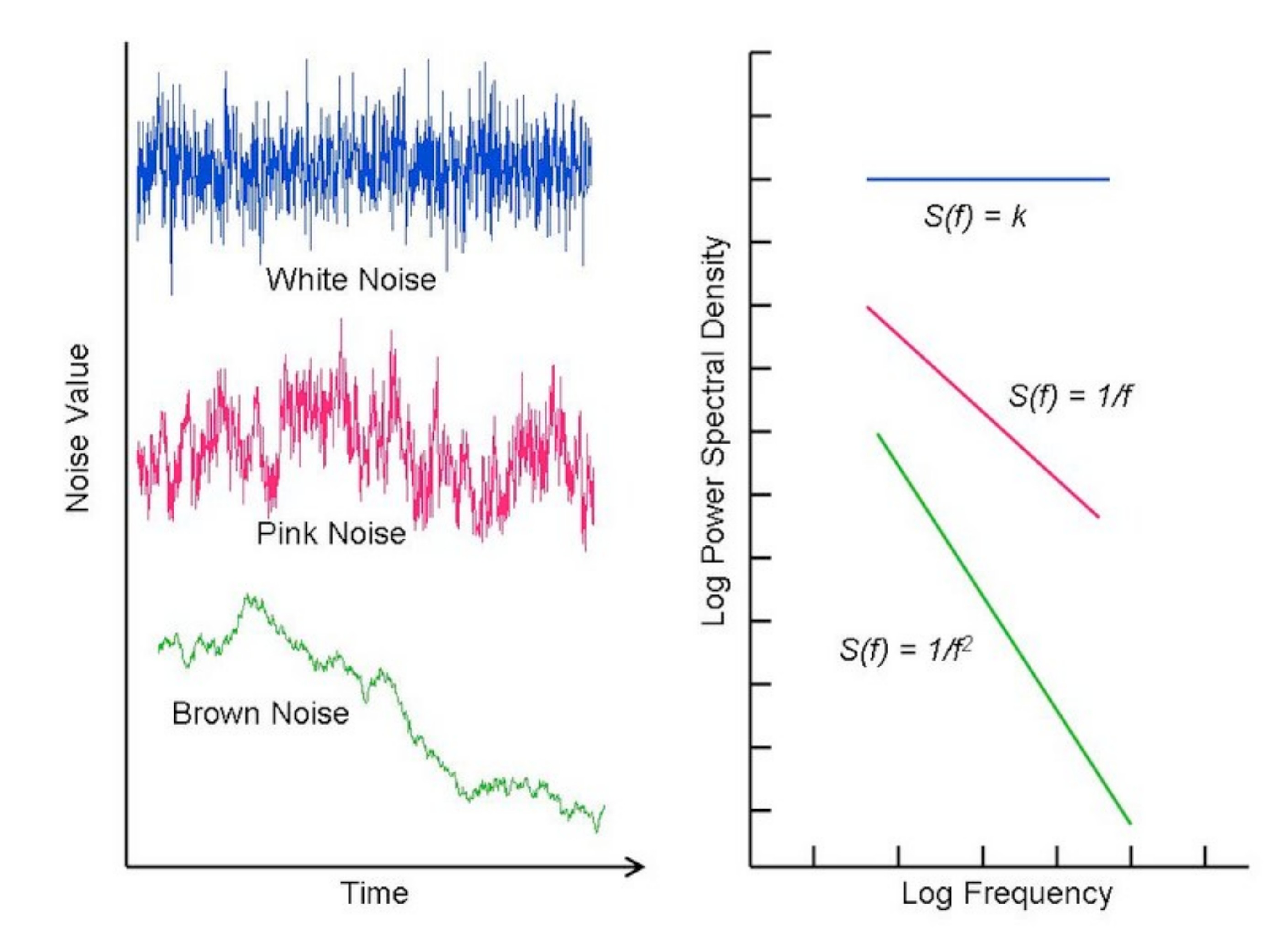}  
\caption{Left: color-coded realizations of time series of various noises. Right: respective power spectra of noises.}
\label{1_f}
\end{figure}

The main source of noise in the PACS data is the so called $1/f$ noise. The $1/f$  noise is ubiquitous in nature. It occurs in many physical, biological and economic systems. In physical systems it is present in some meteorological data series, the electromagnetic radiation output of some astronomical bodies, and in almost all electronic devices (referred to as flicker noise) such as the PACS detectors. The $1/f$ noise is an intermediate between the well understood white noise with no correlation in time and random walk (Brownian motion) noise with no correlation between increments (Fig. \ref{1_f}). The $1/f$ noise spectral density has the form:
\newline
$S(f)=1/f^{\alpha}$
\newline
where $f$ is the frequency, on an interval bounded away from both zero and infinity. The case of $\alpha =1$ , or pink noise, is both the canonical case, and the one of most interest, but the more general form, where $0 < \alpha < 3$, is sometimes referred to simply as $1/f$ noise. The noise power spectrum of the individual PACS timeline exhibits the behavior shown by the magenta line ($1/f$ behavior) in the right panel of Fig. \ref{1_f}. The $1/f$ noise of the PACS photometer, in fact, is roughly $\propto f^{-0.5}$ over the relevant frequencies. 

\subsection{Scan map data reduction}

The {\it{Herschel}} ground segment ({\it{Herschel}} Common Science System - HCSS) has been implemented using Java technology and written in a common effort by the {\it{Herschel}} Science Centre and the three instrument teams. {\it{Herschel}} data are organized in Levels depending on the stage of the data reduction. PACS Level 0 data corresponds to raw data. The data reduction up to Level 1 in the PACS scan map data processing consists only of masking, flagging, adding the instantaneous pointing obtained from the {\it{Herschel}} pointing product, flat fielding and flux calibrating. None of these data reduction steps affect the PACS noise or PSF. After Level 1 the PACS pipeline breaks in two branches to remove the $1/f$ noise in two alternative ways.  As already mentioned, the first method is using full `inversion' algorithms as widely applied by the cosmic microwave background community and the other method uses high-pass filtering of the detector timelines and subsequent direct projection, successfully adopted for Spitzer MIPS 70 or 160~$\mu$m data reductions. An algorithm of the first `inversion' type is available in the HCSS {\it{Herschel}} data processing in the form of an implementation and adaption to {\it{Herschel}} of a version of the MadMap code (Cantalupo et al. 2010). The alternative option is using high-pass filtering of the detector timelines and a direct `naive' map-making.  The former approach is more suited for extended sources since it preserves the extended emission, while it is less suited for point-sources because of algorithm artifacts in the region of bright point-sources and due to the low point-source sensitivity. In the high-pass filtering$+$`naive' map-making (photProject) approach, instead, the $1/f$ noise is removed by subtracting from each timeline the timeline filtered by a running box median filter of a given radius expressed in readouts. The presence of a source within the box boosts the median, thus, leading to the subtraction of part of the source flux (see Fig. 2). This can be overcome by properly ``masking'' the sources on the timeline to exclude the readouts corresponding to the source when estimating the median.  This approach is more suited for point-sources and it is largely used by most of the PACS KP and GT photometric surveys  such as PEP (Lutz et al. 2011), GOODS-{\it{Herschel}} (Elbaz et al. 2011), {\it{Herschel}} Lensing Survey (Egami et al. 2010), and H-Atlas (Eales et al. 2010). Therefore, we study the effect of this route of the pipeline on the PSF and final noise.

\begin{figure}[htp]
\begin{center}
\includegraphics[width=0.2\textwidth]{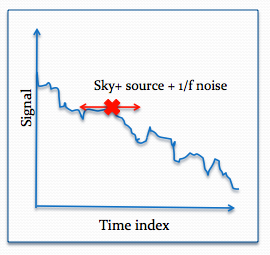}  \includegraphics[width=0.2\textwidth]{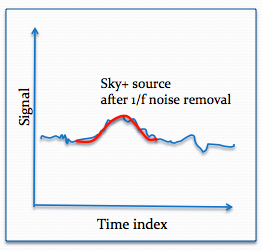}  \includegraphics[width=0.2\textwidth]{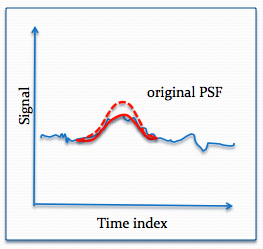} 
\caption{Cartoon of the effect of the high-pass filter on the source flux. The upper panels shows a PACS timeline with a source before and after the median removal. The bottom panel shows the difference between the observed (solid red curve) and the real PSF (dashed red curve) due to the flux loss caused by the high-pass filter. }
\end{center}
\end{figure}

\section{Simulating the timelines}

The $1/f$ noise is very difficult to simulate. Brownian motion is the integral of white noise, and integration of a signal increases the exponent $\alpha$  by 2 whereas the inverse operation of differentiation decreases it by 2. Therefore,  the $1/f$ noise can not be obtained by the simple procedure of integration or of differentiation of such convenient signals. Moreover, there are no simple, even linear stochastic differential equations generating signals with  noise. The widespread occurrence of signals exhibiting such behavior suggests that a generic mathematical explanation might exist. Except for some formal mathematical descriptions like fractional Brownian motion (half-integral of a white noise signal), however, no generally recognized physical explanation of  noise has been proposed. Consequently, simulating completely fake PACS timeline with realistic $1/f$  noise is problematic. To overcome this problem we use the real PACS timeline and add to that fake sources. In this way we do not need to generate artificially $1/f$ noise, telescope and sky background, but we can take advantage of the existing observations thus providing a more realistic simulation. The simulation consists of two main ingredients:
\begin{itemize}
\item [-] real PACS data reduced up to Level 1, that means flat-fielded, flux calibrated and astrometrically calibrated (coordinates of the detector reference pixels and position angle added to the Status table in the Frames class)
\item [-] input PACS PSF map
\end{itemize}

The steps of the simulation are as follows. For each PACS observation (individual AOR), we create a Frames with the same dimension and Status table of the input Frames but with an ``empty'' Signal data cube (set to 0.0 at all pixel and time coordinates). The input PSF of the given band is normalized and scaled to the desired total flux and ``back-projected'' into the timelines of the Frames at the desired sky coordinates. The ``back-projection'' is the inverse process of the projection performed by the PACS pipeline task photProject. After the back-projection the output Frames contains only the simulated sources. The Signal data cube of the output Frames can be simply added to the Signal data cube of the original observation to add the fake sources to the real timelines. However, before doing that, we must take into account the fact that, due to the ``bit-rounding'', the PACS signal is digitized (quantized), while the signal of the timelines containing only the fake sources is continuous. In order to add the fakes sources to the real time line, we need to digitized also their signal. We call this step ``digitization''.  We provide below a description of the photProject algorithm, of the back-projection' and of the digitization.

\subsection{photProject: drizzling method}

\begin{figure}
\includegraphics[width=\columnwidth]{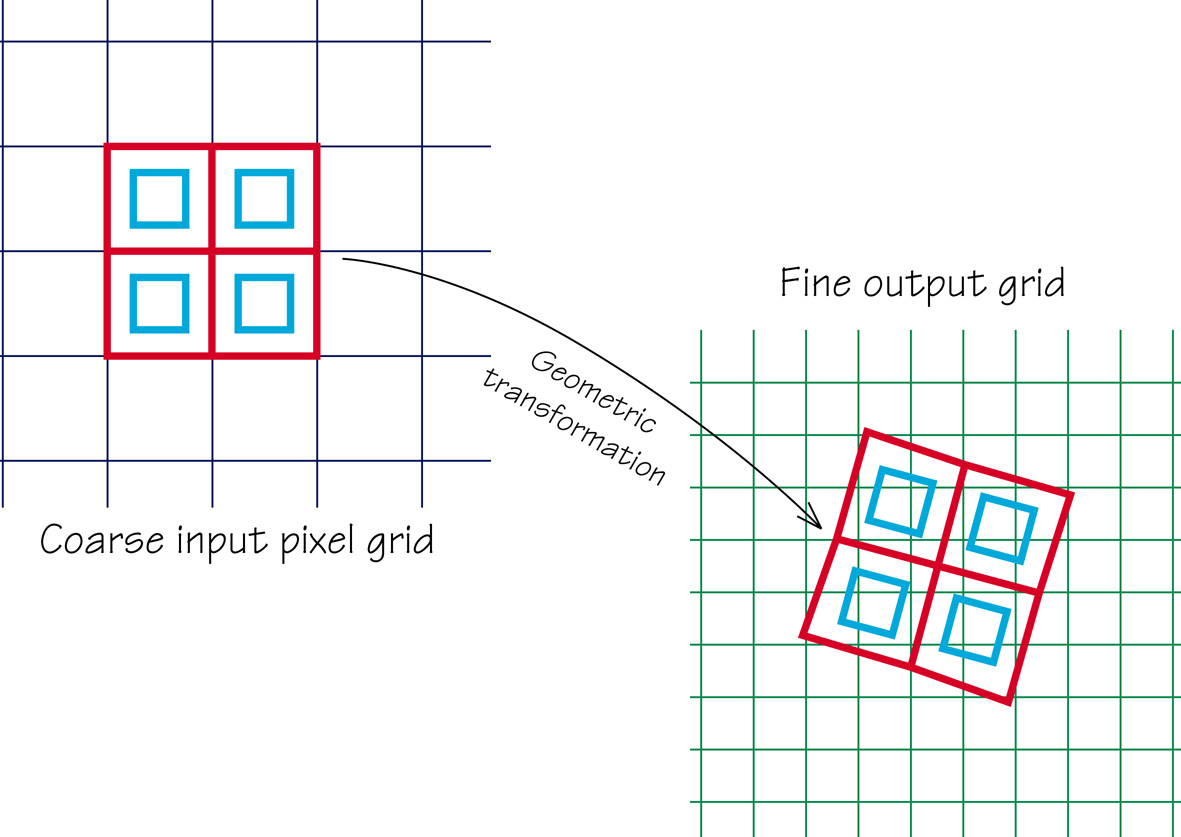}
\caption{Drizzle scheme: the input pixel is shrunken into a drop which contributes to an output pixel.}
\label{fig:projmed}
\end{figure}

The photProject task performs a simple coaddition of images, by using the drizzle method (Fruchter and Hook 2002). There is no particular treatment of the signal in terms of noise removal. The $1/f$ noise is supposed to be removed by the high-pass filtering task.  The drizzle algorithm is conceptually simple. Pixels in the original input images are mapped into pixels in the sub-sampled output image, taking into account shifts and rotations between images and the optical distortion of the camera. However, in order to avoid convolving the image with the large pixel "footprint" of the camera, the user is allowed to shrink the pixel before it is averaged into the output image, as shown in the Fig. \ref{fig:projmed}.


\begin{figure}
\begin{center}
\includegraphics[width=0.4\textwidth]{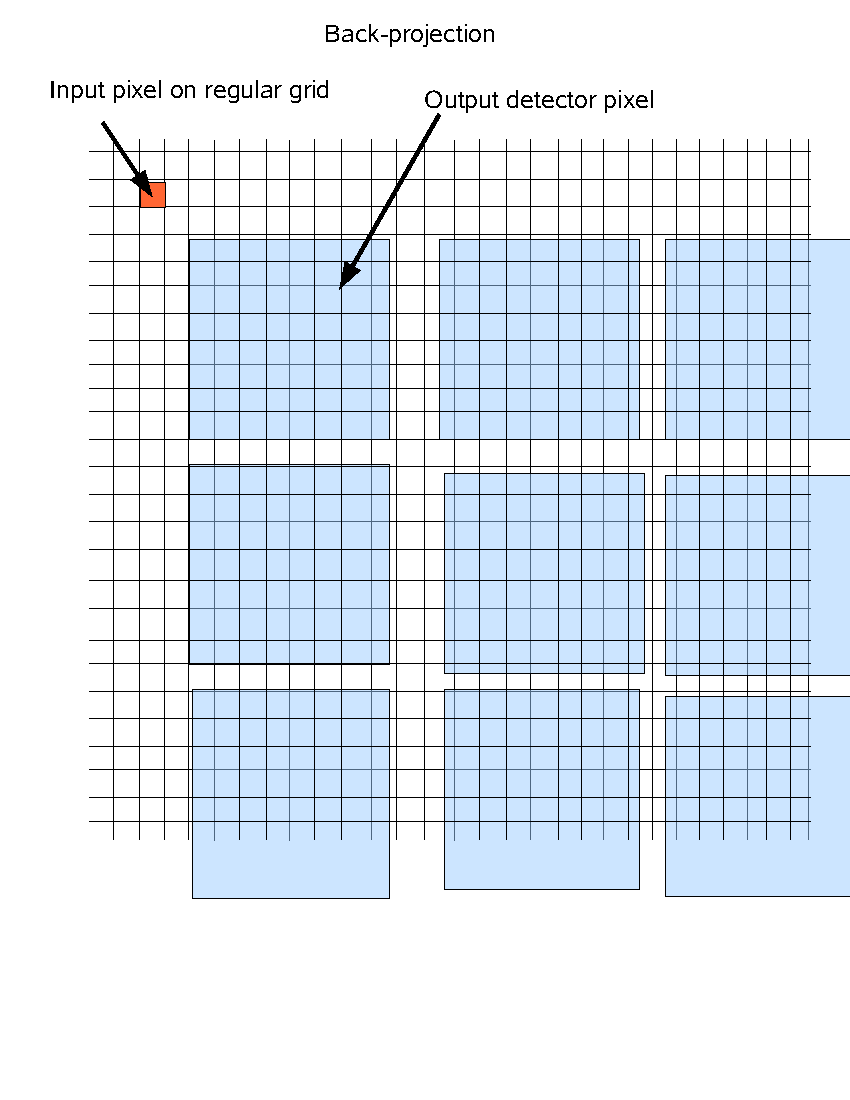}
\caption{Scheme for back-projection. The input regular grid is back-projected into the irregular PACS detector grid.}
\end{center}
\label{back}
\end{figure}

The new shrunken pixels, or "drops", rain down upon the sub-sampled output image. The flux in each drop is divided up among the overlapping output pixels in proportion to the areas of overlap (Fig. \ref{fig:projmed}). Note that if the drop size is sufficiently small not all output pixels have data added to them from each input image. One must therefore choose a drop size that is small enough to avoid degrading the image, but large enough that after all images are "dripped" the coverage is fairly uniform.  Due to the very high redundancy of PACS scan map data, even in the mini-map case, a very small drop size can be chosen (1/10 of the detector pixel size).  Indeed, a small drop size can help in reducing the cross correlated noise due to the projection itself (see for a quantitative treatment the appendix in Casertano et al. 2000). The size of the drop size is usually fixed through the ``pixfrac'' parameter, which gives ratio between the drop and the input pixel size.

\begin{figure*}
\begin{center}
\includegraphics[width=0.9\textwidth]{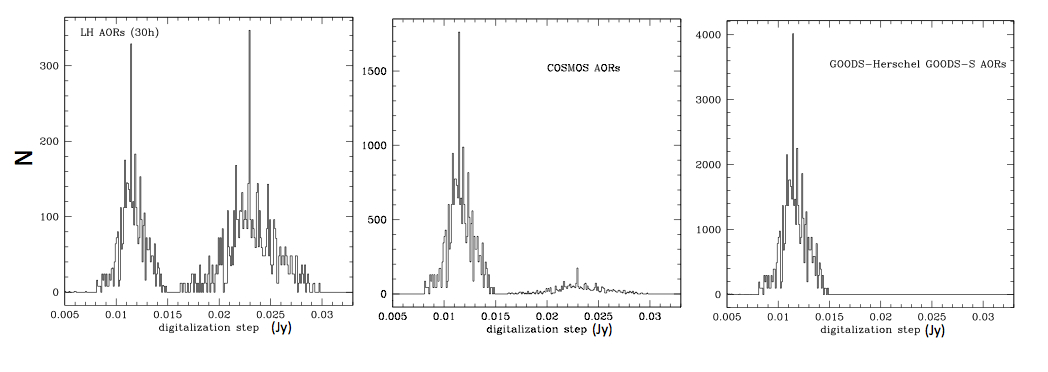} 
\caption{Distribution of the digitization steps for all timelines of the Lockman Hole, Cosmos and GOODS-S fields observation. The bi-modal distribution in the left and central panel is due to the fact that part of the observations were performed in 1-bit rounding (after OD173) and part in 2-bit rounding (before OD173). The GOODS-S filed has been observed later completely in 1-bit rounding.}
\label{digi}
\end{center}
\end{figure*}

The mathematical Formulation of the drizzling method is described below. The flux $I_{x^{'}y^{'}}$ of the output pixel $x^{'},y^{'}$ is given by:
\begin{equation}
I_{x^{'}y^{'}}=\frac{\sum\limits_{i=1}^{N_p} a_{xy}w_{xy}I_{xy} /A_{xy,x^{'}y^{'}} }{W_{x^{'}y^{'}} }
\end{equation}
and the coverage is given by:
\begin{equation}
W_{x^{'}y^{'}}=\sum\limits_{i=1}^{N_p}a_{xy}w_{xy}
\end{equation}
where:
\newline
\begin{itemize}
\item $I_{xy}$ = intensity of the projected input pixel (x,y)
\item $w_{xy}$ = weight of this pixel
\item $a_{xy}$ = fraction of the drop projected in a cell of the output grid (fractional pixel overlap $0 <  a < 1$)
\item $A_{xy,x^{'}y^{'}}$ = ratio of input and output pixel area for flux conservation
\item $I_{x^{'}y^{'}}$ = current intensity in the output pixel
\item $W_{x^{'}y^{'}}$ = current average weight in the output pixel
\end{itemize}

The weight $w_{xy}$ of the pixel can be zero if it is a bad pixel (hot pixels, dead pixels, cosmic rays event), or can be adjusted according to the local noise (the value is then inversely proportional to the variance maps of the input image).

The Drizzle algorithm has a number of desirable characteristics.
\begin{itemize}
\item It preserves surface photometry; one can measure flux density in a combined Frames with an aperture with a fixed number of pixels.
\item It is well suited to handle missing data due to cosmic ray hits and hot pixels.
\item It uses a linear weighting scheme which is statistically optimum among all linear combination schemes, when inverse variance maps are fed in as weights. These weights may even vary spatially in order to accommodate a possibly changing signal-to-noise ratio across different input Frames (e.g. due to variable scattered light).
\item It automatically produces an inverse variance map (the weight map) along with the combined output Frames.
\item It preserves resolution very well.
\item due to its weighting scheme, it largely eliminates the distortion of absolute photometry produced by the flat-fielding of the geometrically distorted images.
\end{itemize}

\begin{figure*}
\begin{center}
\includegraphics[width=1.0\textwidth]{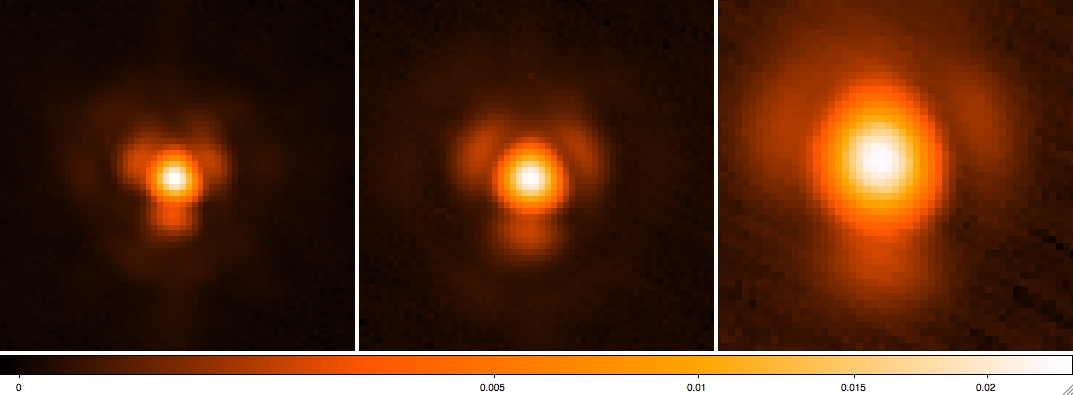}\\
\includegraphics[width=1.0\textwidth]{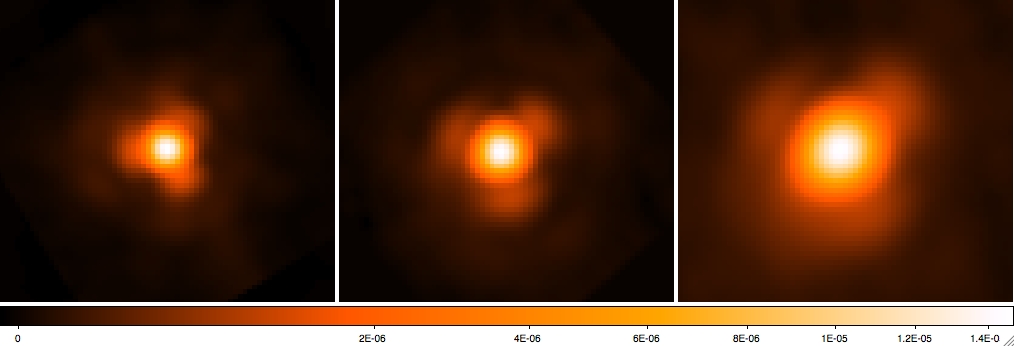}
\caption{Comparison of the PSF of the original
  Vesta PSF (upper panel) and the back-projected Vesta PSF (lower panel) at 70 (left panels), 100 (central panels) and 160 $\mu$m (right panels) at medium speed. The orientation of the original and back-projected PSF is different due to the different position angle.}
\label{or_back}
\end{center}
\end{figure*}

\subsection{The back-projection}

The back-projection is done by reversing the drizzle method. The ingredients needed to do that are:
\begin{itemize}
\item[-] an input map, possibly with very high spatial resolution
\item[-] an existing flux calibrated and astrometrically calibrated Frames
\end{itemize}

In the back-projection the map pixel is considered the input pixel and the PACS detector pixel is considered the output pixel. The IDL code used for this exercise performs the projection of a given input map over any individual image of the input (astrometrically calibrated) Frames, following the same scheme given above. A task in HIPE is available now ({\it{map2signalcube}} or {\it{mapindex2signalcube}}) to perform the back-projection, although it is not implementing the signal digitization (see next section). The main difference with respect to the drizzle method described above, is that the total flux of the PACS detector pixel is reconstructed by summing up the fluxes of all the contributing input map pixels, rescaled by their geometrical weights, $a_{xy}$. The contributing pixels are the ones with $a_{xy} > 0$ in the previous formalism. The flux of the contributing pixels is also rescaled by the ratio between input and output pixel area (see the sketch in Fig. 4) for flux conservation. The weight $w_{xy}$ of the previous formalism is set to 1 since all input pixels are considered good pixels. The pixfrac parameter is set to 1 since no drizzling can be used in this context. However, to mimic the drizzling method, it is preferable to use a input map with a pixel scale much smaller than the PACS pixel scale. In fact, this trick ensures that likely any input pixel contributes to just one PACS detector pixel (see a sketch in fig. 3).  This allows to minimize the cross-correlated noise and the smearing effect due to the projection as in the standard drizzling method.

The PACS data comes with a native drop which is the active area of the detector pixel ($640 \times 640$ $\mu m^2$ over a nominal area of $750 \times 750$ $\mu m^2$): the input map pixel have to be back-projected into the drop and not the nominal area of the PACS pixel. This is already taken into account in the PACS pipeline and in the IDL version of the PACS tasks used for this exercise.

\subsection{The digitization}
The PACS signal is quantized due to the supplementary on-board compression stage ``bit rounding''. Fig. \ref{digi} shows the distribution of the quantization step in Jy of all timelines in all AORs of the different blank field, the Lockman Hole and the COSMOS fields observed by the PEP consortium (PI. D. Lutz, Lutz et al. 2011) and the GOODS-S field observed by the GOODS-{\it{Herschel}} consortium (PI D. Elbaz, Elbaz et al. 2010). The two former fields show a bimodal distribution due to the change in the bit-rounding strategy from 2-bit-rounding to 1-bit-rounding during the winter of 2010 (OD173). The GOODS-S field was observed completely during the summer 2010 and shows a single peak distribution. This shows that the quantization step of the PACS signal can vary from 10 to 20 mJy depending on the bit rounding and so on the observation period.

In order to take into account the quantization of the PACS signal, we adopt the following approach:
\begin{itemize}
\item [1] before adding the original Signal data cube to the fake sources Signal, we transformed the original digitized signal into a continuous signal. This is done by assigning randomly to the signal S(x,y,t) an increment $\Delta S$ of the type $-\Delta_{br}/2 <  \Delta S < \Delta_{br}/2$, where $\Delta_{br}$ is the digitization step (due to Bit Rounding) of the PACS signal of the considered timeline.
\item [2] the fake sources Signal is then added to the continues Signal of the original Frames
\item [3] the new continuous Signal cube with added fake sources is then quantized again by using the same digitization step $\Delta_{br}$ of the original timelines.
\end{itemize}

This process fully preserves the original noise power spectrum since the re-quantization applied in the step 3 of the process provides exactly the same Signal of the input Frames if no artificial sources are added.

\section{The simulated PSF}

As input PSF for our investigation we use the Vesta PSF at 70, 100 and 160 $\mu$m, which are available in both medium and fast speed, as described in the PACS Technical Note PICC-ME-TN-033.  The Vesta PSF is ‘recentered’ by a posteriori correcting for pointing variations. By definition, the procedure produces small PSF images reducing the smearing effect produced the the pointing jitter.  A masked high-pass filter is used by masking a region centered on the source with 60 arcsec radius. A high-pass filter radius of 100 arcsec on sky is used. All maps were projected on a 1 arcsec pixel grid with a map position angle based on the spacecraft position angle at the time of the observation. To ensure a clearly defined zero point in flux, all PSFs are explicitly corrected to “Daophot”-background 0 in an annular aperture of width 10 arcsec just outside the masking radius used. All the data were taken in standard PACS prime scan mode. The blue/green PSF for parallel mode at 60 arcsec/sec and 20 arcsec/sec are partly simulated as described in the Technical Note. 

\begin{figure}
\begin{center}
\includegraphics[width=0.4\textwidth]{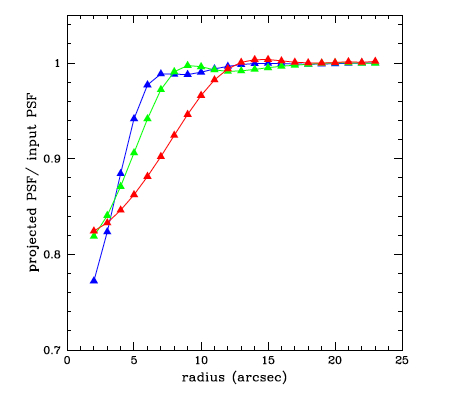} 
\caption{Comparison of the PSF profiles of the original Vesta PSF and the back-projected Vesta PSF in medium speed at 70 (blue sysmbols), 100 (green symbols) and 160 $\mu$m (red sysmbols). The comparison is done by showing the ratio of the curves of growth of the PSF. Due to the projection, part of the flux of the PSF core is spread towards the wings. The flux is completely recovered beyond 4-5 arcsec in the blue and green band and beyond 8-10 arcsec in the red band. The flux is conserved at $\sim$ 1-2\% level.}
\end{center}
\label{or_back_cg}
\end{figure}

\begin{figure*}
\begin{center}
\includegraphics[width=0.95\textwidth]{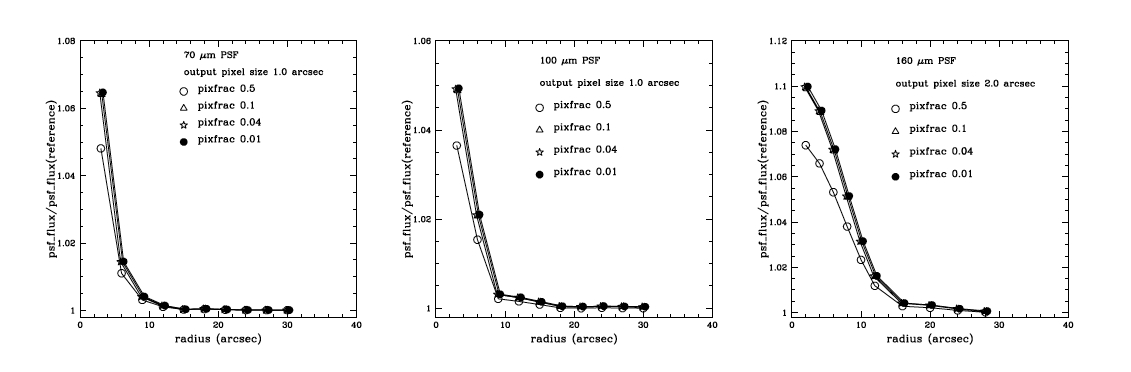}  
\caption{Effect of pixfrac in the PSF projection. The panels show the ratio of the curves of growth of the PSF obtained by varying the pixfrac value in the projection with respect to the case of pixfrac=1 (reference PSF) and at fixed output pixel size at 70 $\mu$m (left panel), 100 $\mu$m (central panel) and 160 $\mu$m (right panel) in medium speed.}
\label{cg_pro}
\end{center}
\end{figure*}

As explained in previous section, the PACS Vesta PSF observed in different bands and scan speeds have been back-projected, digitized and added to the timelines of a real observation taken in the same band and at the same scan speed.  A comparison between original and back-projected PSF in medium speed is shown in Fig. \ref{or_back}. The back-projected PSF is re-projected with the same projection parameters (output pixel scale and drop size) of the original PSF.  Fig. 7 shows the ratio of the curve of growth of the original and back-projected PSF of Fig. \ref{or_back}. The smearing effect due to the back-projection, which is a convolution of the observed PSF and the detector pixel and native drop, is visible within the core of the PSF.  The back-projected PSF has a less pronounced peak. The smearing effect reaches a maximum $\sim$ 25-20\% in all bands and it disappears below 4-6'' at 70 and 100 $\mu$m and 9-10'' at 160 $\mu$m. The same results is obtained for the PSF observed in fast speed.

The naive projection offered by the PACS pipeline through the photProject task is an implementation of the drizzling method. Thus, any observed PSF represents the convolution of the telescope PSF with the PACS pixel size and drop size- PRFs in Spitzer speak. Thus in order to estimate the amount of damage caused by the high-pass filtering removal to the PSF, we need first to disentangle the projection effects from the high-pass filtering effects.

\section{The effect of projection on the PACS PSF}

The shape of the observed PSF in a map obtained via photProject, after excluding the effect of the high pass filtering, is regulated by two parameters: the output pixel size and the drop size (see previous section for a detailed explanations about the meaning of these parameters). In order to determine this effect and to disentangle it from the high-pass filtering effect, we adopt the following approach. We back project the Vesta PSF into  the timeline of relatively short observations taken in the same band and scan speed without adding the signal of the original timelines. This creates  fake timelines that contains only the fake source without telescope and background noise. This allows us to project the fake timelines without applying the high-pass-filtering and to isolate the effect due to the projection itself.

The drop size in the drizzling method implemented by photProject is set by the ``pixfrac'' parameter and is expressed as the ratio between the drop size and the PACS detector pixel size. The default value of pixfrac is 1.  
Fig. 8 shows the effect of reducing the drop size with respect to the default value for the three bands in medium speed and for 3 different output pixel sizes (different symbols in the figures). The effect is shown as the ratio between the PSF curve of growths and the reference PSF curve of growth obtained with the same output pixel size and pixfrac$=$1. In all cases reducing the pixfrac affects only the core of the PSF within the inner 8-10''. A smaller pixfrac value makes the PSF more peaked with a maximum of 6, 5  and 10\% at 70, 100 and 160 $\mu$m respectively, for pixfrac values below 0.1.  Fig. 9  shows, instead, the effect of reducing the output pixel size at fixed pixfrac value at 160 $\mu$m. In this case the reference PSF curve of growth is the one with output pixel size of 2''. We compare to that the cases of 1.0 and 0.5'' output pixel sizes within the same apertures. The different symbols in the figure refer to different pixfrac values. In all cases reducing the output pixel size at fixed pixfrac affects the PSF core by making the PSF more peaked. The effect is negligible (only 2\%) between 2 and 1'' and more significant 6\% between 2 and 0.5''. 
The same quantitative results are obtained also for the fast speed case. We stress here that we analyze this effect as a relative effect between back-projected PSF. As a relative effect, it is independent on the input PSF shape.

In order to take into account these not negligible effects in the further analysis, we created a library of back-projected PSF for the values of high-pass filter width, output pixel size and pixfrac used in the next sections. This allows us to fully isolate the high-pass filtering effect from the projection effect.

\begin{figure}
\begin{center}
\includegraphics[width=0.4\textwidth]{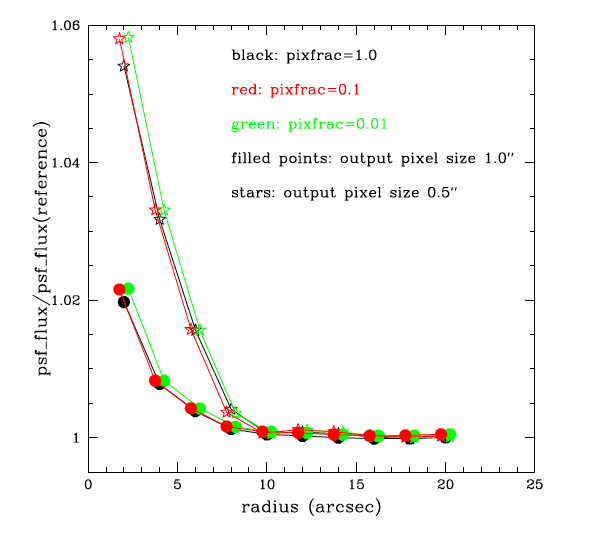} 
\caption{Effect of output pixel size in the PSF projection. The panel shows the ratio of the curves of growth of the PSF obtained by varying the output pixel size in the projection with respect to the case of an output pixel size of 2 arcsec (reference PSF) at fixed pixfrac at 100 $\mu$ in medium speed. The 1 arcsec (filled points) and 0.5 arcsec (stars) are considered. The color-coding is according to the pixfrac values.}
\end{center}
\label{or_back_cg1}
\end{figure}

\begin{figure*}
\begin{center}
\includegraphics[width=0.8\textwidth]{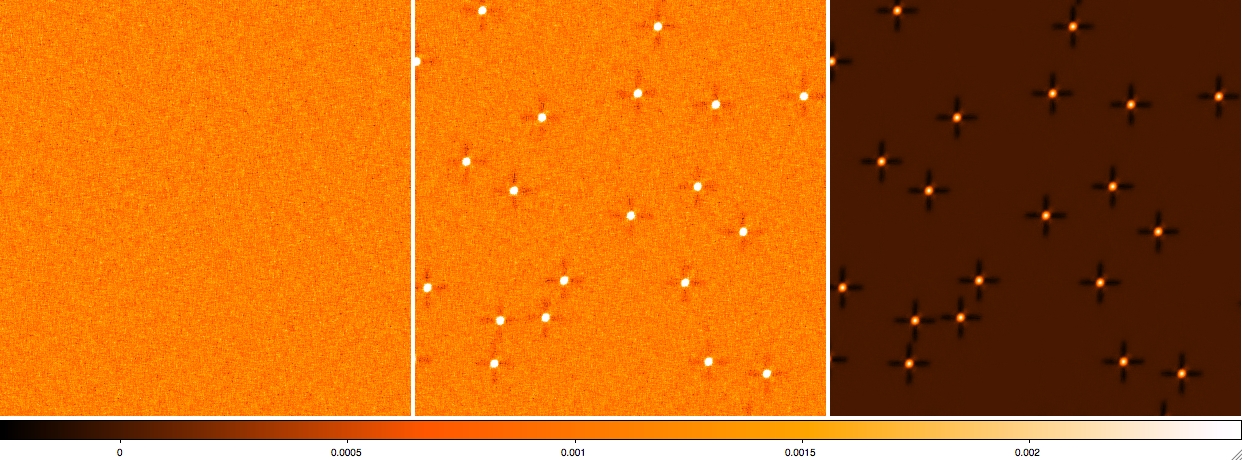}
\caption{Combination of scan and cross scan of the real 160 $\mu$ Lockman Hole observations (left panel). The same map ``polluted'' with bright fake sources is shown in the central panel. The residual map given by the subtraction of ``polluted'' and real maps (left panel) shows only the fake sources with the significant high pass filtering residuals and without background.}
\end{center}
\label{before_after}
\end{figure*}

\section{The effect of high-pass filtering on the PACS PSF}

\subsection{The method}
Blank fields observations of cosmological surveys are the ideal test cases where to add fake sources because they provide a sufficient large area to add a reasonable number of fake sources to test clustering effects and they contain, by design, almost only relatively faint point-sources well below the noise level of the individual timelines.  We identify this type of observation for each band and scan speed: at 70 $\mu$m medium speed the observation of the GOODS-S field of the PEP survey (PI D. Lutz), at 100 and 160 $\mu$m medium speed the observations of the Lockman Hole of the PEP survey, and at 100 and 160 $\mu$m fast speed (parallel mode) the observations of one field of the H-Atlas survey (PI S. Eales). We use two obsid, scan and cross scan, for each field in order to capture the effect of the high-pass filtering in both scan direction. We follow the following approach:

\begin{figure}
\begin{center}
\includegraphics[width=0.4\textwidth]{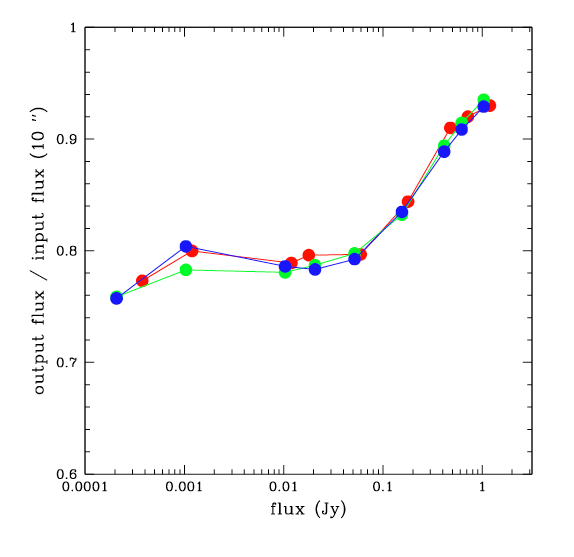}
\caption{Ratio of input and output flux estimated within an aperture of 6 arcsec at 70 and 100 $\mu$m and 10 arcsec at 160 $\mu$m as a function of the input flux. Blue points refer to the 70 $\mu$m fluxes, green points to the 100 $\mu$m fluxes and red points to the 160 $\mu$m fluxes. }
\end{center}
\label{primo_plot}
\end{figure}

\begin{itemize}

\item[-] a postage stamp of the Vesta PSF at the given wavelength and scan speed is back-projected at 40-60 different positions into the original timeline following the method described above. The positions are chosen in order to  avoid any overlap with existing PACS detections and with the position  of 24 $\mu$m priors (this was possible only for the GOODS-S and Lockman Hole fields).

\item[-] in order to test the dependence of the high-pass filtering on the clustering, the Vesta PSF is projected in relatively isolated and relatively crowded region of the maps.

\item[-] any simulation is performed by adding fake sources with the same flux levels. 

\item[-] the original timelines (without fake sources) are reduced in parallel with the same data reduction used for the simulated timelimes (with fake sources). This is done to subtract the original map from altered map in order to get rid of any background and noise. The result is a ``residual'' map that contains only fake sources showing the effect of the data reduction. This allows us to add fake sources at flux level much below the noise level of the final map since the S/N of the fake sources in the residual map (without noise)  is nearly flux independent.

\end{itemize}

Fig 10 shows an example of ``before- after'' the high-pass filter treatment. The left panel in the figure shows the real map without additional fake sources, the central panel shows the real map with the added fake sources and the right panel shows the residual map with only fake sources affected by the high-pass filtering effect. The negative residuals due to the removal of the running median along the scan directions are quite visible. In the following analysis we quantify the significance of these residuals as a function of the input flux and as a function of the masking strategy. The comparison is done in the following way:

\begin{itemize}

\item[-] the 40-60 fake sources are stacked in each residual map to obtained a mean PSF;

\item[-] the flux of the PSF is estimated by doing aperture photometry
  in several apertures with radius ranging from 2-4 to 22 arcsec and by PSF fitting with
PSF of different apertures.

\item[-] we estimate the flux loss by comparing the curve of growth of the PSF obtained from the residual map with the one of the back-projected PSF taken from our library and projected with the same output pixel size and pixfrac value. This allows us to take into account the projection effect and to isolate the high-pass filtering effect. Since the PSF have the same background, we do not remove any background. This allow us to avoid the uncertainties due to the background subtraction. 

\item[-] to estimate the effect neighbors on the high-pass filtering, we measure the flux of each fake source by doing either aperture photometry or PSF fitting. We, then, relate the output/input flux ratio to the number of neighbors along a column in the scan direction as long as the spatial length of the high-pass filter width (see Section 6.2.4 for a detailed explanation of the method).

\end{itemize}

\begin{figure*}
\begin{center}
\includegraphics[width=0.99\textwidth]{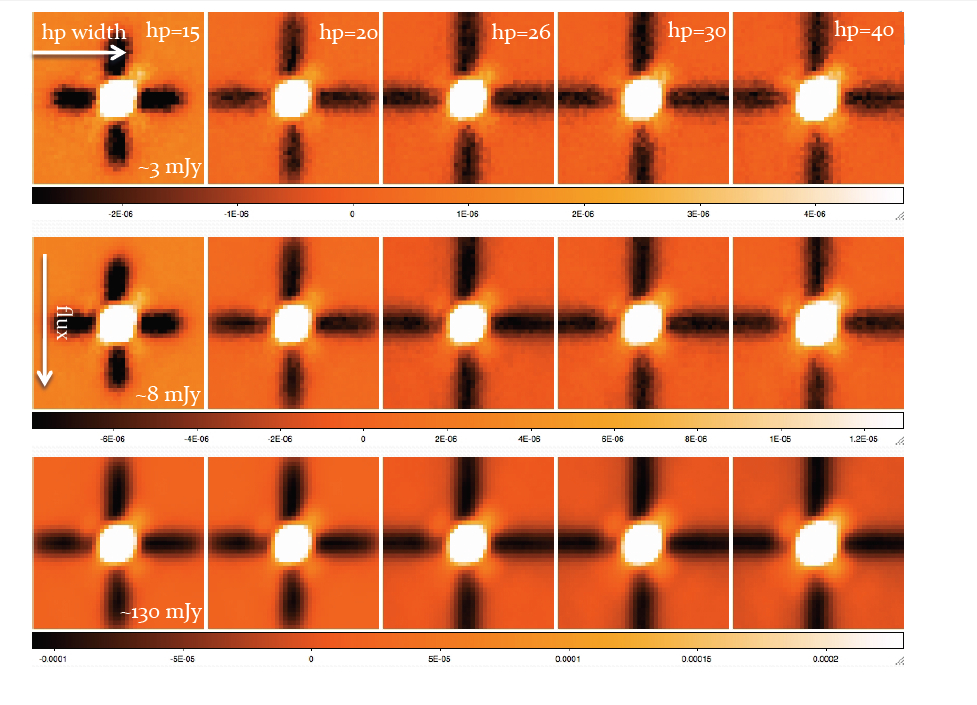}
\includegraphics[width=0.99\textwidth]{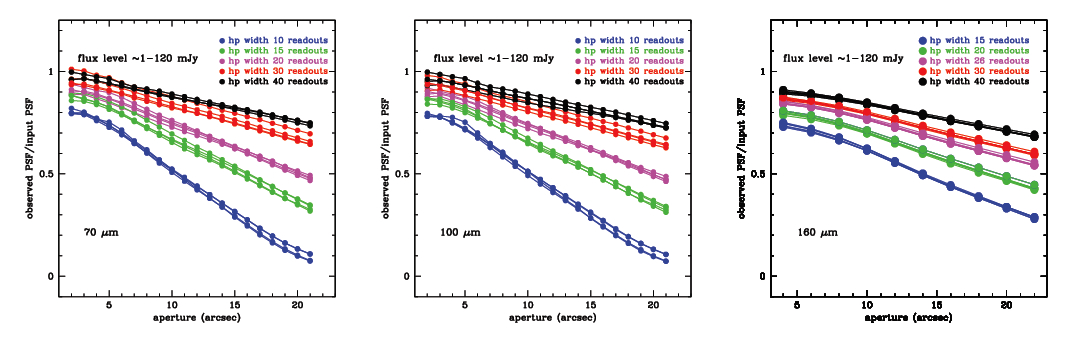}  
\caption{Upper panels: stacked PSF in the residual map as a function of flux and high-pass filter width as indicated in figure: The HPF width is increasing towards the right, the flux is increasing towards the bottom. Lower panels: ratio of the output/input curves of growth at 70 $\mu$m (left bottom panel), 100 $\mu$m (central bottom panel) and 160 $\mu$m (right bottom panel) in medium speed in the no-masking case. The two curves with the same color represent the minimum and maximum source fluxes.}
\label{cg_nomask}
\end{center}
\end{figure*}

\begin{figure*}
\begin{center}
\includegraphics[width=0.8\textwidth]{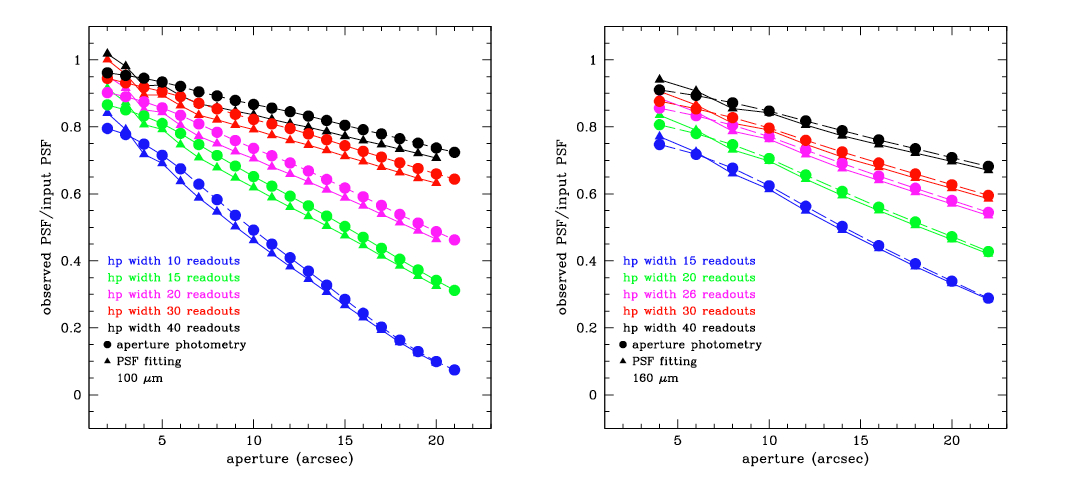}  
\caption{Ratio of the output/input curves of growth at 100 $\mu$m (left panel) and 160 $\mu$m (right panel) in medium speed and no-masking case via aperture photometry (filled points) and PSF fitting (filled triangles). The two methods agree within the uncertainties. The two curves with the same color represent the minimum and maximum source fluxes.}
\label{psf_fitting}
\end{center}
\end{figure*}

In order to ``protect'' the source flux from the damage of the high-pass filtering, a mask is often used to flag the timeline readouts corresponding to the sources and exclude them from the median estimate. We explore here also the effect/efficiency of three masking strategies:

\begin{itemize}

\item[-] no-masking strategy (stacking): fake sources are not masked at all. This exercise is done to show if there is a differential flux dependence of the high pass filtering effect. In addition, the result of this exercise gives a calibration of the amount of flux losses due to the high-pass filtering when doing stacking analysis on PACS maps, since the stacking is usually done for sources much below the detection limit and which are usually not masked.

\item[-] a S/N based mask, where the S/N is the one obtained in the final map. This strategy can not be generally calibrated because the S/N threshold is strictly related to the quality of the data reduction and the definition of the noise, which is not univocal. Thus, the results presented about this method and the threshold used can be taken only as indicative and cannot be considered as a calibration.

\item[-] a masking strategy based on putting circular patches around sources on the basis of a prior catalog. This strategy is, instead,  more ``objective'' since it does not rely on the quality of data reduction and consequent noise level definition but only on a prior  catalog. We analyze also the effect of different ``masking radius''  by putting patches of different size. When doing  so, we use the same approach also for the real sources. 

\end{itemize}

We point out here that the effect of the high-pass filtering is independent from the noise level of the final map and from the S/N level of the source in the final map, which in turn depends on the number of repeated AOR or repetition within the same AOR. The effect is local and it depends on the local S/N of the source on the individual timeline. Assuming that the main source of noise in the timeline is the $1/f$ noise and that the noise power spectrum is similar from observation to observation, the high-pass filter effect on the PSF depends only on the absolute flux of the source. This means that the results obtained here can be used as calibration to any other scan map data independently from the depth of the observation. 

\begin{figure*}
\begin{center}
\includegraphics[width=0.8\textwidth]{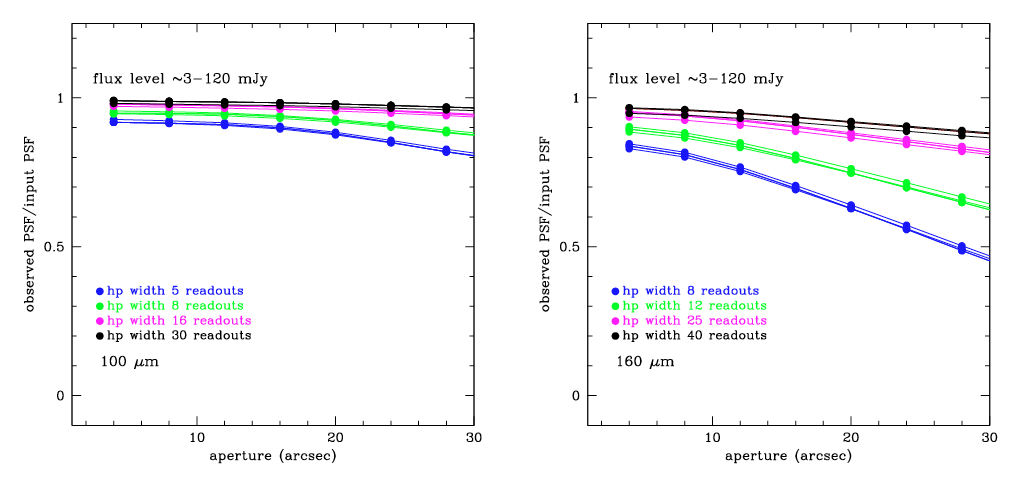}  
\caption{Ratio of the output/input curves of growth at 100 $\mu$m (left panel) and 160 $\mu$m (right panel) in parallel mode for different values of high-pass filtering values (color-coded in the figure) in the no-masking case. The two curves with the same color represent the minimum and maximum source fluxes.}
\label{cg_nomask_para}
\end{center}
\end{figure*}

This exercise is repeated for a combination of 5 values of high-pass filter width for two AOR (scan and cross scan) for each band and scan speed. We use the following parameter values

\begin{itemize}
\item{ 70 and 100 $\mu$m medium speed}

\begin{itemize}
\item high-pass filter width': 10, 15, 20, 30 and 40 readouts
\item output pixel size: 3.2 and 1.0 arcsec
\item pixfrac: 1.0, 0.1, 0.01
\item flux levels: 0.05,0.2,1, 10, 20, 50, 150, 400, 600, 1000 mJy
\end{itemize}

\item{ 160 $\mu$m medium speed}
\begin{itemize}
\item high-pass filter width': 15, 20, 25, 30 and 40 readouts
\item output pixel size: 6.4 and 2.0 arcsec
\item pixfrac: 1.0, 0.1, 0.01
\item flux levels: 0.05,0.2,1, 10, 20, 50, 150, 400, 600, 1000 mJy
\end{itemize}

\item{ 100 $\mu$m fast speed-parallel mode}

\begin{itemize}
\item high-pass filter width': 5, 8, 16 and 30 readouts
\item output pixel size: 3.2 and 2.0 arcsec
\item pixfrac: 1.0, 0.1, 0.01
\item flux levels: 3, 5, 15, 50, 150, 400 mJy
\end{itemize}

\item{ 160 $\mu$m fast speed-parallel mode}

\begin{itemize}
\item high-pass filter width': 8, 12, 25, and 40 readouts
\item output pixel size: 6.4 and 4.0 arcsec
\item pixfrac: 1.0, 0.1, 0.01
\item flux levels: 3, 5, 15, 50, 150, 400 mJy
\end{itemize}

\end{itemize}

\subsection{Results}

Here we analyze the results of our analysis per masking strategy.

\subsubsection{No-masking strategy}

We first analyze the effect of the high-pass filtering in case of no-masking. In order to understand the dependence of the effect on the source flux, we first investigate the relation between the percentage of flux loss versus the input source flux for a single case. We choose to use the high-pass filter widths suggested in the ``IPIPE'' script provided in the HIPE environment for the Deep Survey case of 15 readouts at 70 and 100 $\mu$m and 26 readouts at 160 $\mu$m in the medium speed case. We estimate the percentage of flux loss as the ratio between the flux of the PSF in the residual map and the projected PSF with no high-pass filtering effect, estimated within 6 $arcsec$ at 70 and 100 $\mu$m and 10 $arcsec$ at 160 $\mu$m. Fig. 11 shows the dependence of the perceptual flux loss as a function of the input total PSF flux for the three PACS bands at medium speed. The effect of the high-pass filtering is nearly independent in the range between 0.5-150 mJy. Beyond 150 mJy we see a flux dependence for which the higher the flux the lower the perceptual flux loss. Below 0.5 mJy, well within or close to the confusion limits of the three bands, the flux loss seems to be higher with respect to the 0.5-150 mJy range. Thus, in the following analysis we treat separately the 0.5-150 mJy and the $<0.5$ mJy case. The region of bright fluxes ($> 150$ mJy) is not treated here since it refers to sources bright enough to be detectable even in a single mini scan map and, thus, easy to mask.

Fig. 12 shows the qualitative and quantitative effect of the high pass filtering on the PACS PSF in the 0.5-150 mJy flux range. The upper panel shows the PSF at 160 $\mu$m on the residual maps  at different flux levels and for different HPF widths. The lower panel of the figure shows, instead, the quantitative effect expressed by  the ratio between the curve of growth of the PSF affected by the high-pass filtering effect and the unaffected PSF. The results, obtained via aperture photometry, are shown for the three PACS bands in medium speed, for different flux levels and for different HPF widths. As seen in previous figure, the relative high-pass filtering effect is nearly flux independent for any high filter width used in this flux range. The second evidence is that the smaller the high-pass filtering width, the higher the percentage of flux loss. The third evidence is that  most of the flux loss is due to the removal of the flux of the PSF lobes, e.g. at large radii. Indeed, the percentage of flux loss is much higher in the PSF outer region than in the core.  The results are confirmed also if the fluxes within different apertures are estimated via PSF fitting (see Fig. 13, the PSF fitting is performed by using IDL IAS Stacking Library, $\rm{http://www.ias.u-psud.fr/irgalaxies/}$).

The same result is confirmed in the fast speed parallel mode case although the PSF damage has a much lower significance than in the medium speed case. Fig. 14 shows the results for the 100 and 160 $\mu$m case. The level of damage is nearly constant at 100 $\mu$m at any aperture and it is never higher than $\sim$13\% (smallest high-pass filter width). The 160 $\mu$m case is more similar to the medium speed case. The effect is more significant at larger aperture. The differential effect between extreme flux levels is on average of the order of few percent, thus, confirming the flux independency of the high-pass filtering in the considered flux range even at fast speed in parallel mode.

\begin{figure}
\centerline{\includegraphics[width=0.4\textwidth]{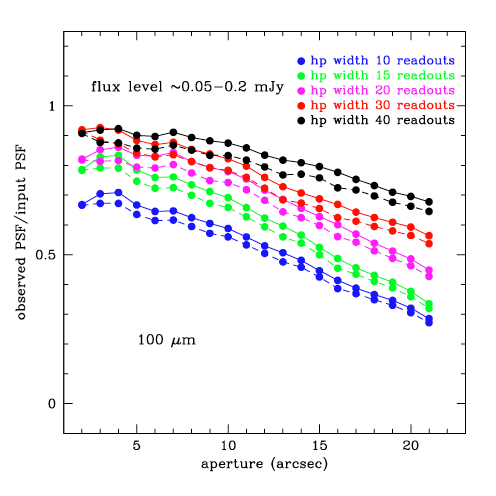}}
\centerline{\includegraphics[width=0.4\textwidth]{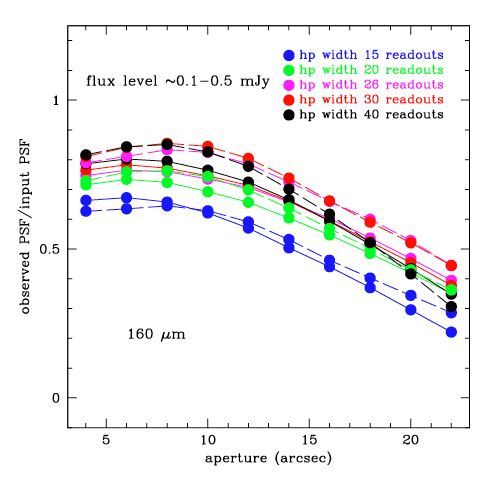}}
\caption{Ratio of the output/input curves of growth at 100 $\mu$m (upper panel) and 160 $\mu$m (bottom panel) in medium speed for different values of high-pass filtering values at a flux level below the confusion noise: $0.05-0.1$ mJy at 100 $\mu$m and $0.1-0.5$ at 160 $\mu$m. The 70$\mu$m case is not shown since it is fully consistent with the 100 $\mu$m case. The two curves with the same color represent the minimum and maximum source fluxes.}
\label{cn}
\end{figure}

We further check also how much the perceptual flux loss is higher at flux lower than 0.5 mJy. For this purpose, we performed the same simulation and analysis on fake sources at 0.2 and 0.05 mJy in each band. Fig. 15 shows the ratio of the curve of growth for the green and the red case. The blue band is consistent with the green band case. The percentage of flux loss is higher than at higher flux level at the PSF core but it reaches the same level beyond 10 $arcsec$ in each band . This effect is likely due to the presence of many faint sources within the confusion level.

\begin{figure}
\centerline{\includegraphics[width=0.4\textwidth]{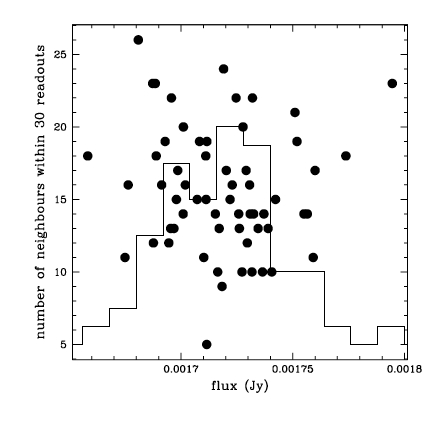}}
\caption{Number of neighbors within the high-pass filtering width of 30 readouts along the scan direction versus the fake source output flux estimated via aperture photometry on the residual map, within an aperture of 10 arcsec at 160 $\mu$m. The rescaled histogram of the flux distribution is over-plotted.}
\label{clust}
\end{figure}

The HPF effect is due to the fact that the flux source itself enhances the median estimated within the HPF width and causes then the flux removal. However, the presence of neighbor sources within the HPF width can in principle increase the effect. In order to take into account the effect of close neighbors entering the high-pass filter width, we estimate the density of neighbors around each source in the following way. Since the GOODS-S and LH map used for our simulations have a different depth and the corresponding PACS catalog have different detection limit, we use the 24 $\mu$m catalogs available for these fields with a cut at the same flux density of 50 $\mu$Jy. We estimate the number of neighbors around each fake source as the density of 24 $\mu$m sources brighter than this cut within a radius equal to the HPF width and within a column along the scan direction. The width of the column is defined by eye as the region where the high-pass filtering effect is well visible in the residual map. We do this estimate for any considered value of the HPF width and at any flux level. The flux of the individual fake sources is measured  through aperture photometry within an aperture of 8 arcsec in the blue and green and within 10 arcsec in the red.  We, then, perform a Spearman test to the flux - number of neighbors relation to look for possible correlations. If the presence of neighbors affect the estimate of the median in the high-pass filter and thus the source flux, we should observe an anti-correlation between the number of neighbors and the source flux. As shown in the example of Fig. 16,  we do not find any significant anti-correlation between the two considered quantities at any flux level and at any value of the HPF width. In all cases the probability of correlation is lower than $10^{-6}$. In addition the dispersion around the mean flux at any flux level and for any HPF value is very small, varying from 1\% at the larger HPF width to 2\% at the smallest HPF width. Thus the presence of neighbors along the scan direction does not affect on average the source flux in the high-pass filtering, although we can not exclude that a more significant effect should be observed in case of blended sources that we do not simulate in the present analysis.

\subsubsection{S/N based masking}

In this masking strategy all sources above a given S/N threshold are masked. The efficiency of this masking strategy depends on the amount of noise of the map and how it is estimated. Thus it depends on the data reduction quality, on the definition of the noise level of the final map and on the flux level of the source considered. Since it only works for detections, this masking strategy can be tested only on sources above the detection level of the considered  maps. 

We follow the same strategy described in the PACS Photometer Interactive Pipeline (IPIPE) scripts (the Deep Extragalactic map or the Bright Point Source scientific cases contained in the HIPE {\it{pipeline}} menu). The map is smoothed with a Gaussian filter. This allows to smooth individual noisy pixels that would  otherwise be masked out. The noise is defined as the standard deviation of the signal over the region with coverage higher than the median coverage. This ensures that the noisy region with very low coverage at the map edges is excluded from the noise estimate, otherwise the noise would be artificially too high and not representative of the map region of interest. We lower the S/N threshold until all visible or detected sources in the real map are masked out. We use this threshold for all the maps with the fake sources added with different flux level. Since we are using only two AOR (scan and cross scan), this strategy allows us to mask fully only the very bright sources. For relatively low faint sources above the detection threshold, only few central pixels are masked. For sources below the detection threshold, the flux loss is the same as for the no-masking strategy since those sources are undetected and, thus, can not be masked with a S/N masking strategy. 

\begin{figure}[h]
\centerline{\includegraphics[width=0.4\textwidth]{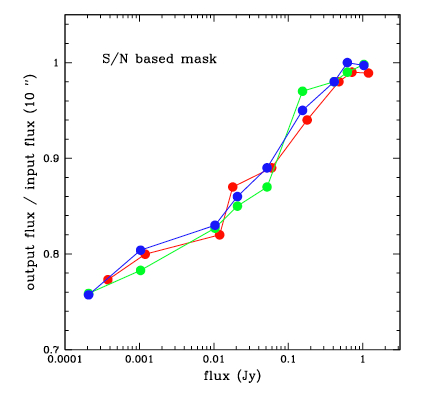}}
\caption{Ratio of input and output flux estimated within an aperture of 10 arcsec as a function of the input flux in the case of a S/N based masking strategy. Blue points refer to the 70 $\mu$m fluxes, green points to the 100 $\mu$m fluxes and red points to the 160 $\mu$m fluxes.}
\label{s_n}
\end{figure}

Fig. 17 shows the ratio between output and input flux estimated via aperture photometry within an aperture of 10 $arcsec$ in each band. The difference between Fig. 17 and the similar Fig. 11 obtained without masking the sources is clear. The effect of masking on the basis of the S/N causes a flux dependence also in the region  at fluxes below 150 mJy, where there was no dependence in Fig. 11. This is due to the fact that the extension of the masked region per source depends on the number of source pixels above the noise threshold chosen. Thus, the higher the source flux, the more extended the masked region. 

We can not provide a calibration for this kind of masking strategy, since it depends on how the noise is estimated and which threshold is chosen for masking.

\subsubsection{Circular patch masking}

This method relies only on an input coordinates catalog. The efficiency of this masking strategy depends on the quality of the prior catalog and on the size of the circular patches. In this exercise the prior catalog contains the known positions of the fake sources. In the reality it can be a preliminary extracted catalog or a catalog based on a different band (e.g. Spitzer MIPS 24 $\mu$m catalog). Since the idea is to mask all sources in the same region independently of the flux, this method provides results consistent with the ones shown in the Fig. 11 obtained in the case of no-masking. The same region of flux independency is maintained although at a different level of flux loss percentage.

As for the no-masking case, we make a separate analysis for the two different regions above and below the 150 mJy limit. Sources below this threshold are treated in the following way. In any band we analyze the impact of different patch sizes: 4, 6 and 8 arcsec at 70 and 100 $\mu$m and 4, 6 and 10 arcsec at 160 $\mu$m. Fig. 18 shows which portion of the PSF is masked within the three concentric apertures: the largest aperture captures all the flux of the PSF core which counts more than 90\% of the Encircled Energy Distribution (EED), but it does not cover the PSF lobes. Fig. 19 shows the ratio of output/input curves of growth for the different patch radii at 70 and 100 $\mu$m and at 160 $\mu$m for medium and fast speed. The results are shown only for the two extreme flux levels ($\sim$1-100 mJy at 70-100 $\mu$m and $\sim$ 1.5 and 100 mJy at 160 $\mu$m) for the usual five different values of high-pass filter widths. For sources above the 150 mJy limit, we use a masking strategy similar to the one used for the data reduction of the PACS primary calibrators, which is a masking radius of 20 and 60 $arcsec$. The analysis leads to the following results:

\begin{itemize}

\item[-] as for the former cases, the larger the HPF width, the better the input PSF is preserved

\item[-] the larger the circular patch size, the smaller the  difference between small and large HPF widths. In the 10 arcsec circular patch mask, there is a lack of flux at maximum of 10\%  within a 8 (10) arcsec aperture for any HPF width in the medium speed case and at maximum 15\% in the fast speed case for the smallest HPF width values. The effect increases to a maximum of $\sim$ 20\% with a small 4 arcsec patch size.

\item[-] the advantage of this method is that it masks at all flux level the same region of EED, thus, at a constant fraction of flux. This preserves the flux independency of the high-pass filtering effect already observed in the no-masking case. As show in Fig. 19 the difference of the curves of growth between the two extreme flux levels is marginal. The ratio of the curves of growth of intermediate flux levels are lying between these two extreme cases and for this reason are not drawn. 

\item[-] the difference in flux suppression between the two extreme  flux levels, though marginal, is increasing in the 160 $\mu$m case at smaller circular  patch sizes: is about 1\% in the case of a 10 arcsec size and 5\% at  4 arcsec size. Instead, it is nearly constant in the 70 and 100 $\mu$m case.

\item[-] for very bright sources, the use of a mask patch of 20 $arcsec$ is sufficient to fully cover the PSF lobes and fully retrieve at any aperture the input flux. Indeed, a fluctuation of 1-2\% around the ratio 1 between output and input flux is observed at any flux level (400-1000 mJy) considered in the experiment.

\item[-] consistent results are obtained also in parallel mode at fast speed and by using a PSF fitting technique instead of aperture photometry for estimating the flux (as shown already for the no-masking case in Fig. 13).

\end{itemize}

\begin{figure}[h]
\centerline{\includegraphics[width=0.4\textwidth]{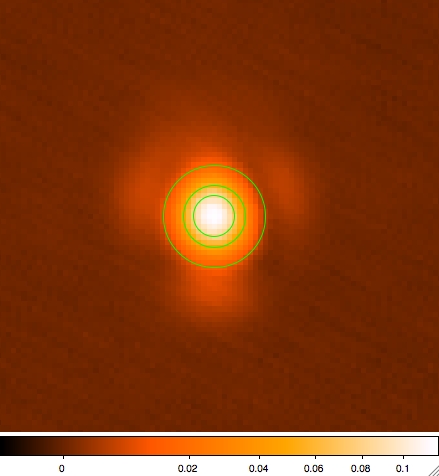}}
\caption{Original Vesta PSF at 160 $\mu$m with superposition of three
  different masking radii (green circles) used for the analysis: 10, 6
  and 4 arcsec.}
\label{rad}
\end{figure}

\begin{figure*}
\begin{center}
\includegraphics[width=0.97\textwidth]{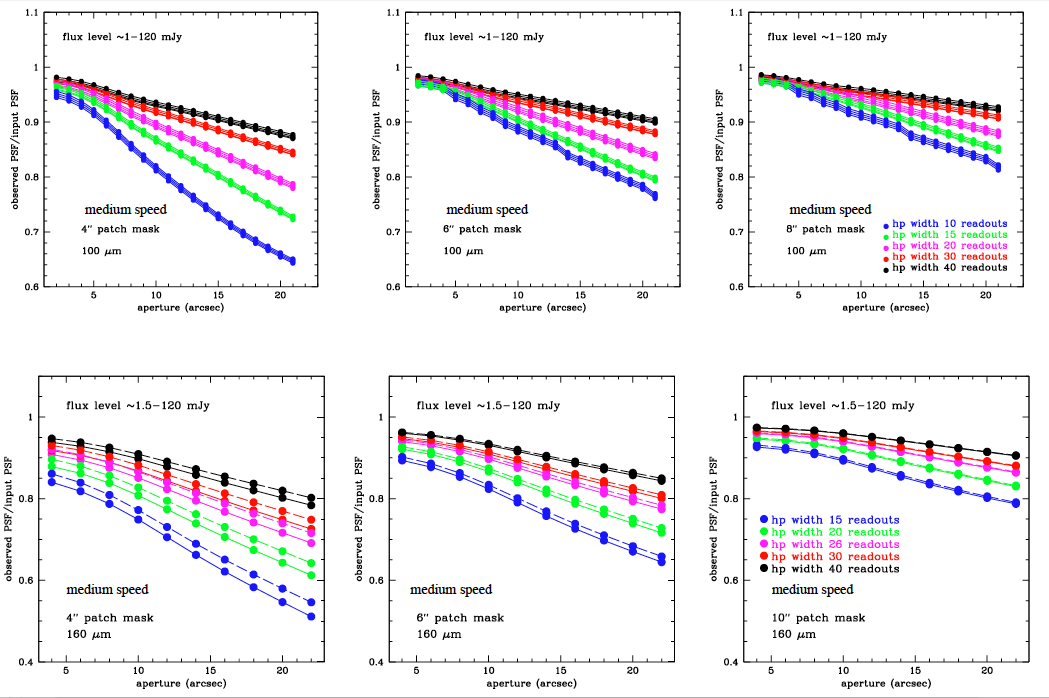} \includegraphics[width=0.99\textwidth]{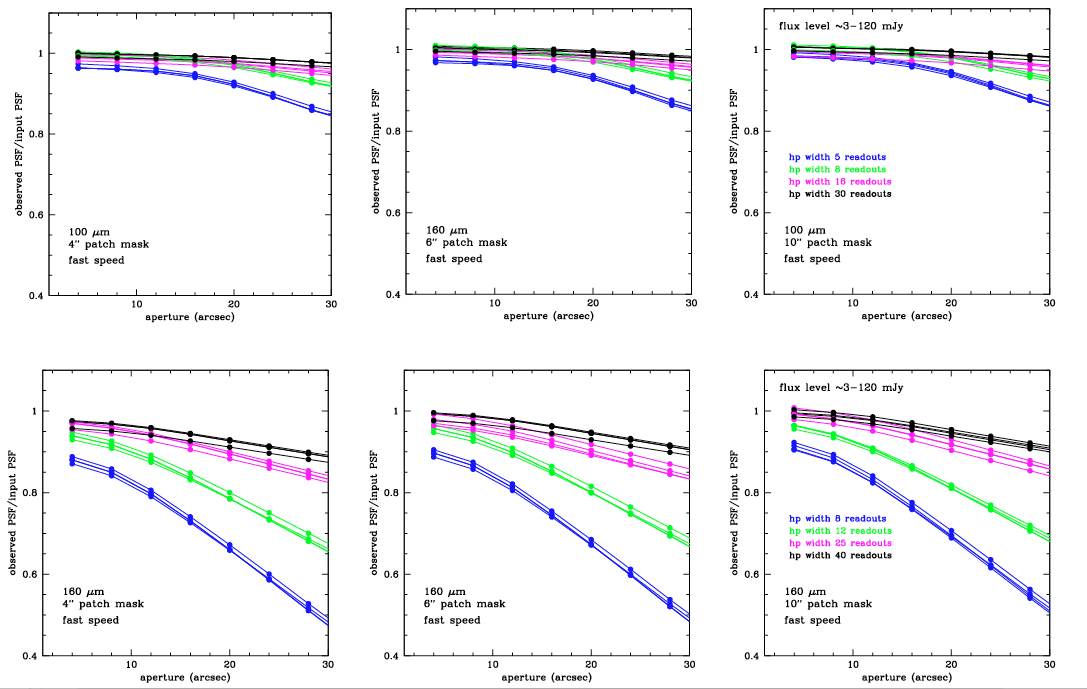}
\caption{Ratio of output/input curves of growth for the circular patch masking strategy at 100 $\mu$m and 160 $\mu$m for medium and fast speed for different circular patch radii. The two curves with the same color represent the minimum and maximum source fluxes.}
\label{mask}
\end{center}
\end{figure*}

\subsubsection{The clustering effect}

In this section we consider a particular effect which affects all flux estimates based on the stacking analysis. This analysis is used to bit the noise by stacking many postage stamps of undetected sources in a PACS maps, centered at the source position provided by a prior catalog. The stacked postage stamp provides a mean PSF, and , thus, an estimate of the mean PACS flux of the considered source population. In this technique one has to consider the so called ``clustering effect''. This effect causes a broadening of the  PSF and an enhancement of the mean flux due to the contribution of many neighbors to the outer regions of the stacked PSF. Thus, the more clustered the sources of the considered map, the more significant the broadening of the PSF and the enhancement of the flux of the stacked image. The aim of this paragraph is not to provide a calibration of this effect, which depends on the clustering properties of the considered galaxy population, but to show how the high-pass filter effect can affect the PSF shape and thus the clustering effect. Indeed, the main effect of the median removal is to remove most of the flux in the PSF outer regions (see lower panel of Fig. 12). This should reduce the contribution of neighbor sources to the outer region of the stacked image. To check this point we modify our simulation in the following way:

\begin{itemize}
\item after the step 1 of the method (see section 3.3) we add an additional step. This step consists in taking the Fast Fourier Transform (FFT) of the real timelines, randomizing the phase of the FFT of each timeline and re-obtaining a new timeline by taking the inverse FFT. Randomizing the phase of the FFT of the timelines has the effect of reshuffling the signal along the timeline without affecting the noise properties since the Noise Power Spectrum remains unchanged. The new timelines have, thus, the same noise properties of the original ones but no sources.
\item instead of adding the usual 60 sources, we add to the timeline fake sources at the positions of the Spitzer 24 $\mu$m prior catalog used for the neighbor effect analysis (see section 6.2.1). 
\item the Signal cube with fake sources is then digitized as in the previous analysis and the data are reduced following the same procedure.
\item as done for the previous analysis, the sources are added all at the same flux level. A full calibration of the clustering effect would require to add sources at different fluxes according to a given luminosity function, but this is beyond the scope of this experiment. Indeed the aim of the this exercise is just to show how much the median removal can reduce the clustering effect. 
\item we create two PSFs: one obtained by stacking the $\sim$ 1500 clustered sources affected by the high-pass filtering effect and one obtained by stacking the same sources just projected and without data reduction effect. We compare each PSF with the PSF (reference) taken from the library described in section 5, at the corresponding output pixel size and pixfrac value, not affected by either high-pass filtering effect or clustering effect.
\end{itemize}

\begin{figure}
\centerline{\includegraphics[width=0.4\textwidth]{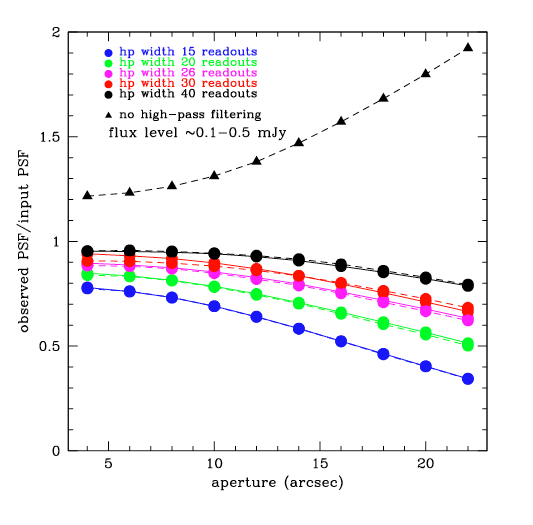}}
\caption{The filled triangles show the ratio of curves of growth of the PSF affected by clustering effect and no high-pass filtering effect and the reference PSF. The filled points (color-codes as explained in the figure)  show the ratio of the PSF affected by both clustering and high-pass filtering effect with respect to the reference PSF. The two curves with the same color represent the minimum and maximum source fluxes.}
\label{clust1}
\end{figure}

Fig. 20 shows the ratio of the curves of growth of the two mentioned PSF with respect to the reference PSF. The stars shows the ratio between the PSF affected by the clustering effect and not by high-pass filtering effect. The enhancement of the flux is present within any aperture but it is particularly significant in the outer region of the PSF. The result is confirmed by doing either aperture photometry or PSF fitting. The filled points, color coded as explained in the figure, are the ratio of the curves of growth affected by both high-pass filtering and clustering effects. The clustering effects is remarkably reduced at any high-pass filtering width used. The smaller the high-pass filtering width, the higher the flux removal in the PSF lobes and thus the lower the flux enhancement caused by clustering.  Thus, the results points out that in any simulation or analysis of the clustering effect in PACS maps, it is necessary to account for the effect of the high-pass filter on the PSF.

\begin{figure}
\centerline{\includegraphics[width=0.4\textwidth]{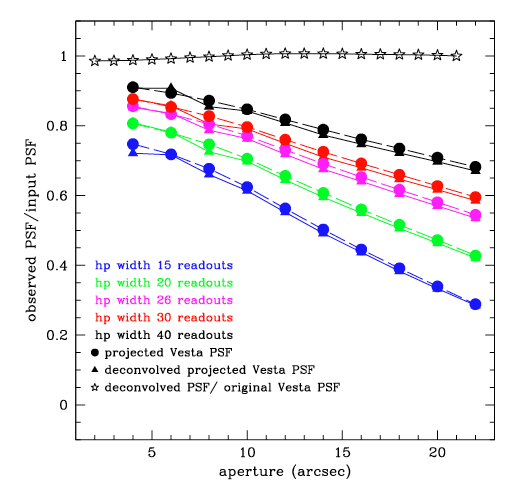}}
\caption{Ratio of the curves of growth (empty stars) between the original Vesta PSF at 160 $\mu$m and the projection of the parent PSF, $\mathcal{P}_{\textrm{parent}}$, projected with the same projection parameters used for obtaining the Vesta PSF. The filled triangles show the ratio of the curves of growth of the projected $\mathcal{P}_{\textrm{parent}}$ with and without high-pass filtering effect. The ratios are consistent with the ones obtained by using the Vesta PSF as in previous results.}
\label{trimmed}
\end{figure}

\subsection{How important is the input PSF}

As seen above, when introducing synthetic sources onto the timeline 
and producing full maps, the PSF extracted from the final product 
differs from the original input Vesta PSF. 

In order to test if these projection and map-making effect could affect the
estimate of flux losses due to high-pass filtering, we have proceeded to produce
a {\em parent} PSF which, once introduced onto the timeline and processed in the
usual manner, will resemble the original Vesta.

The problem to be solved consist into a double de-convolution or 
reconstruction in Fourier space. 

Given the original Vesta PSF (simply called {\em Vesta} or 
$\mathcal{P}_1$ here) and the result of projection and map making (called
$\mathcal{P}_2$ from now on), we first need to find a convolution
kernel (called $\mathcal{K}$) that transforms $\mathcal{P}_1$ into 
$\mathcal{P}_2$ by means of a convolution:

\begin{equation}
\mathcal{P}_1\ast\mathcal{K} = \mathcal{P}_2
\end{equation}

Given the kernel $\mathcal{K}$, we would then like to find a new parent PSF
$\mathcal{P}_{\textrm{parent}}$, such that:

\begin{equation}
\mathcal{P}_{\textrm{parent}}\ast\mathcal{K} = \mathcal{P}_1\textrm{.}
\end{equation}

Both $\mathcal{K}$ and $\mathcal{P}_{\textrm{parent}}$ can be derived using the 
the algorithm developed independently by Lucy (1974,  Astron. J. 79, 745) 
and Richardson (1972, J. Opt. Soc. Am.  62, 55). We adopted the
IRAF\footnote{IRAF is distributed by the National Optical Astronomy Observatory,
which is operated by the Association of Universities for Research in Astronomy
(AURA) under cooperative agreement with the National Science Foundation.} 
incarnation of this algorithm, implemented in the {\tt lucy} task, included in
the {\tt stsdas} package. We defer to
the IRAF/{\tt lucy} documentation\footnote{the documentation of the {\tt lucy}
task in the IRAF environment is available in the command line help, or at 
http://stsdas.stsci.edu/cgi-bin/gethelp.cgi?lucy} for further details.

In the first step, {\tt lucy} is used with $\mathcal{P}_2$ and $\mathcal{P}_1$
as inputs, with the aim to produce $\mathcal{K}$. In the second step, {\tt lucy}
is used with $\mathcal{P}_1$ and $\mathcal{K}$ as inputs, with the aim to
produce $\mathcal{P}_{\textrm{parent}}$.

Fig. 21 shows the ratio (stars) between the original Vesta PSF at 160 $\mu$m and the projection of the parent PSF, $\mathcal{P}_{\textrm{parent}}$, projected with the same projection parameters used for obtaining the Vesta PSF. The ratio of the fluxes is only 1-2\% below 1 in the inner 10 $arcsec$ instead of 20\% as observed in Fig. \ref{or_back_cg}. Nevertheless, even using this more realistic representation of the Vesta PSF does not change at all the results obtained so far, as shown by the consistency of the ratio of the curves of growth obtained with the projected Vesta PSF and the projected parent PSF $\mathcal{P}_{\textrm{parent}}$. Indeed, the inner 8-10 arcsec of the Vesta PSF at 160 $\mu$m or the inner 4-5 arcsec at 70 and 100 $\mu$m are under-sampled by the PACS detectors, since they affect just 1 or 2 readouts, which unlikely affect significantly the estimate of the median within the box chosen in the high-pass filter task. Thus, we conclude that the precise shape of the core of the input PSF chosen for this experiment  does not have a role in determining  the effect of the high-pass filtering.

\section{Suggestions for a proper source extraction}
Whatever combination of masking strategy and parameter setting is adopted to reduce the PACS data, the final extracted source fluxes must be corrected for the flux loss due to the high-pass filtering effect. The results described above show that this correction is practicable only in the case of a circular patch masking strategy. Indeed, only in this case the extracted source fluxes might be corrected by a multiplicative correction factor which depends only on the data reduction parameter settings (high-pass filter width, drop size and output pixel size, circular patch radius) and which is independent on the source flux over a very large flux range (see Fig. 11). This correction is not doable in the S/N masking strategy because the correction factor is flux dependent (Fig. 17) and it is strongly affected by the way the S/N is estimated and thus it is calibratable only case by case and per flux bin.

On the assumption that the PACS user performs the data reduction by following the IPIPE scripts and that the user adopts the option of a circular patch masking strategy, we suggest the following steps to retrieve the total final flux of the extracted sources:
\begin{itemize}
\item  after setting PACS filter and circular patch radius, the user should identify which panel in Fig. 19 is appropriate for her/his case 
\item in case of {\it{aperture photometry}} the user must measure the source flux, $f_{ap}$, within an aperture radius, $r_{ap}$, using the favorite aperture photometry tool

\item in case of {\it{PSF fitting photometry}} the user must measure the source flux, $f_{ap}$, by using a PSF reduced in the same way as the data (same data reduction parameter setting and circular patch radius) and, thus, exhibiting the same distortions/losses/residues as the reduced data. This can be achieved by constructing the PSF from the real map or by reducing the publicly available PACS observations of the Vesta PSF with the same technique used for the real data. As common, the PSF is assumed to be cut at some radius, $r_{ap}$, and normalized to total 1 inside that radius

\item the user must, then, identify in the appropriate panel the correction factor $corr_{HPF}(r_{ap})={\rm{(observed flux/input flux)}}$ for the adopted values of high-pass filter width and aperture $r_{ap}$
\item determine from the encircled energy curves released with PICC-ME-TN-033, the aperture correction $corr_{ap}(r_{ap})$
\item estimate the total final flux as: $flux_{true}=f_{ap}/(corr_{HPF}(r_{ap})*corr_{ap}(r_{ap}))$
\end{itemize}

\section{The PACS map noise components}

Two different components can be used to describe the noise in a PACS xmap: the noise per pixel and the cross-correlated noise. The former is given by the error map and it should scale as the inverse of the square root of the observing time or the coverage, if the confusion noise is not reached. The latter depends on the mapping itself and on other sources of cross-correlation such as the $1/f$ noise left after running the HighPass filter task. In principle the noise per pixel should be estimated via error propagation. However, the high-pass filter task completely changes the noise power spectrum of the PACS timeline, thus making it impossible to propagate the errors. To overcome this problem we use two different methods to estimate the error map depending on the data redundancy. We also provide here a way to estimate and take into account the cross-correlated noise.

\subsection{The high redundancy case}
\subsubsection{The noise per pixel}
In the high redundancy case we follow the method outlines in Lutz et al. (2011). In many PACS observations, the same part of the sky is repeatedly observed in a large number of AORs or in few AORs with a large repetition number. This allows to obtain several maps of the same field either by reducing individually each AOR or by splitting the AOR in chunks by repetition number or by odd and even scan legs, to increase the number of individual maps. If all the maps, then, are mapped into the same World Coordinate System (WCS), the final map can be obtained by (weighted) averaging the flux of each output pixel. In the same way, the error of each pixel can be obtained as the error of the (weighted) mean of the flux distribution of each output pixel. This procedure allows to overcome any issue related by the noise propagation. Since the coverage of an individual pixel can vary map by map depending of the number of pixels removed due to flagging (bad detector pixels, saturated pixels, glitches), it is preferable to estimate the final map as a mean weighted by the coverage. In this case the error of each pixel is estimated as the error of a weighted mean as follows:

\begin{equation}
\sigma{^2}=\frac{1}{1+\sum{\omega_i^2}}\sum_{i=1}^{N} {\omega_i (x_i-{\mu})^2}
\end{equation}

where N, $\omega_i$ and $x_i$  are, respectively, the number of repeated observations,  the coverage  and the flux of the $i^{th}$ map pixel, and $\mu$ is the weighted mean. It is clear that the higher the number of observations, the better the mean is known, thus the smaller the noise per pixel. Indeed, the left panel of Fig. 22 shows the tight anti-correlation in the log-log space between the noise per pixel and its final coverage, which is proportional to the number (N) of repeated observations. The relation is tight, with a dispersion of 16-18\% in medium speed, and, as expected, the noise per pixel scales as the inverse of the square root of the coverage.  The Fig. 22 shows also how the different setting of high-pass filter width (central panel) and projection parameter (right panel) affects the normalization of the relation but not its slope.

\begin{figure*}
\begin{center}
\includegraphics[width=0.9\textwidth]{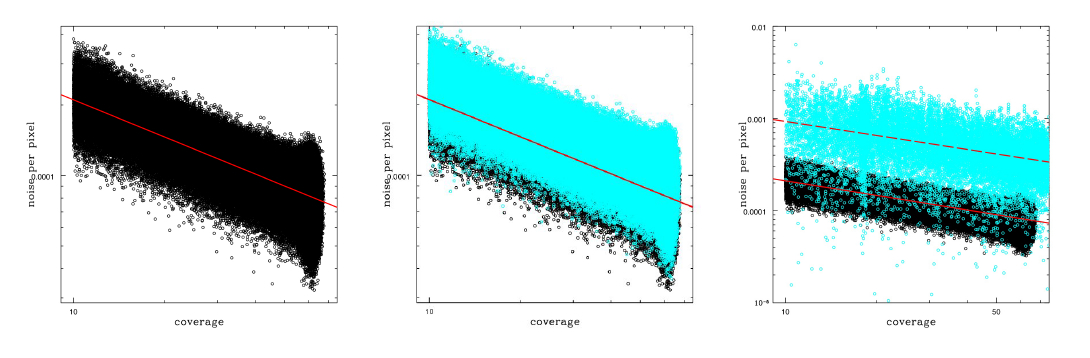}  
\caption{Examples of the tight anti-correlation of the noise per pixel versus the pixel coverage (left plot). The central panel shows the same anti-correlation  for two different values of the high-pass filter width (cyan and black points). The right panel shows the anti-correlation for two different values of the pixfrac parameter (cyan and black).}
\label{npp}
\end{center}
\end{figure*}

\subsubsection{Cross-correlated noise}
There are two sources of correlation in the pixel noise: the projection itself and the residual $1/f$ not removed by the high-pass filtering task, in particular along the scan direction.  The nature of the cross-correlated noise due to the projection in the drizzle method, implemented in photProject, is nicely explained in Fruchter \& Hook (2002). In particular, in the case of the PACS scan map, one can approximate that the dither pattern is entirely uniform and continuously fills the output plane. In this case, the ratio between the total noise (uncorrelated and cross correlated) and  un-correlated noise can be retrieved by two simple formulas:
\begin{equation}
f = \left\{ \begin{array}{ll}
\frac{r}{1-\frac{1}{3r}} & \textrm{for}  r\ge 1\\
\frac{1}{1-\frac{r}{3}} & \textrm{for}   r\le1
\end{array} \right. \label{}
\end{equation}

where $r$ is the ratio between drop and output pixel size. In addition to this source of cross correlated noise, one has to consider also the contribution of the residual $1/f$ noise not completely removed by the high-pass filtering task. Due to this contribution, the final cross-correlated noise of a map can not be simply retrieved via formulas as in the previous case but it must be estimated from the data itself. In case of high redundancy data this is feasible with the following procedure. We consider the series of individual AOR maps used to create the final (weighted) mean map. Following Lutz et al. (2011), a correlation map can be constructed by considering series of pairs of pixels in each individual map with the same relative distance from a reference pixel. We call $\Delta$f the flux deviation of an output pixel in a AOR map with respect to the (weighted) mean flux of the final coadded map. We estimate $\Delta$f for a large number of pairs taken in a square grid of $n \times n$ positions in each AOR map. The reference pixel is the one at the center of the grid. We estimate, then, the correlation coefficient, $\rho(i,j)$, between the $i^{th}$ and the $j^{th}$ pixels in the following way:

\begin{figure}
\centerline{\includegraphics[width=0.4\textwidth]{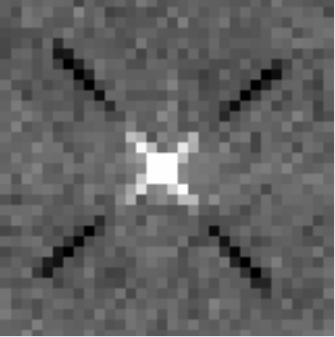}}
\caption{Example of cross-correlation matrix. The negative cross-correlation due to the high-pass filtering along the scan direction is visible.}
\label{corri}
\end{figure}

\begin{equation} 
\rho(i,j)=\frac{\sum{\Delta f_1 \Delta f_2}}{\sqrt{\sum{(\Delta f_1)^2}} \sqrt{\sum{(\Delta f_2)^2}}} 
\end{equation}

The correlation coefficients are stored in a correlation map with $n \times n$ dimension (see Fig. 23 for an example of cross correlation matrix). The total flux of a given source can be expressed as the sum 
\begin{equation}
g(x_1..x_n) = \sum_{k=1}^n{a_k \cdot x_k}
\end{equation} 
where $a_k$ is the weight of the pixel $k^{th}$ and $x_k$ is its flux. The weight $a_k$ can be expressed by the PSF in case of PSF fitting or by the fractional area of the contributing pixel in case of aperture photometry. According to error propagation, the total noise of such sum can be expressed by the formula:

\begin{equation}
 \sigma_g^2 = \displaystyle\sum_{k,l=1}^{n}{a_k \sigma_k \cdot a_l \sigma_l \cdot \rho(k,l)}
\end{equation} 

where $\sigma_k$ and $\sigma_l$ are the errors and $\rho(k,l)$ is the correlation coefficient of the pixels $k^{th}$ and $i^{th}$. In case of no correlation, the correlation matrix is diagonal and $\rho(k,l)=1$ if $k=l$ and is 0 elsewhere. In this case the uncorrelated noise coincides with the total noise and it is given by the sum in quadrature of the error of the individual pixels. In the assumption that the errors per pixel are uniform, $\sigma_k  \sim \sigma_l $, across the area of interest (the PSF or the aperture), as it is reasonable in the case of PACS maps, the ratio between total noise and purely uncorrelated noise is expressed by the formula:
\begin{equation}
f^2 = \frac{\displaystyle\sum\limits_{i,j}{a_i a_j \rho(i,j)}}{\displaystyle\sum\limits_{k}{a_k^2}}
\end{equation}
The ratio $f$ can be used to correct the purely uncorrelated noise to retrieve the total noise for a given flux estimate. As mentioned above, the amount of correction depends on how the flux is measured (PSF fitting or aperture photometry) and the area used for the estimate (see Eq. 7). Measures of the $f$ ratio in 30 different regions of a large PACS map in a regular grid of $30\times 30$ pixels, show that $f$ varies only by 3\% across the map. This allows to measure a mean $f$ for a given map and to use it to correct any uncorrelated noise associated to a flux estimate.

\begin{figure}
\centerline{\includegraphics[width=0.4\textwidth]{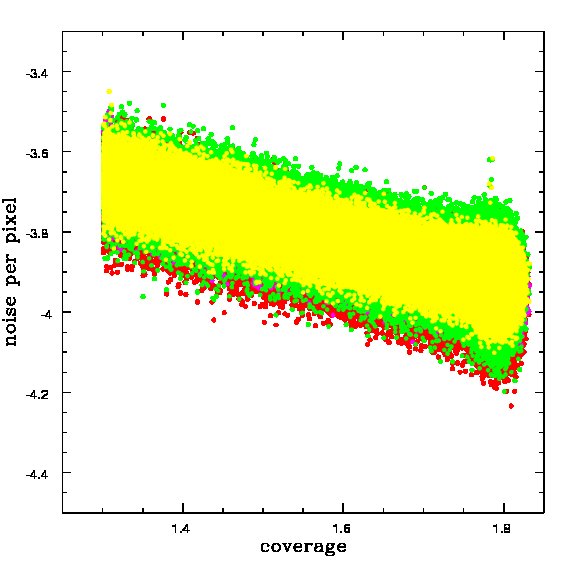}}
\caption{Example of noise per pixel and pixel coverage anti-correlations obtained by splitting the individual AORs in several chunks by repetition and odd and even scan legs.}
\label{diff_calib}
\end{figure}

\subsection{The low redundancy case}

In case of low redundancy the previous method can not be applied. To overcome this problem, we propose here a calibration of the two noise components, noise per pixel and cross-correlated noise, based on a set of data with high redundancy and as a function of different kinds of data reduction. The calibration is performed as follows:

\begin{itemize}
\item[-] the observations of several blank fields with high data redundancy (more than 20 AOR, each with repetition number $N \ge 2$ in medium speed and 5 AOR with repetition $N=2$ in parallel mode) are reduced following the interactive (IPIPE) scripts provided in the HIPE environment for the PACS Photometer. 
\item[-] the data reduction is performed for different values of high-pass filter width, output pixel size and pixfrac in a rather large parameter space
\item[-] given the tight relation between coverage map and error map (noise per pixel), as shown in previous section, a calibration of the best fit relation between coverage and error per pixel is provided for the parameter space considered
\item[-] the cross-correlation matrix is estimated for all the considered cases
\item[-] this library of matrices is used to estimate the cross-correlation correction factor as a function of different PSF shapes and apertures
\item[-] fitting functions are, then, estimated to retrieve the best calibration parameters for any given settings of high pass filter width, output pixel size and pixfrac value.
\end{itemize}

\begin{figure*}
\centering
\includegraphics[width=0.9\textwidth]{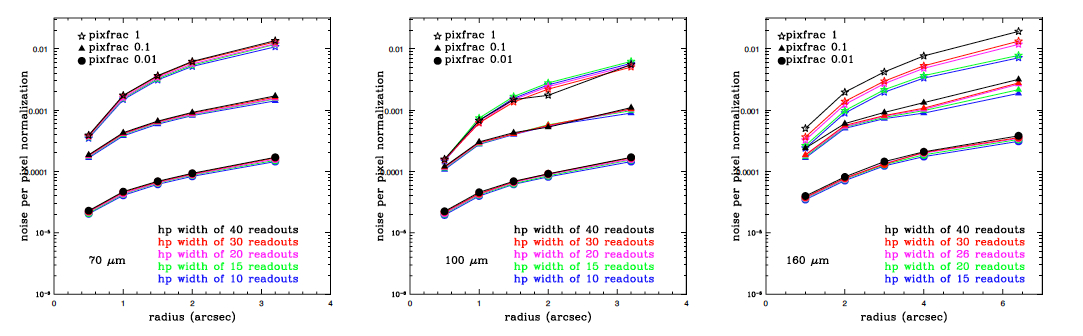}  
\caption{Relation between the intercept of the best-fit noise-coverage anti-correlation as a function of the output pixel size and pixfrac values for different high-pass filter widths at 70 $\mu$m (left panel), 100 $\mu$m (central panel) and 160 $\mu$m (right panel)}
\label{intercept}
\end{figure*}

We considered the following parameter space at 70 and 100 $\mu$m in medium speed:

\begin{itemize}
\item high-pass filter width: 10, 15, 20, 30 and 40 readouts
\item output pixel size: 3.2, 2.0, 1.5, 1.0 and 0.5''
\item pixfrac: 1.0, 0.5, 0.1, 0.04 and 0.01
\end{itemize}

and at 160 $\mu$m in medium speed:

\begin{itemize}
\item high-pass filter width': 15, 20, 25, 30 and 40 readouts
\item output pixel size: 6.4, 4.0, 3.0, 2.0 and 1.0''
\item pixfrac: 1.0, 0.5, 0.1, 0.04 and 0.01
\end{itemize}

For the parallel mode case, we limit our analysis at the 100 and 160 $\mu$m bands since we could not find observations with enough data redundancy at 70 $\mu$m. For the parallel mode case we use the same values of output pixel size and pixfrac as listed above, but we limit the analysis to only 3 values of high-pass filter width: 5,  8 and 16 readouts at 100 $\mu$m and 5, 8 and 12 readouts at 160 $\mu$m.

The PACS Photometer IPIPE scripts proposes two different methods to estimate Photproject maps: either with a ``simple'' mean or with a weighted mean, where the weights are estimated as the inverse of the square of the standard deviation of the flux in a box, along the timeline, $\sim$ 100-200 times the high-pass filter width. The use of one of these methods does not have any effect on the final relation between coverage and error map. Indeed, as shown in Fig. 24, reducing the same datasets with different mapping methods and by changing also the width of the box for estimating the weights in the weighted mean (different color coding), does not impact the coverage map-error map relation. Only the scatter of the relation slightly changes of few percent.  All the tabulated results in the Appendix are obtained by using the ``simple'' mean in photProject, which is the default behavior of the task. 

\subsubsection{The behavior of the noise and its components}

\begin{figure*}
\begin{center}
\includegraphics[width=1.0\textwidth]{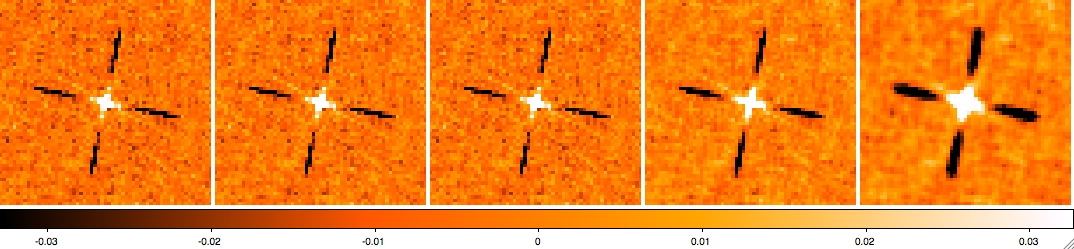} \includegraphics[width=1.0\textwidth]{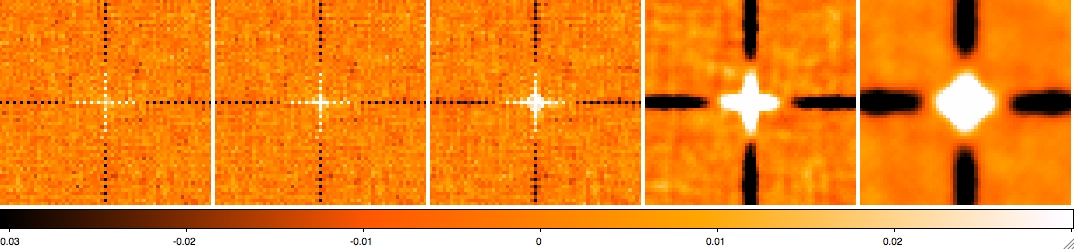} \includegraphics[width=1.0\textwidth]{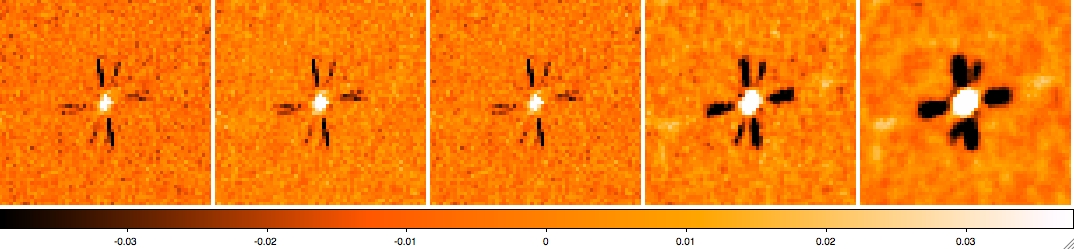} 
\caption{Example of cross-correlation maps as a function of the ratio between drop and output pixel size (from the left to the right) for the 100 $\mu$m medium speed (upper panels), the 160 $\mu$m medium speed (central panels) and the 160 $\mu$m fast speed (bottom panels).}
\label{corr_maps}
\end{center}
\end{figure*}

In this paragraph we analyse the behavior of the global noise and the individual noise components as a function of the data reduction parameter settings to guide the user in the choice of the parameter values. The noise components are estimated as explained above in the high redundancy case. The global noise of the final map is estimated in the following way. A large number of source extractions is performed at random positions in the map. The dispersion of the flux distribution of the random extractions is taken as the global mean 1$\sigma$ noise of the map.  The relation between the noise per pixel and the coverage map is estimated in the log-log space via a linear regression. The slope of the best fit line is, in all cases, consistent with $\alpha =-0.5$ in the blue and the green band, as expected if the noise is proportional to the inverse square root of the time spent on the pixel (coverage). The slope deviates from the value $-0.5$ for several parameter settings of the red band case in prime and in parallel mode, although the slope $\alpha=-0.5$ is still consistent with the data. In particular in the parallel mode case, the larger rms of the relation ($\sim 25\%$ with respect to the $\sim 17\%$ of the prime mode case) prevents us from measuring in an accurate way the slope of the relation. This is probably due to the much less uniform coverage of the maps considered in this exercise with respect to the prime mode case. Since in all considered parallel mode cases the slope $\alpha=-0.5$ is a good representation of the data (according to a ${\chi}^2$ test), we fix the slope to this value when doing the linear regression and we fit only the value of the intercept $\beta$.

In order to understand how the noise per pixel is changing as a function of the data reduction parameter settings, we choose as indicator of the mean absolute value of the noise per pixel the intercept $\beta$ of the linear regression. Fig. 25 shows the relation of $\beta$ with the output pixel size and the pixfrac values for several high-pass filter widths. In all bands, the noise per pixel is increasing at larger output pixel values and at larger pixfrac values. In other words, since the ratio between drop (expressed via pixfrac) and output pixel size is decreasing from bottom to up of Fig. 25, we can also conclude that the noise per pixel is decreasing as a function of this ratio. 

The cross-correlation correction factor $f$ is estimated in two ways depending on the extraction method. Indeed, in case of PSF fitting, the value of $f$ depends on the shape of the PSF chosen, as shown in Eq. 8. For a realistic estimate of $f$, in this experiment we choose the PSF estimated by using the high-pass filter width, output pixel size and pixfrac value used to derive the cross-correlation matrix. In this way we can take into account the effect of the high-pass filter width and projection on the PSF shape. In case of aperture photometry, instead, the same weight is assigned to all pixels contributing to the source flux. So in this case we adopt $a_i=1/N$ in Eq. 8, where N is the number of pixels within the aperture used for the photometry. In both cases we used PSF and apertures of three increasing sizes to check also the dependency of $f$ on the extension of the region used for source extraction. Fig. 26 shows an example of the change of cross-correlation matrix as a function of the drop/output pixel size ratio (increasing from the left to the right) for the 100 $\mu$m and 160 $\mu$m medium speed and 160 $\mu$m fast speed case. The negative correlation along the scan direction is visble in all cases and, the higher the drop/output pixel size ratio, the stronger the cross-correlation. This is due to the fact, that at higher value of the drop/output pixel size ratio the cross-correlated noise due to the projection is dominating with respect to the contribution given by the $1/f$ noise not completely removed by the high-pass filter.  The left panel of Fig. 27 shows the comparison of the cross-correlation correction factor estimated within the same aperture by PSF fitting and by aperture photometry. The deviations from the 1 to 1 relation are minor (of the order of few percent). Thus, the shape of the PSF does not affect significantly the estimate of $f$.  This is confirmed also in the central panel of the same figure, which shows the comparison of $f$ estimated via PSF fitting with PSF of different dimensions. Even in this plot, the deviations from the 1 to 1 relation are of the order of few percent. More interestingly, we see in the right panel of Fig. 27 that cross-correlated correction factor varies much more as a function of the drop/output pixel size ratio than the high-pass filter width (as color coded in the figure). As already shown visually in the cross-correlation matrix of Fig. 26 we see in this plot quantitatively, that till the drop/output pixel size ratio is below 1, $f$ is nearly constant with small variations (about 8-10\%) due to the high pass filter width. At drop/output pixel size ratio higher than 1, the cross correlation noise introduced by the projection (see Fruchter \& Hook for a nice explanation of this contribution) is dominating over any other source of cross correlation. 

Since noise per pixel and cross correlation correction factor $f$ have opposite dependences with respect to the output pixel size and pixfrac values  (drop/output pixel size), they tend to cancel out the effect of each other. Indeed, once we consider the global noise as a function of those parameters and for various high-pass filter widths (Fig. 28), we see that the variations due to the different output pixel size and pixfrac values at fixed high-pass filter width, are minor. What is is left is only the impact of the high-pass filter width.

\begin{figure*}
\begin{center}
\includegraphics[width=0.9\textwidth]{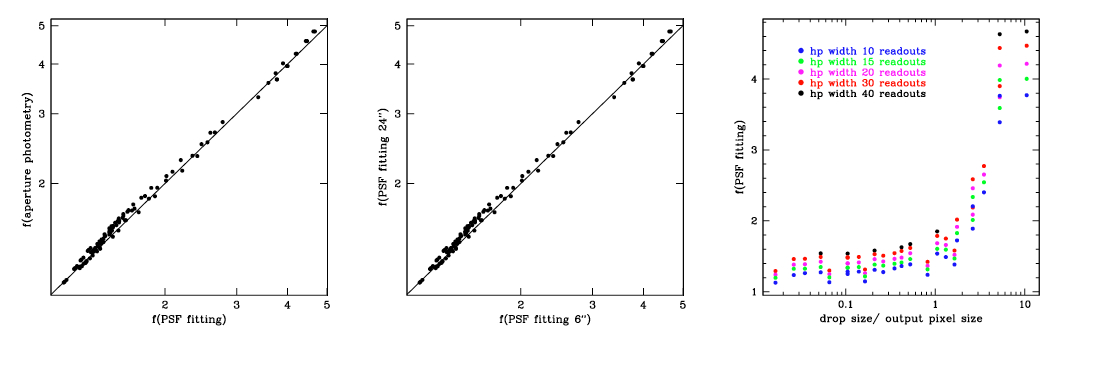} 
\caption{Cross-correlated noise correction factor measured by PSF photometry versus PSF fitting (left panel), measured within two different apertures (central panel) and as a function of the ratio between drop and output pixel size (right panel).}
\label{ff}
\end{center}
\end{figure*}

\begin{figure}
\centerline{\includegraphics[width=0.4\textwidth]{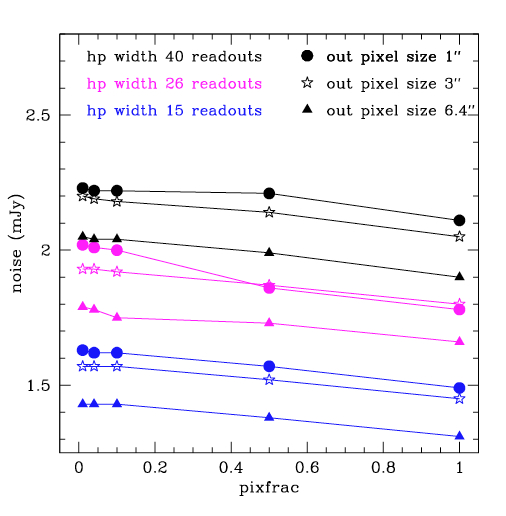}}
\caption{Mean global noise as a function of the output pixel size and the pixfrac value for different value of the high-pass filter width.}
\label{global}
\end{figure}

\subsubsection{Fitting the parameter space}

The exercize illustrated in the previous sections, provides the best fit parameters of the corverage-error map relation and the factor $f$ in a huge parameter space. The tables in the appendix, which list these values, can be used to interpolate in the parameter space to retrieve the calibration parameters to estimate the error map and the cross correlated noise contribution for a given parameter setting. However, in order to help the user in handling more easily this info, we investigate the possibility to use a fitting function to retrieve the slope $\alpha$ and the intercept $\beta$ of the coverage-error map relation, and the factor $f$. The fitting function must provide the value of $\alpha$, $\beta$ and $f$ for any set of high-pass filter width ($hp$), output pixel size ($out$) and pixfrac ($PIX$) values. We perform the fit by using the 125 points in the $hp,out,pix$ space for which we estimated $\alpha$, $\beta$ and $f$. Fig. 25 and the right panel of Fig. 27, suggest that the dependence of $\beta$ and $f$ from the parameter settings could be well expressed by a polynomial function, while $\alpha$ is in the majority of the cases constant. Indeed,  the best fit for $\alpha$, $\beta$ and $f$, in any band  in the medium speed case, is given by a third order polynomial in three variables $hp,out,pix$:

\begin{equation}
\alpha= \displaystyle\sum_{i,j,k=1}^{n}{a_{ijk}hp^iout^jpix^k}   
\end{equation}
\begin{equation}
\beta= \displaystyle\sum_{i,j,k=1}^{n}{b_{ijk}hp^iout^jpix^k}   
\end{equation}
\begin{equation}
f= \displaystyle\sum_{i,j,k=1}^{n}{c_{ijk}hp^iout^jpix^k}   
\end{equation}
where $i+j+k \le n$ and $n=3$. A third order polynomial in three variables contains 20 coefficients. The coefficient $a_{ijk}$, $b_{ijk}$ and $c_{ijk}$ are listed in the Appendix for each band of the medium speed case. In all cases the best fit is performed via a linear regression and by using the error in the estimates of $\alpha$ and $\beta$ derived via bootstrapping in any single coverage-error map linear regression. The error in the coefficient $f$ is estimated to be of the order of 3\%, as given by the dispersion of the distribution of 20 $f$ values estimated in different regions of the same map. The ${\chi}^2$ of the best fit is varying from 0.7 to 3 among all considered cases.
The smaller parameter space (75 points) of the parallel mode case with respect to the medium speed case, prevents us from fitting the 20 parameters of the third order polynomial function. However, to overcome this problem, we use the following trick. The sampling rate of the blue and green band in parallel mode is 5 Hz (frames are averaged $8 \times 8$ on board of {\it{Herschel}}), thus half of the sampling rate (10 Hz, since the frames are averaged $4 \times 4$ on board) of the prime mode in the same bands, while the scan speed is three times the medium speed. This leads to the fact that the spatial length corresponding to a parallel mode blue/green readout is 6 times the spatial length of a readout in prime mode. In the red band, the sampling rate is the same in prime and parallel mode and only the scan speed is different. Thus, in this case, the spatial length corresponding to a parallel mode red readout is 3 times the spatial length of a readout in prime mode.  This means that the parallel mode cases can be considered as points in the parameter space defined for the prime mode in medium speed, if the parallel mode green high-pass filter width is multiplied by 6 and the red one is multiplied by 3. With this trick we can perform the fit of the 20 parameters of the polynomial functions with 200 points in the parameter space and, thus, with higher accuracy. The best fit parameters obtained with and without including the parallel mode points are in agreement within the errors, confirming that the fitting functions provided here can be used to derive also the parallel mode noise information if the parallel mode high-pass filter width is expressed in terms of prime mode medium speed high-pass filter width. This method is implemented in the task {\it{photCoverage2Noise}} in the HIPE environment.

\section{Conclusions}

We investigate the effect of the data reduction via the ``high-pass filter$+$ naive projection'' route of the PACS Photometer pipeline on the PSF and noise of the PACS maps in any band and in two scan speeds (medium and fast). After disentangling the effect of the projection from the effect of the high-pass filter, we show that the running median removal adopted by the PACS pipeline can cause significant flux losses at any flux level. We provide a calibration of the perceptual flux loss in a representative portion of the PACS data reduction parameter settings and flux levels for several masking strategies. We analyse the advantages and disadvantages of the considered masking strategies and suggest that a mask based on putting circular patches on prior positions is the best solution to reduce the amount of flux loss and to preserve a flux independency of the high-pass filtering effect in large part of the flux range.  This allows to create a final catalog easy to correct via a simple correction factor almost at all fluxes with the exclusion of extremely bright sources which require a different or no calibration. We also show that, for stacking analysis, the impact of the high-pass filtering effect is to reduce significantly the clustering effect. Thus for any stacking analysis, the proper shape of the PSF affected by high-pass filtering should be considered. 
We used the method described in Lutz et al. (2011) for measuring the PACS Photometer error map and cross correlated noise correction factor for several observations with high data redundancy. We use these observations to provide a calibration of the error and coverage maps and of the cross correlated noise correction factor as a function of the PACS data reduction parameter settings. We conclude that, in terms of noise, the output pixel size and pixfrac values determine the behavior of the individual components of the PACS photometer map noise. However, since the dependences of the noise per pixel and $f$ with respect to output pixel size and pixfrac values are opposite, they cancel out each other. Thus, the global noise is determined almost only by the choice of the high-pass filter width.

\begin{acknowledgements}
PACS has been developed by a consortium of institutes led by MPE 
(Germany) and including UVIE (Austria); KUL, CSL, IMEC (Belgium); CEA, 
OAMP (France); MPIA (Germany); IFSI, OAP/AOT, OAA/CAISMI, LENS, SISSA 
(Italy); IAC (Spain). This development has been supported by the funding 
agencies BMVIT (Austria), ESA-PRODEX (Belgium), CEA/CNES (France),
DLR (Germany), ASI (Italy), and CICYT/MCYT (Spain.)

\end{acknowledgements}

\footnotesize{
\clearpage
\begin{table*}
\begin{minipage}{10.in}
\begin{sideways}
\begin{tabular}[b]{l | c c c | c c c | c c c | c c c | c c c}
\hline
\hline
\multicolumn{1}{c}{} \\
\multicolumn{1}{c}{output pixel size} &  \multicolumn{3}{c}{pixfrax=0.01} &  \multicolumn{3}{c}{pixfrax=0.04} &  \multicolumn{3}{c}{pixfrax=0.1} &  \multicolumn{3}{c}{pixfrax=0.5} &  \multicolumn{3}{c}{pixfrax=1}\\
\multicolumn{1}{c}{} \\
\hline
\multicolumn{1}{c}{} \\
\multicolumn{16}{c}{HPF width 10 readouts} \\
\multicolumn{1}{c}{} \\
\hline
\renewcommand{\arraystretch}{0.2}\renewcommand{\tabcolsep}{0.05cm}
 & $\alpha$& $\beta$ & rms & $\alpha $ & $\beta $ & rms & $\alpha$ & $\beta $ & rms & $\alpha $ & $\beta$ & rms & $\alpha $ & $\beta $ & rms   \\
\hline
0.5 &  -0.50 & -4.69 & 0.074 & -0.49 & -4.11 & 0.074 & -0.49 & -3.78 & 0.074 & -0.50 & -3.46 & 0.073 & -0.50 & -3.46 & 0.073 \\
1.0 & -0.50 & -4.39 & 0.073 & -0.50 & -3.79 & 0.073 & -0.50 & -3.42 & 0.073 & -0.50 & -2.93 & 0.073 & -0.50 & -2.83 & 0.074 \\
1.5 & -0.50 & -4.21 & 0.073 & -0.50 & -3.61 & 0.073 & -0.50 & -3.23 & 0.073 & -0.50 & -2.65 & 0.073 & -0.50 & -2.51 & 0.074 \\
2.0 & -0.50 & -4.08 & 0.073 & -0.50 & -3.48 & 0.073 & -0.50 & -3.09 & 0.073 & -0.50 & -2.47 & 0.074 & -0.49 & -2.29 & 0.074 \\
3.2 & -0.50 & -3.84 & 0.074 & -0.50 & -3.24 & 0.074 & -0.50 & -2.85 & 0.074 & -0.50 & -2.20 & 0.075 & -0.49 & -1.97 & 0.075 \\

\hline
\multicolumn{1}{c}{} \\
\multicolumn{16}{c}{HPF width 15 readouts} \\
\multicolumn{1}{c}{} \\
\hline
\renewcommand{\arraystretch}{0.2}\renewcommand{\tabcolsep}{0.05cm}
 & $\alpha$& $\beta$ & rms & $\alpha $ & $\beta $ & rms & $\alpha$ & $\beta $ & rms & $\alpha $ & $\beta$ & rms & $\alpha $ & $\beta $ & rms   \\
\hline
0.5 & -0.50 & -4.68 & 0.074 & -0.49 & -4.10 & 0.074 & -0.49 & -3.76 & 0.074 & -0.50 & -3.44 & 0.073 & -0.50 & -3.44 & 0.073 \\
1.0 & -0.50 & -4.37 & 0.073 & -0.50 & -3.78 & 0.073 & -0.50 & -3.40 & 0.073 & -0.50 & -2.91 & 0.073 & -0.50 & -2.81 & 0.074 \\
1.5 & -0.50 & -4.19 & 0.073 & -0.50 & -3.59 & 0.073 & -0.50 & -3.21 & 0.073 & -0.50 & -2.63 & 0.073 & -0.50 & -2.49 & 0.074 \\
2.0 & -0.50 & -4.06 & 0.073 & -0.50 & -3.47 & 0.073 & -0.50 & -3.08 & 0.073 & -0.50 & -2.45 & 0.073 & -0.50 & -2.27 & 0.074 \\
3.2 & -0.50 & -3.82 & 0.074 & -0.50 & -3.22 & 0.074 & -0.50 & -2.82 & 0.074 & -0.50 & -2.17 & 0.074 & -0.49 & -1.93 & 0.075 \\

\hline
\multicolumn{1}{c}{} \\
\multicolumn{16}{c}{HPF width 20 readouts} \\
\multicolumn{1}{c}{} \\
\hline
\renewcommand{\arraystretch}{0.2}\renewcommand{\tabcolsep}{0.05cm}
 & $\alpha$& $\beta$ & rms & $\alpha $ & $\beta $ & rms & $\alpha$ & $\beta $ & rms & $\alpha $ & $\beta$ & rms & $\alpha $ & $\beta $ & rms   \\
\hline
0.5 & -0.50 & -4.66 & 0.074 & -0.49 & -4.09 & 0.074 & -0.49 & -3.75 & 0.074 & -0.50 & -3.43 & 0.073 & -0.50 & -3.43 & 0.073 \\
1.0 & -0.50 & -4.36 & 0.073 & -0.50 & -3.76 & 0.073 & -0.50 & -3.39 & 0.073 & -0.50 & -2.90 & 0.073 & -0.50 & -2.80 & 0.074 \\
1.5 & -0.50 & -4.18 & 0.073 & -0.50 & -3.58 & 0.073 & -0.50 & -3.20 & 0.073 & -0.50 & -2.62 & 0.073 & -0.50 & -2.47 & 0.074 \\
2.0 & -0.50 & -4.05 & 0.073 & -0.50 & -3.45 & 0.073 & -0.50 & -3.07 & 0.073 & -0.50 & -2.44 & 0.073 & -0.49 & -2.25 & 0.074 \\
3.2 & -0.50 & -3.80 & 0.074 & -0.50 & -3.20 & 0.074 & -0.50 & -2.81 & 0.074 & -0.50 & -2.15 & 0.074 & -0.49 & -1.92 & 0.074 \\

\hline
\multicolumn{1}{c}{} \\
\multicolumn{16}{c}{HPF width 30 readouts} \\
\multicolumn{1}{c}{} \\
\hline
\renewcommand{\arraystretch}{0.2}\renewcommand{\tabcolsep}{0.05cm}
 & $\alpha$& $\beta$ & rms & $\alpha $ & $\beta $ & rms & $\alpha$ & $\beta $ & rms & $\alpha $ & $\beta$ & rms & $\alpha $ & $\beta $ & rms   \\
\hline
0.5 & -0.50 & -4.65 & 0.074 & -0.49 & -4.07 & 0.074 & -0.49 & -3.74 & 0.074 & -0.50 & -3.42 & 0.073 & -0.50 & -3.42 & 0.073 \\
1.0 & -0.50 & -4.34 & 0.073 & -0.50 & -3.75 & 0.073 & -0.50 & -3.38 & 0.073 & -0.50 & -2.88 & 0.073 & -0.50 & -2.77 & 0.073 \\
1.5 & -0.50 & -4.17 & 0.073 & -0.50 & -3.57 & 0.073 & -0.50 & -3.19 & 0.073 & -0.50 & -2.60 & 0.073 & -0.50 & -2.45 & 0.073 \\
2.0 & -0.50 & -4.04 & 0.073 & -0.50 & -3.44 & 0.073 & -0.50 & -3.05 & 0.073 & -0.50 & -2.42 & 0.073 & -0.50 & -2.22 & 0.074 \\
3.2 & -0.50 & -3.78 & 0.074 & -0.50 & -3.18 & 0.074 & -0.50 & -2.79 & 0.074 & -0.50 & -2.13 & 0.074 & -0.50 & -1.89 & 0.074 \\

\hline
\multicolumn{1}{c}{} \\
\multicolumn{16}{c}{HPF width 40 readouts} \\
\multicolumn{1}{c}{} \\
\hline
\renewcommand{\arraystretch}{0.2}\renewcommand{\tabcolsep}{0.05cm}
 & $\alpha$& $\beta$ & rms & $\alpha $ & $\beta $ & rms & $\alpha$ & $\beta $ & rms & $\alpha $ & $\beta$ & rms & $\alpha $ & $\beta $ & rms   \\
\hline
0.5 & -0.50 & -4.64 & 0.074 & -0.49 & -4.06 & 0.074 & -0.49 & -3.73 & 0.074 & -0.50 & -3.41 & 0.073 & -0.50 & -3.41 & 0.073 \\
1.0 & -0.50 & -4.33 & 0.073 & -0.50 & -3.74 & 0.073 & -0.50 & -3.37 & 0.073 & -0.50 & -2.87 & 0.073 & -0.50 & -2.76 & 0.073 \\
1.5 & -0.50 & -4.16 & 0.073 & -0.50 & -3.56 & 0.073 & -0.50 & -3.18 & 0.073 & -0.50 & -2.59 & 0.073 & -0.50 & -2.44 & 0.073 \\
2.0 & -0.50 & -4.03 & 0.073 & -0.50 & -3.43 & 0.073 & -0.50 & -3.04 & 0.073 & -0.50 & -2.41 & 0.073 & -0.49 & -2.21 & 0.073 \\
3.2 & -0.50 & -3.77 & 0.073 & -0.50 & -3.17 & 0.073 & -0.50 & -2.77 & 0.073 & -0.50 & -2.12 & 0.074 & -0.49 & -1.87 & 0.074 \\

\hline
\end{tabular}
\end{sideways}  
\end{minipage}
\caption{The table lists the values of the best fit parameters, slope $\alpha$ and the intercept $\beta$, of the relation $lg(noise per pixel)=\alpha \times lg(coverage)+\beta$ and the rms of the relation in dex. The values are listed for different values of HPF width, output pixel size and pixfrac values for the 70 $\mu$m band in medium speed.}
\label{t1}
\begin{minipage}{0.5\hsize}
\end{minipage}
\end{table*}}

\footnotesize{
\clearpage
\begin{table*}
\begin{minipage}{10.in}
\begin{sideways}
\begin{tabular}[b]{l | c c c | c c c | c c c | c c c | c c c}
\hline
\hline
\multicolumn{1}{c}{} \\
\multicolumn{1}{c}{output pixel size} &  \multicolumn{3}{c}{pixfrax=0.01} &  \multicolumn{3}{c}{pixfrax=0.04} &  \multicolumn{3}{c}{pixfrax=0.1} &  \multicolumn{3}{c}{pixfrax=0.5} &  \multicolumn{3}{c}{pixfrax=1}\\
\multicolumn{1}{c}{} \\
\hline
\multicolumn{1}{c}{} \\
\multicolumn{16}{c}{HPF width 10 readouts} \\
\multicolumn{1}{c}{} \\
\hline
\renewcommand{\arraystretch}{0.2}\renewcommand{\tabcolsep}{0.05cm}
 & $\alpha$& $\beta$ & rms & $\alpha $ & $\beta $ & rms & $\alpha$ & $\beta $ & rms & $\alpha $ & $\beta$ & rms & $\alpha $ & $\beta $ & rms   \\
\hline
0.5 & -0.49 & -4.71 & 0.073 & -0.46 & -4.17 & 0.078 & -0.39 & -3.97 & 0.073 & -0.39 & -3.83 & 0.072 & -0.39 & -3.83 & 0.072 \\
1.0 & -0.47 & -4.40 & 0.072 & -0.45 & -3.86 & 0.072 & -0.43 & -3.56 & 0.072 & -0.38 & -3.30 & 0.072 & -0.41 & -3.17 & 0.072 \\
1.5 & -0.43 & -4.21 & 0.072 & -0.42 & -3.71 & 0.072 & -0.41 & -3.40 & 0.072 & -0.40 & -2.98 & 0.072 & -0.42 & -2.81 & 0.072 \\
2.0 & -0.42 & -4.09 & 0.072 & -0.42 & -3.58 & 0.072 & -0.42 & -3.26 & 0.072 & -0.41 & -2.77 & 0.072 & -0.42 & -2.59 & 0.072 \\
3.2 & -0.40 & -3.84 & 0.072 & -0.39 & -3.36 & 0.072 & -0.39 & -3.05 & 0.073 & -0.41 & -2.49 & 0.073 & -0.43 & -2.24 & 0.073 \\

\hline
\multicolumn{1}{c}{} \\
\multicolumn{16}{c}{HPF width 15 readouts} \\
\multicolumn{1}{c}{} \\
\hline
\renewcommand{\arraystretch}{0.2}\renewcommand{\tabcolsep}{0.05cm}
 & $\alpha$& $\beta$ & rms & $\alpha $ & $\beta $ & rms & $\alpha$ & $\beta $ & rms & $\alpha $ & $\beta$ & rms & $\alpha $ & $\beta $ & rms   \\
\hline
0.5 & -0.49 & -4.69 & 0.073 & -0.45 & -4.16 & 0.073 & -0.39 & -3.95 & 0.073 & -0.39 & -3.81 & 0.072 & -0.39 & -3.81 & 0.072 \\
1.0 & -0.46 & -4.38 & 0.072 & -0.45 & -3.84 & 0.072 & -0.43 & -3.55 & 0.072 & -0.39 & -3.28 & 0.072 & -0.42 & -3.13 & 0.072 \\
1.5 & -0.43 & -4.20 & 0.072 & -0.42 & -3.69 & 0.072 & -0.41 & -3.39 & 0.072 & -0.40 & -2.96 & 0.072 & -0.42 & -2.78 & 0.072 \\
2.0 & -0.42 & -4.07 & 0.072 & -0.42 & -3.56 & 0.072 & -0.42 & -3.24 & 0.072 & -0.41 & -2.76 & 0.072 & -0.42 & -2.56 & 0.072 \\
3.2 & -0.40 & -3.81 & 0.072 & -0.40 & -3.33 & 0.072 & -0.40 & -3.01 & 0.072 & -0.41 & -2.47 & 0.072 & -0.43 & -2.21 & 0.073 \\

\hline
\multicolumn{1}{c}{} \\
\multicolumn{16}{c}{HPF width 20 readouts} \\
\multicolumn{1}{c}{} \\
\hline
\renewcommand{\arraystretch}{0.2}\renewcommand{\tabcolsep}{0.05cm}
 & $\alpha$& $\beta$ & rms & $\alpha $ & $\beta $ & rms & $\alpha$ & $\beta $ & rms & $\alpha $ & $\beta$ & rms & $\alpha $ & $\beta $ & rms   \\
\hline
0.5 & -0.49 & -4.68 & 0.073 & -0.45 & -4.15 & 0.073 & -0.39 & -3.94 & 0.073 & -0.39 & -3.81 & 0.072 & -0.39 & -3.81 & 0.072 \\
1.0 & -0.46 & -4.37 & 0.072 & -0.45 & -3.83 & 0.072 & -0.42 & -3.54 & 0.072 & -0.38 & -3.29 & 0.072 & -0.39 & -3.21 & 0.074 \\
1.5 & -0.42 & -4.18 & 0.072 & -0.41 & -3.69 & 0.072 & -0.40 & -3.39 & 0.072 & -0.39 & -2.98 & 0.072 & -0.41 & -2.83 & 0.072 \\
2.0 & -0.41 & -4.05 & 0.072 & -0.40 & -3.57 & 0.072 & -0.40 & -3.26 & 0.072 & -0.39 & -2.80 & 0.072 & -0.40 & -2.62 & 0.072 \\
3.2 & -0.40 & -3.80 & 0.072 & -0.40 & -3.31 & 0.072 & -0.40 & -2.99 & 0.072 & -0.40 & -2.49 & 0.072 & -0.41 & -2.27 & 0.072 \\

\hline
\multicolumn{1}{c}{} \\
\multicolumn{16}{c}{HPF width 30 readouts} \\
\multicolumn{1}{c}{} \\
\hline
\renewcommand{\arraystretch}{0.2}\renewcommand{\tabcolsep}{0.05cm}
 & $\alpha$& $\beta$ & rms & $\alpha $ & $\beta $ & rms & $\alpha$ & $\beta $ & rms & $\alpha $ & $\beta$ & rms & $\alpha $ & $\beta $ & rms   \\
\hline
0.5 & -0.49 & -4.66 & 0.073 & -0.45 & -4.13 & 0.073 & -0.39 & -3.93 & 0.073 & -0.38 & -3.81 & 0.072 & -0.38 & -3.81 & 0.072 \\
1.0 & -0.46 & -4.36 & 0.072 & -0.45 & -3.82 & 0.072 & -0.42 & -3.53 & 0.072 & -0.37 & -3.30 & 0.072 & -0.39 & -3.21 & 0.072 \\
1.5 & -0.42 & -4.17 & 0.072 & -0.40 & -3.68 & 0.072 & -0.39 & -3.39 & 0.072 & -0.38 & -2.98 & 0.077 & -0.39 & -2.87 & 0.072 \\
2.0 & -0.40 & -4.04 & 0.072 & -0.40 & -3.56 & 0.072 & -0.39 & -3.25 & 0.072 & -0.38 & -2.81 & 0.072 & -0.39 & -2.66 & 0.072 \\
3.2 & -0.40 & -3.78 & 0.072 & -0.40 & -3.29 & 0.072 & -0.40 & -2.98 & 0.072 & -0.39 & -2.49 & 0.072 & -0.39 & -2.30 & 0.072 \\

\hline
\multicolumn{1}{c}{} \\
\multicolumn{16}{c}{HPF width 40 readouts} \\
\multicolumn{1}{c}{} \\
\hline
\renewcommand{\arraystretch}{0.2}\renewcommand{\tabcolsep}{0.05cm}
 & $\alpha$& $\beta$ & rms & $\alpha $ & $\beta $ & rms & $\alpha$ & $\beta $ & rms & $\alpha $ & $\beta$ & rms & $\alpha $ & $\beta $ & rms   \\
\hline

0.5 & -0.49 & -4.65 & 0.073 & -0.45 & -4.12 & 0.073 & -0.39 & -3.92 & 0.073 & -0.38 & -3.80 & 0.072 & -0.38 & -3.80 & 0.072 \\
1.0 & -0.46 & -4.34 & 0.072 & -0.44 & -3.81 & 0.072 & -0.42 & -3.52 & 0.072 & -0.37 & -3.28 & 0.072 & -0.39 & -3.17 & 0.072 \\
1.5 & -0.42 & -4.16 & 0.072 & -0.41 & -3.67 & 0.072 & -0.40 & -3.37 & 0.072 & -0.39 & -2.96 & 0.072 & -0.44 & -4.03 & 0.079 \\
2.0 & -0.41 & -3.54 & 0.072 & -0.40 & -3.22 & 0.072 & -0.39 & -2.78 & 0.072 & -0.40 & -2.61 & 0.072 & -0.41 & -3.76 & 0.072 \\
3.2 & -0.40 & -3.27 & 0.072 & -0.40 & -2.96 & 0.072 & -0.39 & -2.46 & 0.072 & -0.40 & -2.26 & 0.072 & -0.40 & -2.26 & 0.072 \\

\hline
\end{tabular}
\end{sideways}  
\end{minipage}
\caption{The table lists the values of the best fit parameters, slope $\alpha$ and the intercept $\beta$, of the relation $lg(noise per pixel)=\alpha \times lg(coverage)+\beta$ and the rms of the relation in dex. The values are listed for different values of HPF width, output pixel size and pixfrac values for the 100 $\mu$m band in medium speed.}
\label{t1}
\begin{minipage}{0.5\hsize}
\end{minipage}
\end{table*}}

\footnotesize{
\clearpage
\begin{table*}
\begin{minipage}{10.in}
\begin{sideways}
\begin{tabular}[b]{l | c c c | c c c | c c c | c c c | c c c}
\hline
\hline
\multicolumn{1}{c}{} \\
\multicolumn{1}{c}{output pixel size} &  \multicolumn{3}{c}{pixfrax=0.01} &  \multicolumn{3}{c}{pixfrax=0.04} &  \multicolumn{3}{c}{pixfrax=0.1} &  \multicolumn{3}{c}{pixfrax=0.5} &  \multicolumn{3}{c}{pixfrax=1}\\
\multicolumn{1}{c}{} \\
\hline
\multicolumn{1}{c}{} \\
\multicolumn{16}{c}{HPF width 15 readouts} \\
\multicolumn{1}{c}{} \\
\hline
\renewcommand{\arraystretch}{0.2}\renewcommand{\tabcolsep}{0.05cm}
 & $\alpha$& $\beta$ & rms & $\alpha $ & $\beta $ & rms & $\alpha$ & $\beta $ & rms & $\alpha $ & $\beta$ & rms & $\alpha $ & $\beta $ & rms   \\
\hline
1.0 &    -0.48   &   -4.46  &   0.073  &  -0.44  &    -3.95  &   0.073  &  -0.37  &    -3.78   &  0.073  &  -0.37  &    -3.62  &   0.072  &  -0.36   &   -3.62   &  0.072 \\
2.0 &    -0.45   &   -4.15  &   0.072  &  -0.43  &    -3.30  &   0.072  &  -0.43  &    -3.30   &  0.072  &  -0.34  &    -3.14  &   0.072  &  -0.36   &   -3.05   &  0.072 \\
3.0 &    -0.41   &   -3.91  &   0.072  &  -0.40  &    -3.43  &   0.072  &  -0.39  &    -3.14   &  0.072  &  -0.32  &    -2.91  &   0.072  &  -0.36   &   -2.71   &  0.072 \\
4.0 &    -0.36   &   -3.76  &   0.072  &  -0.36  &    -3.05  &   0.072  &  -0.36  &    -3.05   &  0.072  &  -0.36  &    -2.48  &   0.072  &  -0.36   &   -2.48   &  0.072 \\
6.4 &    -0.39   &   -3.51  &   0.072  &  -0.38  &    -3.05  &   0.072  &  -0.39  &    -2.73   &  0.072  &  -0.37  &    -2.29  &   0.072  &  -0.36   &   -2.15   &  0.072 \\

\hline
\multicolumn{1}{c}{} \\
\multicolumn{16}{c}{HPF width 20 readouts} \\
\multicolumn{1}{c}{} \\
\hline
\renewcommand{\arraystretch}{0.2}\renewcommand{\tabcolsep}{0.05cm}
 & $\alpha$& $\beta$ & rms & $\alpha $ & $\beta $ & rms & $\alpha$ & $\beta $ & rms & $\alpha $ & $\beta$ & rms & $\alpha $ & $\beta $ & rms   \\
\hline
1.0 &     -0.48   &   -4.44  &   0.073  &  -0.44  &    -3.93  &   0.073  &  -0.37  &    -3.76   &  0.073  &  -0.37  &    -3.60  &   0.072  &  -0.37   &   -3.58   &  0.072 \\
2.0 &     -0.45   &   -4.13  &   0.072  &  -0.45  &    -3.60  &   0.072  &  -0.43  &    -3.28   &  0.072  &  -0.34  &    -3.12  &   0.072  &  -0.36   &   -3.01   &  0.072 \\
3.0 &     -0.41   &   -3.89  &   0.072  &  -0.40  &    -3.41  &   0.072  &  -0.39  &    -3.12   &  0.072  &  -0.32  &    -2.88  &   0.072  &  -0.36   &   -2.67   &  0.072 \\
4.0 &     -0.36   &   -3.73  &   0.072  &  -0.36  &    -3.29  &   0.072  &  -0.36  &    -3.01   &  0.072  &  -0.35  &    -2.61  &   0.072  &  -0.37   &   -2.44   &  0.072 \\
6.4 &     -0.39   &   -3.48  &   0.072  &  -0.39  &    -3.00  &   0.072  &  -0.41  &    -2.67   &  0.072  &  -0.38  &    -2.23  &   0.072  &  -0.36   &   -2.11   &  0.072 \\

\hline
\multicolumn{1}{c}{} \\
\multicolumn{16}{c}{HPF width 26 readouts} \\
\multicolumn{1}{c}{} \\
\hline
\renewcommand{\arraystretch}{0.2}\renewcommand{\tabcolsep}{0.05cm}
 & $\alpha$& $\beta$ & rms & $\alpha $ & $\beta $ & rms & $\alpha$ & $\beta $ & rms & $\alpha $ & $\beta$ & rms & $\alpha $ & $\beta $ & rms   \\
\hline
1.0 &     -0.48   &   -4.43  &   0.073  &  -0.44  &    -3.91  &   0.073  &  -0.37  &    -3.74   &  0.073  &  -0.37  &    -3.60  &   0.072  &  -0.38   &   -3.49   &  0.072 \\
2.0 &     -0.45   &   -4.12  &   0.072  &  -0.45  &    -3.58  &   0.072  &  -0.43  &    -3.27   &  0.072  &  -0.35  &    -3.08  &   0.072  &  -0.38   &   -2.91   &  0.072 \\
3.0 &     -0.41   &   -3.87  &   0.072  &  -0.40  &    -3.39  &   0.072  &  -0.39  &    -3.10   &  0.072  &  -0.33  &    -2.84  &   0.072  &  -0.38   &   -2.57   &  0.072 \\
4.0 &     -0.35   &   -3.70  &   0.072  &  -0.35  &    -3.28  &   0.072  &  -0.36  &    -2.99   &  0.072  &  -0.36  &    -2.56  &   0.072  &  -0.39   &   -2.32   &  0.072 \\
6.4 &     -0.43   &   -3.46  &   0.072  &  -0.43  &    -2.94  &   0.072  &  -0.44  &    -2.58   &  0.072  &  -0.41  &    -2.10  &   0.072  &  -0.40   &   -1.93   &  0.072 \\

\hline
\multicolumn{1}{c}{} \\
\multicolumn{16}{c}{HPF width 30 readouts} \\
\multicolumn{1}{c}{} \\
\hline
\renewcommand{\arraystretch}{0.2}\renewcommand{\tabcolsep}{0.05cm}
 & $\alpha$& $\beta$ & rms & $\alpha $ & $\beta $ & rms & $\alpha$ & $\beta $ & rms & $\alpha $ & $\beta$ & rms & $\alpha $ & $\beta $ & rms   \\
\hline
1.0 &     -0.48   &   -4.42  &   0.073  &  -0.44  &    -3.90  &   0.073  &  -0.37  &    -3.73   &  0.073  &  -0.37  &    -3.55  &   0.072  &  -0.39   &   -3.44   &  0.072 \\
2.0 &     -0.45   &   -4.11  &   0.072  &  -0.45  &    -3.57  &   0.072  &  -0.43  &    -3.25   &  0.072  &  -0.36  &    -3.05  &   0.072  &  -0.39   &   -2.86   &  0.072 \\
3.0 &     -0.40   &   -3.38  &   0.072  &  -0.40  &    -3.38  &   0.072  &  -0.39  &    -3.09   &  0.072  &  -0.33  &    -2.83  &   0.072  &  -0.39   &   -2.53   &  0.072 \\
4.0 &     -0.35   &   -3.69  &   0.072  &  -0.36  &    -3.26  &   0.072  &  -0.36  &    -2.97   &  0.072  &  -0.36  &    -2.52  &   0.072  &  -0.40   &   -2.28   &  0.072 \\
6.4 &     -0.44   &   -3.45  &   0.072  &  -0.44  &    -2.92  &   0.072  &  -0.44  &    -2.56   &  0.072  &  -0.41  &    -2.09  &   0.072  &  -0.41   &   -1.88   &  0.072 \\

\hline
\multicolumn{1}{c}{} \\
\multicolumn{16}{c}{HPF width 40 readouts} \\
\multicolumn{1}{c}{} \\
\hline
\renewcommand{\arraystretch}{0.2}\renewcommand{\tabcolsep}{0.05cm}
 & $\alpha$& $\beta$ & rms & $\alpha $ & $\beta $ & rms & $\alpha$ & $\beta $ & rms & $\alpha $ & $\beta$ & rms & $\alpha $ & $\beta $ & rms   \\
\hline
1.0 &     -0.30   &   -3.62  &   0.175  &  -0.38  &    -3.49  &   0.072  &  -0.30  &    -3.62   &  0.175  &  -0.38  &    -3.49  &   0.072  &  -0.42   &   -3.30   &  0.072 \\
2.0 &     -0.46   &   -4.09  &   0.072  &  -0.46  &    -3.54  &   0.072  &  -0.44  &    -3.22   &  0.072  &  -0.37  &    -2.99  &   0.072  &  -0.42   &   -2.71   &  0.072 \\
3.0 &     -0.43   &   -3.84  &   0.072  &  -0.41  &    -3.34  &   0.072  &  -0.40  &    -3.04   &  0.072  &  -0.35  &    -2.73  &   0.072  &  -0.42   &   -2.38   &  0.072 \\
4.0 &     -0.39   &   -3.68  &   0.072  &  -0.40  &    -3.19  &   0.072  &  -0.40  &    -2.88   &  0.072  &  -0.40  &    -2.39  &   0.072  &  -0.43   &   -2.12   &  0.072 \\
6.4 &     -0.45   &   -3.42  &   0.072  &  -0.45  &    -2.87  &   0.072  &  -0.46  &    -2.50   &  0.072  &  -0.44  &    -1.96  &   0.072  &  -0.44   &   -1.72   &  0.072 \\

\end{tabular}
\end{sideways}  
\end{minipage}
\caption{The table lists the values of the best fit parameters, slope $\alpha$ and the intercept $\beta$, of the relation $lg(noise per pixel)=\alpha \times lg(coverage)+\beta$ and the rms of the relation in dex. The values are listed for different values of HPF width, output pixel size and pixfrac values for the 160 $\mu$m band in medium speed.}
\label{t1}
\begin{minipage}{0.5\hsize}
\end{minipage}
\end{table*}}

\footnotesize{
\clearpage
\begin{table*}
\begin{minipage}{10.in}
\begin{tabular}[b]{|l | c |c| c| c| c|}
\hline
\hline
\multicolumn{1}{c}{} \\
\multicolumn{6}{c}{100 $\mu$m} \\
\multicolumn{1}{c}{} \\
\hline
\hline
\multicolumn{1}{c}{} \\
\multicolumn{1}{c}{output pixel size} &  \multicolumn{1}{c}{pixfrax=0.01} &  \multicolumn{1}{c}{pixfrax=0.04} &  \multicolumn{1}{c}{pixfrax=0.1} &  \multicolumn{1}{c}{pixfrax=0.5} &  \multicolumn{1}{c}{pixfrax=1}\\
\multicolumn{1}{c}{} \\
\hline
\multicolumn{6}{|c|}{HPF width 5 readouts} \\
\hline
     0.5  &  -4.58  &   -4.30   &  -3.87  &  -3.78  &  -3.82 \\
     1.0  &  -4.26  &   -3.97   &  -3.51  &  -3.30  &  -3.23 \\
     1.5  &  -4.08  &   -3.79   &  -3.31  &  -3.00  &  -2.85 \\
     2.0  &  -3.99  &   -3.69   &  -3.20  &  -2.82  &  -2.62 \\
     3.2  &  -3.73  &   -3.42   &  -2.92  &  -2.47  &  -2.29 \\
\hline
\multicolumn{6}{|c|}{HPF width 8 readouts} \\
\hline
     0.5  &  -4.52  &   -4.24   &  -3.80  &  -3.72  &  -3.78 \\
     1.0  &  -4.20  &   -3.91   &  -3.46  &  -3.25  &  -3.20 \\
     1.5  &  -4.03  &   -3.73   &  -3.25  &  -2.95  &  -2.82 \\
     2.0  &  -3.92  &   -3.62   &  -3.13  &  -2.76  &  -2.59 \\
     3.2  &  -3.62  &   -3.31   &  -2.81  &  -2.38  &  -2.24 \\
\hline
\multicolumn{6}{|c|}{HPF width 16 readouts} \\
\hline
     0.5  &  -2.80  &   -2.51   &  -2.07  &  -1.91  &  -1.93 \\
     1.0  &  -2.45  &   -2.16   &  -1.69  &  -1.42  &  -1.34 \\
     1.5  &  -2.23  &   -1.93   &  -1.45  &  -1.09  &  -0.94 \\
     2.0  &  -2.07  &   -1.76   &  -1.27  &  -0.85  &  -0.67 \\
     3.2  &  -1.57  &   -1.25   &  -0.74  &  -0.29  &  -0.15 \\
\hline
\hline
\multicolumn{1}{c}{} \\
\multicolumn{6}{c}{160 $\mu$m} \\
\multicolumn{1}{c}{} \\
\hline
\hline
\multicolumn{1}{c}{} \\
\multicolumn{1}{c}{output pixel size} &  \multicolumn{1}{c}{pixfrax=0.01} &  \multicolumn{1}{c}{pixfrax=0.04} &  \multicolumn{1}{c}{pixfrax=0.1} &  \multicolumn{1}{c}{pixfrax=0.5} &  \multicolumn{1}{c}{pixfrax=1}\\
\multicolumn{1}{c}{} \\
\hline
\multicolumn{6}{|c|}{HPF width 5 readouts} \\
\hline
     1.0  &  -4.32  &   -4.11   &  -3.80  &  -3.70  &  -3.66 \\
     2.0  &  -4.02  &   -3.79   &  -3.43  &  -3.13  &  -3.03 \\
     3.0  &  -3.74  &   -3.49   &  -3.10  &  -2.65  &  -2.56 \\
     4.0  &  -3.51  &   -3.26   &  -2.84  &  -2.31  &  -2.31 \\
     6.4  &  -3.45  &   -3.18   &  -2.75  &  -2.26  &  -2.79 \\
\hline
\multicolumn{6}{|c|}{HPF width 8 readouts} \\
\hline
     1.0  &  -4.08  &   -3.91   &  -3.64  &  -3.73  &  -3.59 \\
     2.0  &  -4.01  &   -3.81   &  -3.48  &  -3.30  &  -3.02 \\
     3.0  &  -3.86  &   -3.63   &  -3.27  &  -2.87  &  -2.53 \\
     4.0  &  -3.67  &   -3.42   &  -3.02  &  -2.48  &  -2.15 \\
     6.4  &  -3.30  &   -3.03   &  -2.59  &  -1.94  &  -1.94 \\
\hline
\multicolumn{6}{|c|}{HPF width 12 readouts} \\
\hline
     1.0  &  -2.97  &   -2.84   &  -2.67  &  -3.23  &  -3.25 \\
     2.0  &  -3.39  &   -3.23   &  -2.99  &  -3.19  &  -2.96 \\
     3.0  &  -3.60  &   -3.41   &  -3.12  &  -3.02  &  -2.61 \\
     4.0  &  -3.64  &   -3.43   &  -3.09  &  -2.76  &  -2.25 \\
     6.4  &  -3.32  &   -3.06   &  -2.65  &  -2.00  &  -1.57 \\
\hline
\hline
\end{tabular}
\end{minipage}
\caption{The table lists the values of the best fit parameters, $\beta$, which is the intercept of the relation $lg(noise per pixel)=\alpha \times lg(coverage)+\beta$. Due to the higher noise of the parallel mode error-coverage map relation, le slope of the regression is fixed to $\alpha=-0.5$. The values are listed for different values of HPF width, output pixel size and pixfrac values for the 100 $\mu$m and 160 $\mu$m bands in parallel mode.}
\label{t1}
\begin{minipage}{0.5\hsize}
\end{minipage}
\end{table*}}


\footnotesize{
\begin{table*}
\begin{tabular}[b]{l | c c c c c | c c c c c | c c c c c  }
\hline
\hline
\multicolumn{1}{c|}{ } & \multicolumn{5}{c|}{6 $\arcsec$ aperture} & \multicolumn{5}{c|}{12 $\arcsec$ aperture} & \multicolumn{5}{c}{15 $\arcsec$ aperture}\\
\hline
\multicolumn{1}{c|}{output } &  \multicolumn{5}{c|}{pixfrax} & \multicolumn{5}{c|}{pixfrax} &  \multicolumn{5}{c}{pixfrax}\\
\multicolumn{1}{c|}{pixel size} &  \multicolumn{1}{c}{0.01} &  \multicolumn{1}{c}{0.04} &  \multicolumn{1}{c}{0.1} &  \multicolumn{1}{c}{0.5} &  \multicolumn{1}{c|}{1} &  \multicolumn{1}{c}{0.01} &  \multicolumn{1}{c}{0.04} &  \multicolumn{1}{c}{0.1} &  \multicolumn{1}{c}{0.5} &  \multicolumn{1}{c|}{1} &  \multicolumn{1}{c}{0.01} &  \multicolumn{1}{c}{0.04} &  \multicolumn{1}{c}{0.1} &  \multicolumn{1}{c}{0.5} &  \multicolumn{1}{c}{1}\\
\hline
\multicolumn{16}{c}{HPF width 10 readouts} \\
\hline
0.5  &    1.27  & 1.34 & 1.52 &  3.66 & 3.66 & 1.19  & 1.26 &  1.42 &  3.38 &  3.38 & 1.28 & 1.36 &  1.5 &  3.75  & 3.76 \\
1.0  &    1.26  & 1.29 & 1.36 &  2.15 & 3.30 & 1.16  & 1.19 &  1.26 &  2.02 &  3.14 & 1.27 & 1.30 &  1.3 &  2.20  & 3.41 \\
1.5  &    1.25  & 1.27 & 1.31 &  1.69 & 2.34 & 1.15  & 1.17 &  1.21 &  1.56 &  2.21 & 1.25 & 1.28 &  1.3 &  1.72  & 2.41 \\
2.0  &    1.22  & 1.23 & 1.26 &  1.46 & 1.85 & 1.11  & 1.12 &  1.14 &  1.34 &  1.72 & 1.22 & 1.24 &  1.2 &  1.48  & 1.89 \\
3.2  &    1.12  & 1.13 & 1.14 &  1.23 & 1.38 & 1.01  & 1.02 &  1.03 &  1.13 &  1.28 & 1.12 & 1.13 &  1.1 &  1.25  & 1.41 \\
\hline
\multicolumn{16}{c}{HPF width 15 readouts} \\
\hline
0.5  &    1.34  & 1.42 & 1.61 &  3.95 & 3.95 & 1.42  & 1.50 &  1.71 &  4.03 &  4.03 & 1.35 & 1.43 &  1.6 &  4.04  & 4.06 \\
1.0  &    1.37  & 1.41 & 1.48 &  2.35 & 3.58 & 1.44  & 1.48 &  1.56 &  2.48 &  3.82 & 1.38 & 1.41 &  1.4 &  2.39  & 3.70 \\
1.5  &    1.34  & 1.36 & 1.40 &  1.83 & 2.54 & 1.37  & 1.40 &  1.44 &  1.90 &  2.70 & 1.34 & 1.37 &  1.4 &  1.86  & 2.61 \\
2.0  &    1.34  & 1.36 & 1.39 &  1.61 & 2.03 & 1.37  & 1.38 &  1.41 &  1.66 &  2.11 & 1.34 & 1.36 &  1.3 &  1.62  & 2.06 \\
3.2  &    1.21  & 1.22 & 1.23 &  1.34 & 1.49 & 1.26  & 1.27 &  1.28 &  1.41 &  1.59 & 1.23 & 1.24 &  1.2 &  1.37  & 1.54 \\
\hline
\multicolumn{16}{c}{HPF width 20 readouts} \\
\hline
0.5  &    1.43  & 1.52 & 1.72 &  4.24 & 4.24 & 1.66  & 1.76 &  2.00 &  4.76 &  4.76 & 1.45 & 1.53 &  1.7 &  4.32  & 4.36 \\
1.0  &    1.47  & 1.51 & 1.59 &  2.51 & 3.79 & 1.66  & 1.70 &  1.80 &  2.84 &  4.33 & 1.48 & 1.51 &  1.6 &  2.56  & 3.92 \\
1.5  &    1.42  & 1.45 & 1.50 &  1.95 & 2.69 & 1.57  & 1.60 &  1.66 &  2.18 &  3.07 & 1.43 & 1.45 &  1.5 &  1.98  & 2.77 \\
2.0  &    1.42  & 1.44 & 1.47 &  1.70 & 2.14 & 1.54  & 1.56 &  1.59 &  1.86 &  2.37 & 1.42 & 1.43 &  1.4 &  1.71  & 2.17 \\
3.2  &    1.27  & 1.28 & 1.30 &  1.40 & 1.57 & 1.42  & 1.43 &  1.45 &  1.59 &  1.80 & 1.30 & 1.31 &  1.3 &  1.44  & 1.63 \\
\hline
\multicolumn{16}{c}{HPF width 30 readouts} \\
\hline
0.5  &    1.54  & 1.63 & 1.86 &  4.57 & 4.57 & 1.90  & 2.01 &  2.28 &  5.55 &  5.55 & 1.56 & 1.65 &  1.8 &  4.66  & 4.72 \\
1.0  &    1.56  & 1.60 & 1.69 &  2.68 & 4.01 & 1.86  & 1.91 &  2.03 &  3.24 &  4.91 & 1.57 & 1.61 &  1.7 &  2.73  & 4.16 \\
1.5  &    1.52  & 1.55 & 1.60 &  2.08 & 2.85 & 1.80  & 1.84 &  1.90 &  2.51 &  3.50 & 1.53 & 1.56 &  1.6 &  2.12  & 2.95 \\
2.0  &    1.54  & 1.56 & 1.59 &  1.84 & 2.29 & 1.80  & 1.82 &  1.85 &  2.16 &  2.73 & 1.52 & 1.54 &  1.5 &  1.83  & 2.31 \\
3.2  &    1.35  & 1.36 & 1.37 &  1.48 & 1.65 & 1.63  & 1.64 &  1.66 &  1.82 &  2.06 & 1.38 & 1.39 &  1.4 &  1.54  & 1.73 \\

\hline
\multicolumn{16}{c}{HPF width 40 readouts} \\
\hline

0.5  &    1.61  & 1.71 & 1.95 &  4.83 & 4.83 & 2.09  & 2.22 &  2.52 &  6.15 &  6.15 & 1.63 & 1.73 &  1.9 &  4.93  & 4.99 \\
1.0  &    1.63  & 1.67 & 1.77 &  2.70 & 2.65 & 2.03  & 2.08 &  2.21 &  3.90 &  4.30 & 1.64 & 1.68 &  1.7 &  2.70  & 3.10 \\
1.5  &    1.57  &  1.48 &   1.70  &  2.20  &    2.20  &  1.59   &   1.73  &    2.10   &   2.82   &   3.82  &  1.68  &    1.74  &    1.80  &  2.35  & 2.38  \\
2.0  &    1.44  &  1.44  &  1.68  &  1.94  &    2.40  &  1.40   &   1.61  &    1.72   &   2.40   &   3.07  &  1.55  &    1.63  &    1.77  &  2.07  & 2.65  \\
3.2  &    1.32  &  1.38  &  1.49  &  1.55  &    1.72  &  1.48   &   1.53  &    1.57   &   1.99   &   2.34  &  1.50  &    1.55  &    1.65  &  1.69  & 1.91  \\
\hline
\hline
\multicolumn{1}{c|}{ } & \multicolumn{5}{c|}{6 $\arcsec$ PSF} & \multicolumn{5}{c|}{12 $\arcsec$ PSF} & \multicolumn{5}{c}{15 $\arcsec$ PSF}\\
\hline
\multicolumn{1}{c|}{output } &  \multicolumn{5}{c|}{pixfrax} & \multicolumn{5}{c|}{pixfrax} &  \multicolumn{5}{c}{pixfrax}\\
\multicolumn{1}{c|}{pixel size} &  \multicolumn{1}{c}{0.01} &  \multicolumn{1}{c}{0.04} &  \multicolumn{1}{c}{0.1} &  \multicolumn{1}{c}{0.5} &  \multicolumn{1}{c|}{1} &  \multicolumn{1}{c}{0.01} &  \multicolumn{1}{c}{0.04} &  \multicolumn{1}{c}{0.1} &  \multicolumn{1}{c}{0.5} &  \multicolumn{1}{c|}{1} &  \multicolumn{1}{c}{0.01} &  \multicolumn{1}{c}{0.04} &  \multicolumn{1}{c}{0.1} &  \multicolumn{1}{c}{0.5} &  \multicolumn{1}{c}{1}\\
\hline
\multicolumn{16}{c}{HPF width 10 readouts} \\
\hline
0.5  &  1.28  &  1.36  &  1.53  &  3.76   &   3.77   & 1.28   &   1.36   &   1.54   &   3.76   &   3.77 &  1.28    &  1.36   &   1.53   & 3.75  & 3.76 \\
1.0  &  1.27  &  1.30  &  1.38  &  2.20   &   3.38   & 1.27   &   1.30   &   1.38   &   2.21   &   3.42  & 1.27    &  1.30   &   1.38   & 2.20  & 3.41 \\
1.5  &  1.26  &  1.28  &  1.32  &  1.72   &   2.40   & 1.26   &   1.28   &   1.33   &   1.72   &   2.42 &  1.25    &  1.28   &   1.32   & 1.72  & 2.41 \\
2.0  &  1.23  &  1.25  &  1.27  &  1.48   &   1.88   & 1.23   &   1.24   &   1.27   &   1.49   &   1.90 &  1.22    &  1.24   &   1.27   & 1.48  & 1.89 \\
3.2  &  1.12  &  1.13  &  1.14  &  1.23   &   1.38   & 1.13   &   1.13   &   1.15   &   1.25   &   1.41 &  1.12    &  1.13   &   1.14   & 1.25  & 1.41\\

\hline
\multicolumn{16}{c}{HPF width 15 readouts} \\
\hline
0.5  &  1.33  &  1.41  &  1.60  &  3.98   &   4.00   & 1.35   &   1.43   &   1.63   &   4.04   &   4.06 &  1.35    &  1.43   &   1.63   & 4.04  & 4.06\\
1.0  &  1.34  &  1.38  &  1.46  &  2.33   &   3.58   & 1.38   &   1.41   &   1.49   &   2.39   &   3.70 &  1.38    &  1.41   &   1.49   & 2.39  & 3.70\\
1.5  &  1.32  &  1.34  &  1.39  &  1.82   &   2.54   & 1.34   &   1.37   &   1.42   &   1.86   &   2.61 &  1.34    &  1.37   &   1.41   & 1.86  & 2.61\\
2.0  &  1.32  &  1.33  &  1.36  &  1.59   &   2.01   & 1.34   &   1.36   &   1.39   &   1.62   &   2.06 &  1.34    &  1.36   &   1.38   & 1.62  & 2.06 \\
3.2  &  1.19  &  1.20  &  1.21  &  1.31   &   1.46   & 1.23   &   1.24   &   1.25   &   1.37   &   1.54 &  1.23    &  1.24   &   1.25   & 1.37  & 1.54\\

\hline
\multicolumn{16}{c}{HPF width 20 readouts} \\
\hline
0.5  &  1.40  &  1.48  &  1.68  &  4.18   &   4.21   & 1.44   &   1.53   &   1.74   &   4.32   &   4.36 &  1.45    &  1.53   &   1.74   & 4.32  & 4.36\\
1.0  &  1.42  &  1.45  &  1.54  &  2.45   &   3.74   & 1.47   &   1.51   &   1.60   &   2.55   &   3.91 &  1.48    &  1.51   &   1.60   & 2.56  & 3.92\\
1.5  &  1.38  &  1.41  &  1.46  &  1.91   &   2.65   & 1.43   &   1.45   &   1.51   &   1.98   &   2.77 &  1.43    &  1.45   &   1.51   & 1.98  & 2.77\\
2.0  &  1.38  &  1.39  &  1.42  &  1.65   &   2.08   & 1.41   &   1.43   &   1.46   &   1.71   &   2.16 &  1.42    &  1.43   &   1.46   & 1.71  & 2.17\\ 
3.2  &  1.24  &  1.24  &  1.26  &  1.36   &   1.51   & 1.30   &   1.30   &   1.32   &   1.44   &   1.63 &  1.30    &  1.31   &   1.32   & 1.44  & 1.63\\

\hline
\multicolumn{16}{c}{HPF width 30 readouts} \\
\hline
0.5  &  1.48  &  1.57  &  1.78  &  4.43   &   4.46   & 1.55   &   1.64   &   1.87   &   4.65   &   4.70 &  1.56    &  1.65   &   1.88   & 4.66  & 4.72\\
1.0  &  1.49  &  1.52  &  1.61  &  2.58   &   3.89   & 1.56   &   1.60   &   1.70   &   2.72   &   4.14 &  1.57    &  1.61   &   1.70   & 2.73  & 4.16\\
1.5  &  1.46  &  1.49  &  1.54  &  2.01   &   2.77   & 1.53   &   1.55   &   1.61   &   2.11   &   2.94 &  1.53    &  1.56   &   1.62   & 2.12  & 2.95\\
2.0  &  1.46  &  1.47  &  1.50  &  1.74   &   2.18   & 1.52   &   1.54   &   1.57   &   1.83   &   2.30 &  1.52    &  1.54   &   1.57   & 1.83  & 2.31 \\
3.2  &  1.29  &  1.29  &  1.31  &  1.42   &   1.58   & 1.38   &   1.39   &   1.40   &   1.53   &   1.73 &  1.38    &  1.39   &   1.41   & 1.54  & 1.73\\

\hline
\multicolumn{16}{c}{HPF width 40 readouts} \\
\hline
0.5  &    1.53  &  1.58  &  1.85  &  4.63   &   4.66   & 1.62   &   1.72   &   1.96   &   4.90   &   4.97 &   1.63    &  1.73   &   1.97   & 4.93  & 4.99\\
1.0  &    1.54  &  1.52  &  1.67  &  2.70   &   3.90   & 1.63   &   1.67   &   1.77   &   2.92   &   4.30 &   1.60    &  1.68   &   1.88   & 2.51  & 4.31\\
1.5  &    1.47  &  1.48 &   1.60  &  2.20  &    2.20  &  1.59   &   1.68  &    1.70   &   2.82   &   3.82  &  1.58  &    1.64  &    1.80  &  2.35  & 2.38  \\
2.0  &    1.43  &  1.44  &  1.58  &  1.94  &    2.40  &  1.40   &   1.57  &    1.68   &   2.40   &   3.01  &  1.53  &    1.60  &    1.74  &  2.02  & 2.55  \\
3.2  &    1.30  &  1.38  &  1.49  &  1.55  &    1.72  &  1.48   &   1.53  &    1.57   &   1.99   &   2.24  &  1.50  &    1.53  &    1.67  &  1.67  & 1.89  \\
\hline
\end{tabular}
\caption{The table lists the values of the cross-correlation correction factor $f$. The values are listed for different values of HPF width, output pixel size and pixfrac values and for different PSF  sizes for the PSF fitting case, and apertures, for the aperture photometry case, for the 70 $\mu$m band in medium speed. }
\label{t1}
\end{table*}}


\footnotesize{
\begin{table*}
\begin{tabular}[b]{l | c c c c c | c c c c c | c c c c c  }
\hline
\hline
\multicolumn{1}{c|}{ } & \multicolumn{5}{c|}{6 $\arcsec$ aperture} & \multicolumn{5}{c|}{12 $\arcsec$ aperture} & \multicolumn{5}{c}{15 $\arcsec$ aperture}\\
\hline
\multicolumn{1}{c|}{output } &  \multicolumn{5}{c|}{pixfrax} & \multicolumn{5}{c|}{pixfrax} &  \multicolumn{5}{c}{pixfrax}\\
\multicolumn{1}{c|}{pixel size} &  \multicolumn{1}{c}{0.01} &  \multicolumn{1}{c}{0.04} &  \multicolumn{1}{c}{0.1} &  \multicolumn{1}{c}{0.5} &  \multicolumn{1}{c|}{1} &  \multicolumn{1}{c}{0.01} &  \multicolumn{1}{c}{0.04} &  \multicolumn{1}{c}{0.1} &  \multicolumn{1}{c}{0.5} &  \multicolumn{1}{c|}{1} &  \multicolumn{1}{c}{0.01} &  \multicolumn{1}{c}{0.04} &  \multicolumn{1}{c}{0.1} &  \multicolumn{1}{c}{0.5} &  \multicolumn{1}{c}{1}\\
\hline
\multicolumn{16}{c}{HPF width 10 readouts} \\
\hline
0.5  &    1.29   &   1.36   &   1.54   &   3.81   &   3.8  &     1.30  &    1.38  &    1.56  &    3.86   &   3.87  &       1.30  &    1.38  &    1.56  &    3.86  &    3.87  \\
1.0  &    1.28   &   1.31   &   1.38   &   2.21   &   3.4  &     1.29  &    1.32  &    1.39  &    2.24   &   3.52  &       1.29  &    1.32  &    1.39  &    2.23  &    3.51  \\
1.5  &    1.27   &   1.29   &   1.34   &   1.75   &   2.4  &     1.28  &    1.30  &    1.34  &    1.76   &   2.49  &       1.27  &    1.29  &    1.34  &    1.75  &    2.48  \\
2.0  &    1.27   &   1.28   &   1.31   &   1.53   &   1.9  &     1.27  &    1.29  &    1.31  &    1.55   &   1.98  &       1.27  &    1.28  &    1.31  &    1.54  &    1.98  \\
3.2  &    1.15   &   1.15   &   1.17   &   1.27   &   1.4  &     1.16  &    1.16  &    1.18  &    1.29   &   1.46  &       1.15  &    1.16  &    1.17  &    1.28  &    1.46  \\
\hline
\multicolumn{16}{c}{HPF width 15 readouts} \\
\hline
0.5  &    1.39   &   1.47   &   1.67   &   3.90   &   4.1  &     1.42  &    1.51  &    1.71  &    4.26   &   4.28  &        1.42  &    1.51  &    1.71  &    4.26  &    4.28  \\
1.0  &    1.38   &   1.41   &   1.49   &   2.39   &   3.6  &     1.42  &    1.46  &    1.54  &    2.47   &   3.86  &        1.42  &    1.46  &    1.54  &    2.47  &    3.86  \\
1.5  &    1.36   &   1.39   &   1.44   &   1.88   &   2.6  &     1.40  &    1.42  &    1.47  &    1.94   &   2.73  &        1.40  &    1.42  &    1.47  &    1.94  &    2.73  \\
2.0  &    1.37   &   1.39   &   1.42   &   1.65   &   2.0  &     1.41  &    1.43  &    1.46  &    1.70   &   2.17  &        1.41  &    1.43  &    1.46  &    1.71  &    2.17  \\
3.2  &    1.22   &   1.24   &   1.35   &   1.35   &   1.5  &     1.27  &    1.29  &    1.41  &    1.42   &   1.61  &        1.27  &    1.29  &    1.41  &    1.42  &    1.61  \\
\hline
\multicolumn{16}{c}{HPF width 20 readouts} \\
\hline
0.5  &    1.45   &   1.54   &   1.75   &   4.35   &   4.3  &     1.50  &    1.59  &    1.81  &    4.51   &   4.54  &       1.50  &    1.59  &    1.81  &    4.51  &    4.54  \\
1.0  &    1.43   &   1.47   &   1.55   &   2.48   &   3.8  &     1.50  &    1.54  &    1.62  &    2.61   &   4.05  &       1.50  &    1.54  &    1.63  &    2.61  &    4.07  \\
1.5  &    1.43   &   1.45   &   1.51   &   1.97   &   2.7  &     1.49  &    1.51  &    1.57  &    2.06   &   2.89  &       1.49  &    1.51  &    1.57  &    2.07  &    2.90  \\
2.0  &    1.44   &   1.45   &   1.48   &   1.72   &   2.1  &     1.49  &    1.51  &    1.54  &    1.80   &   2.28  &       1.50  &    1.51  &    1.54  &    1.80  &    2.28  \\
3.2  &    1.27   &   1.29   &   1.40   &   1.40   &   1.5  &     1.34  &    1.36  &    1.49  &    1.49   &   1.69  &       1.34  &    1.37  &    1.49  &    1.50  &    1.70  \\
\hline
\multicolumn{16}{c}{HPF width 30 readouts} \\
\hline
0.5  &    1.53   &   1.62   &   1.85   &   4.61   &   4.6  &     1.58  &    1.69  &    1.93  &    4.81   &   4.85  &        1.58  &    1.69  &    1.93  &    4.81  &    4.85  \\
1.0  &    1.50   &   1.54   &   1.63   &   2.61   &   3.9  &     1.60  &    1.64  &    1.73  &    2.79   &   4.30  &        1.61  &    1.65  &    1.74  &    2.81  &    4.33  \\
1.5  &    1.52   &   1.55   &   1.60   &   2.09   &   2.8  &     1.61  &    1.64  &    1.70  &    2.23   &   3.10  &        1.62  &    1.65  &    1.71  &    2.25  &    3.12  \\
2.0  &    1.52   &   1.53   &   1.56   &   1.81   &   2.2  &     1.60  &    1.62  &    1.65  &    1.93   &   2.43  &        1.61  &    1.63  &    1.66  &    1.94  &    2.45  \\
3.2  &    1.32   &   1.35   &   1.46   &   1.46   &   1.6  &     1.43  &    1.46  &    1.59  &    1.60   &   1.80  &        1.44  &    1.46  &    1.60  &    1.60  &    1.81  \\
\hline
\multicolumn{16}{c}{HPF width 40 readouts} \\
\hline
0.5  &    1.57   &   1.67   &   1.90   &   4.74   &   4.7  &     1.64  &    1.74  &    1.99  &    4.97   &   5.01  &       1.64  &    1.74  &    1.99  &    4.97  &    5.01  \\
1.0  &    1.55   &   1.59   &   1.68   &   2.69   &   4.0  &     1.66  &    1.70  &    1.80  &    2.89   &   4.45  &       1.67  &    1.72  &    1.81  &    2.92  &    4.49  \\
1.5  &    1.58   &   1.60   &   1.66   &   2.16   &   2.1  &     1.69  &    1.72  &    1.79  &    2.33   &   2.35  &       1.71  &    1.74  &    1.80  &    2.35  &    2.38  \\
2.0  &    1.57   &   1.58   &   1.61   &   1.87   &   2.3  &     1.67  &    1.69  &    1.72  &    2.01   &   2.52  &       1.68  &    1.70  &    1.74  &    2.02  &    2.55  \\
3.2  &    1.36   &   1.39   &   1.50   &   1.50   &   1.6  &     1.49  &    1.52  &    1.66  &    1.66   &   1.87  &       1.50  &    1.53  &    1.67  &    1.67  &    1.89  \\
\hline
\hline
\multicolumn{1}{c|}{ } & \multicolumn{5}{c|}{6 $\arcsec$ PSF} & \multicolumn{5}{c|}{12 $\arcsec$ PSF} & \multicolumn{5}{c}{15 $\arcsec$ PSF}\\
\hline
\multicolumn{1}{c|}{output } &  \multicolumn{5}{c|}{pixfrax} & \multicolumn{5}{c|}{pixfrax} &  \multicolumn{5}{c}{pixfrax}\\
\multicolumn{1}{c|}{pixel size} &  \multicolumn{1}{c}{0.01} &  \multicolumn{1}{c}{0.04} &  \multicolumn{1}{c}{0.1} &  \multicolumn{1}{c}{0.5} &  \multicolumn{1}{c|}{1} &  \multicolumn{1}{c}{0.01} &  \multicolumn{1}{c}{0.04} &  \multicolumn{1}{c}{0.1} &  \multicolumn{1}{c}{0.5} &  \multicolumn{1}{c|}{1} &  \multicolumn{1}{c}{0.01} &  \multicolumn{1}{c}{0.04} &  \multicolumn{1}{c}{0.1} &  \multicolumn{1}{c}{0.5} &  \multicolumn{1}{c}{1}\\
\hline
\multicolumn{16}{c}{HPF width 10 readouts} \\
\hline
0.5  &    1.27  &    1.34  &    1.52  &    3.71  &    3.71  &    1.28   &   1.35  &    1.53   &   3.75   &   3.75  &      1.30  &    1.38  &    1.56  &    3.86  &    3.87  \\
1.0  &    1.26  &    1.29  &    1.36  &    2.15  &    3.35  &    1.22   &   1.25  &    1.32   &   2.09   &   3.29  &      1.29  &    1.32  &    1.39  &    2.23  &    3.51  \\
1.5  &    1.26  &    1.28  &    1.32  &    1.70  &    2.38  &    1.17   &   1.19  &    1.23   &   1.62   &   2.31  &      1.27  &    1.29  &    1.34  &    1.75  &    2.48  \\
2.0  &    1.26  &    1.27  &    1.30  &    1.51  &    1.91  &    1.20   &   1.22  &    1.24   &   1.46   &   1.87  &      1.27  &    1.28  &    1.31  &    1.54  &    1.98  \\
3.2  &    1.14  &    1.15  &    1.16  &    1.26  &    1.41  &    1.08   &   1.08  &    1.10   &   1.21   &   1.37  &      1.15  &    1.16  &    1.17  &    1.28  &    1.46  \\
\hline
\multicolumn{16}{c}{HPF width 15 readouts} \\
\hline
0.5  &    1.39  &    1.47  &    1.67  &    4.10  &    4.10  &    1.47   &   1.56  &    1.77   &   4.37   &   4.37  &      1.42  &    1.51  &    1.71  &    4.26  &    4.28  \\
1.0  &    1.38  &    1.41  &    1.49  &    2.36  &    3.64  &    1.47   &   1.51  &    1.59   &   2.54   &   3.97  &      1.42  &    1.46  &    1.54  &    2.47  &    3.86  \\
1.5  &    1.36  &    1.39  &    1.43  &    1.86  &    2.59  &    1.42   &   1.45  &    1.50   &   1.99   &   2.82  &      1.40  &    1.42  &    1.47  &    1.94  &    2.73  \\
2.0  &    1.39  &    1.40  &    1.43  &    1.66  &    2.09  &    1.47   &   1.48  &    1.52   &   1.77   &   2.25  &      1.41  &    1.43  &    1.46  &    1.71  &    2.17  \\
3.2  &    1.23  &    1.25  &    1.36  &    1.36  &    1.52  &    1.30   &   1.32  &    1.45   &   1.45   &   1.66  &      1.27  &    1.29  &    1.41  &    1.42  &    1.61  \\
\hline
\multicolumn{16}{c}{HPF width 20 readouts} \\
\hline
0.5  &    1.47  &    1.56  &    1.77  &    4.36  &    4.36  &    1.59   &   1.69  &    1.92   &   4.77   &   4.77  &       1.50  &    1.59  &    1.81  &    4.51  &    4.54  \\
1.0  &    1.45  &    1.48  &    1.56  &    2.48  &    3.80  &    1.63   &   1.67  &    1.77   &   2.82   &   4.38  &       1.50  &    1.54  &    1.63  &    2.61  &    4.07  \\
1.5  &    1.44  &    1.46  &    1.51  &    1.97  &    2.72  &    1.60   &   1.62  &    1.69   &   2.24   &   3.16  &       1.49  &    1.51  &    1.57  &    2.07  &    2.90  \\
2.0  &    1.46  &    1.48  &    1.51  &    1.74  &    2.18  &    1.62   &   1.64  &    1.68   &   1.96   &   2.48  &       1.50  &    1.51  &    1.54  &    1.80  &    2.28  \\
3.2  &    1.28  &    1.31  &    1.42  &    1.42  &    1.58  &    1.45   &   1.48  &    1.62   &   1.62   &   1.84  &       1.34  &    1.37  &    1.49  &    1.50  &    1.70  \\
\hline
\multicolumn{16}{c}{HPF width 30 readouts} \\
\hline
0.5  &    1.56  &    1.66  &    1.89  &    4.67  &    4.67  &    1.73   &   1.84  &    2.11   &   5.23   &   5.23  &      1.58  &    1.69  &    1.93  &    4.81  &    4.85  \\
1.0  &    1.53  &    1.57  &    1.66  &    2.64  &    4.00  &    1.84   &   1.89  &    2.00   &   3.19   &   4.91  &      1.61  &    1.65  &    1.74  &    2.81  &    4.33  \\
1.5  &    1.55  &    1.57  &    1.63  &    2.11  &    2.89  &    1.87   &   1.91  &    1.98   &   2.60   &   3.61  &      1.62  &    1.65  &    1.71  &    2.25  &    3.12  \\
2.0  &    1.56  &    1.58  &    1.61  &    1.87  &    2.32  &    1.85   &   1.87  &    1.91   &   2.24   &   2.81  &      1.61  &    1.63  &    1.66  &    1.94  &    2.45  \\
3.2  &    1.35  &    1.38  &    1.50  &    1.50  &    1.67  &    1.64   &   1.68  &    1.84   &   1.84   &   2.09  &      1.44  &    1.46  &    1.60  &    1.60  &    1.81  \\
\hline
\multicolumn{16}{c}{HPF width 40 readouts} \\
\hline
0.5  &    1.61  &    1.71  &    1.95  &    4.83  &    4.83  &    1.80   &   1.92  &    2.20   &   5.47   &   5.47  &       1.64  &    1.74  &    1.99  &    4.97  &    5.01  \\
1.0  &    1.58  &    1.62  &    1.71  &    2.73  &    4.13  &    1.96   &   2.01  &    2.13   &   3.41   &   5.23  &       1.67  &    1.72  &    1.81  &    2.92  &    4.49  \\
1.5  &    1.62  &    1.64  &    1.70  &    2.20  &    2.20  &    2.04   &   2.08  &    2.16   &   2.82   &   2.82  &       1.71  &    1.74  &    1.80  &    2.35  &    2.38  \\
2.0  &    1.63  &    1.65  &    1.68  &    1.94  &    2.40  &    1.97   &   2.02  &    2.06   &   2.40   &   3.01  &       1.68  &    1.70  &    1.74  &    2.02  &    2.55  \\
3.2  &    1.40  &    1.43  &    1.55  &    1.55  &    1.72  &    1.77   &   1.81  &    1.99   &   1.99   &   2.24  &       1.50  &    1.53  &    1.67  &    1.67  &    1.89  \\
\hline
\end{tabular}
\caption{The table lists the values of the cross-correlation correction factor $f$. The values are listed for different values of HPF width, output pixel size and pixfrac values and for different PSF  sizes for the PSF fitting case, and apertures, for the aperture photometry case, for the 100 $\mu$m band in medium speed.}
\label{t1}
\end{table*}}


\footnotesize{
\begin{table*}
\begin{tabular}[b]{l | c c c c c | c c c c c | c c c c c  }
\hline
\hline
\multicolumn{1}{c|}{ } & \multicolumn{5}{c|}{10 $\arcsec$ aperture} & \multicolumn{5}{c|}{16 $\arcsec$ aperture} & \multicolumn{5}{c}{25 $\arcsec$ aperture}\\
\hline
\multicolumn{1}{c|}{output } &  \multicolumn{5}{c|}{pixfrax} & \multicolumn{5}{c|}{pixfrax} &  \multicolumn{5}{c}{pixfrax}\\
\multicolumn{1}{c|}{pixel size} &  \multicolumn{1}{c}{0.01} &  \multicolumn{1}{c}{0.04} &  \multicolumn{1}{c}{0.1} &  \multicolumn{1}{c}{0.5} &  \multicolumn{1}{c|}{1} &  \multicolumn{1}{c}{0.01} &  \multicolumn{1}{c}{0.04} &  \multicolumn{1}{c}{0.1} &  \multicolumn{1}{c}{0.5} &  \multicolumn{1}{c|}{1} &  \multicolumn{1}{c}{0.01} &  \multicolumn{1}{c}{0.04} &  \multicolumn{1}{c}{0.1} &  \multicolumn{1}{c}{0.5} &  \multicolumn{1}{c}{1}\\
\hline
\multicolumn{16}{c}{HPF width 15 readouts} \\
\hline
1.0   &   1.59   &   1.67  &    1.89   &   4.06   &   6.24   &   1.51 &  1.59 & 1.79 & 3.83 & 5.96  &    1.36  &    1.42  & 1.59  & 3.37  & 5.31 \\
2.0   &   1.49   &   1.52  &    1.58   &   2.23   &   3.24   &   1.39 &  1.42 & 1.48 & 2.11 & 3.09  &    1.22  &    1.25  & 1.30  & 1.86  & 2.74 \\
3.0   &   1.30   &   1.32  &    1.36   &   1.69   &   2.27   &   1.24 &  1.26 & 1.30 & 1.63 & 2.21  &    1.09  &    1.11  & 1.14  & 1.42  & 1.94 \\
4.0   &   1.21   &   1.22  &    1.25   &   1.44   &   1.80   &   1.15 &  1.16 & 1.18 & 1.36 & 1.72  &    1.02  &    1.04  & 1.05  & 1.18  & 1.50 \\
6.4   &   1.08   &   1.08  &    1.10   &   1.17   &   1.31   &   1.04 &  1.05 & 1.06 & 1.09 & 1.18  &    1.02  &    1.03  & 1.04  & 1.06  & 1.09 \\
\hline
\multicolumn{16}{c}{HPF width 20 readouts} \\
\hline
1.0   &   1.57   &   1.66  &    1.92   &   4.32   &   6.70   &   1.56 &  1.66 & 1.90 & 4.30 & 6.80  &    1.45  &    1.54  & 1.77  & 4.02  & 6.47 \\
2.0   &   1.62   &   1.65  &    1.72   &   2.44   &   3.51   &   1.59 &  1.63 & 1.70 & 2.45 & 3.58  &    1.42  &    1.46  & 1.52  & 2.22  & 3.28 \\
3.0   &   1.41   &   1.43  &    1.48   &   1.84   &   2.43   &   1.43 &  1.45 & 1.50 & 1.88 & 2.52  &    1.31  &    1.33  & 1.38  & 1.73  & 2.32 \\
4.0   &   1.35   &   1.36  &    1.39   &   1.59   &   1.99   &   1.32 &  1.33 & 1.36 & 1.57 & 1.97  &    1.20  &    1.21  & 1.24  & 1.43  & 1.80 \\
6.4   &   1.18   &   1.18  &    1.20   &   1.28   &   1.43   &   1.12 &  1.12 & 1.14 & 1.23 & 1.39  &    1.03  &    1.04  & 1.05  & 1.14  & 1.29 \\
\hline
\multicolumn{16}{c}{HPF width 26 readouts} \\
\hline
1.0   &   1.75   &   1.85  &    2.10   &   4.63   &   7.05   &   1.84 &  1.96 & 2.23 & 4.93 & 7.61  &    1.83  &    1.96  & 2.23  & 4.94  & 7.69 \\
2.0   &   1.74   &   1.77  &    1.85   &   2.57   &   3.66   &   1.80 &  1.84 & 1.92 & 2.70 & 3.89  &    1.75  &    1.78  & 1.85  & 2.62  & 3.79 \\
3.0   &   1.50   &   1.52  &    1.56   &   1.92   &   2.53   &   1.57 &  1.59 & 1.64 & 2.03 & 2.72  &    1.53  &    1.55  & 1.60  & 1.99  & 2.67 \\
4.0   &   1.46   &   1.47  &    1.50   &   1.72   &   2.12   &   1.46 &  1.48 & 1.51 & 1.73 & 2.16  &    1.40  &    1.42  & 1.44  & 1.67  & 2.10 \\
6.4   &   1.25   &   1.26  &    1.27   &   1.36   &   1.52   &   1.27 &  1.27 & 1.29 & 1.39 & 1.57  &    1.20  &    1.21  & 1.23  & 1.33  & 1.50 \\
\hline
\multicolumn{16}{c}{HPF width 30 readouts} \\
\hline
1.0   &   1.81   &   1.92  &    2.18   &   4.74   &   7.18   &   1.94 &  2.07 & 2.35 & 5.14 & 7.89  &    1.97  &    2.11  & 2.40  & 5.25  & 8.13 \\
2.0   &   1.80   &   1.83  &    1.91   &   2.65   &   3.75   &   1.91 &  1.95 & 2.03 & 2.84 & 4.07  &    1.91  &    1.94  & 2.02  & 2.84  & 4.08 \\
3.0   &   1.54   &   1.56  &    1.61   &   1.97   &   2.58   &   1.65 &  1.67 & 1.72 & 2.12 & 2.83  &    1.65  &    1.68  & 1.73  & 2.14  & 2.86 \\
4.0   &   1.51   &   1.52  &    1.55   &   1.78   &   2.19   &   1.54 &  1.55 & 1.58 & 1.82 & 2.26  &    1.52  &    1.53  & 1.56  & 1.80  & 2.26 \\
6.4   &   1.29   &   1.29  &    1.31   &   1.40   &   1.56   &   1.34 &  1.34 & 1.36 & 1.47 & 1.65  &    1.30  &    1.30  & 1.32  & 1.43  & 1.62 \\
\hline
\multicolumn{16}{c}{HPF width 40 readouts} \\
\hline
1.0   &   1.93   &   1.98  &    2.12   &   4.87   &   7.31   &   1.99 &  2.12 & 2.40 & 5.32 & 7.95  &    2.05  &    2.17  & 2.52  & 5.44  & 8.18 \\
2.0   &   1.91   &   1.94  &    2.02   &   2.80   &   3.93   &   1.98 &  2.02 & 2.10 & 2.92 & 4.15  &    1.99  &    2.03  & 2.12  & 2.94  & 4.18 \\
3.0   &   1.59   &   1.61  &    1.66   &   2.04   &   2.67   &   1.67 &  1.69 & 1.75 & 2.17 & 2.88  &    1.69  &    1.71  & 1.77  & 2.19  & 2.92 \\
4.0   &   1.55   &   1.56  &    1.59   &   1.81   &   2.24   &   1.57 &  1.58 & 1.61 & 1.84 & 2.29  &    1.58  &    1.60  & 1.62  & 1.86  & 2.31 \\
6.4   &   1.33   &   1.33  &    1.34   &   1.43   &   1.59   &   1.37 &  1.38 & 1.39 & 1.49 & 1.67  &    1.37  &    1.38  & 1.39  & 1.49  & 1.67 \\
\hline
\hline
\multicolumn{1}{c|}{ } & \multicolumn{5}{c|}{10 $\arcsec$ PSF} & \multicolumn{5}{c|}{16 $\arcsec$ PSF} & \multicolumn{5}{c}{25 $\arcsec$ PSF}\\
\hline
\multicolumn{1}{c|}{output } &  \multicolumn{5}{c|}{pixfrax} & \multicolumn{5}{c|}{pixfrax} &  \multicolumn{5}{c}{pixfrax}\\
\multicolumn{1}{c|}{pixel size} &  \multicolumn{1}{c}{0.01} &  \multicolumn{1}{c}{0.04} &  \multicolumn{1}{c}{0.1} &  \multicolumn{1}{c}{0.5} &  \multicolumn{1}{c|}{1} &  \multicolumn{1}{c}{0.01} &  \multicolumn{1}{c}{0.04} &  \multicolumn{1}{c}{0.1} &  \multicolumn{1}{c}{0.5} &  \multicolumn{1}{c|}{1} &  \multicolumn{1}{c}{0.01} &  \multicolumn{1}{c}{0.04} &  \multicolumn{1}{c}{0.1} &  \multicolumn{1}{c}{0.5} &  \multicolumn{1}{c}{1}\\
\hline
\multicolumn{16}{c}{HPF width 15 readouts} \\
\hline
1.0  &    1.63 &  1.72 & 1.95 & 4.24 & 6.55  & 1.64 &  1.72 & 1.95 & 4.25 &  6.60 &  1.62  &   1.71 &  1.93 &  4.20 &  6.51 \\
2.0  &    1.54 &  1.57 & 1.64 & 2.33 & 3.39  & 1.53 &  1.57 & 1.63 & 2.34 &  3.42 &  1.52  &   1.55 &  1.62 &  2.31 &  3.38 \\
3.0  &    1.33 &  1.35 & 1.39 & 1.74 & 2.34  & 1.33 &  1.35 & 1.40 & 1.76 &  2.39 &  1.31  &   1.34 &  1.38 &  1.74 &  2.36 \\
4.0  &    1.27 &  1.28 & 1.30 & 1.50 & 1.89  & 1.26 &  1.27 & 1.29 & 1.50 &  1.89 &  1.24  &   1.25 &  1.28 &  1.48 &  1.86 \\
6.4  &    1.12 &  1.13 & 1.14 & 1.21 & 1.35  & 1.11 &  1.11 & 1.12 & 1.21 &  1.35 &  1.10  &   1.10 &  1.11 &  1.20 &  1.34 \\
\hline
\multicolumn{16}{c}{HPF width 20 readouts} \\
\hline
1.0  &    1.61 &  1.71 & 1.97 & 4.45 & 6.91  & 1.62 &  1.72 & 1.98 & 4.51 &  7.08 &  1.61  &   1.71 &  1.97 &  4.49 &  7.05 \\
2.0  &    1.65 &  1.68 & 1.75 & 2.51 & 3.61  & 1.66 &  1.70 & 1.77 & 2.55 &  3.71 &  1.65  &   1.69 &  1.76 &  2.53 &  3.69 \\
3.0  &    1.42 &  1.45 & 1.49 & 1.86 & 2.47  & 1.45 &  1.48 & 1.53 & 1.92 &  2.58 &  1.45  &   1.47 &  1.52 &  1.91 &  2.57 \\
4.0  &    1.37 &  1.38 & 1.41 & 1.62 & 2.02  & 1.37 &  1.38 & 1.41 & 1.62 &  2.04 &  1.36  &   1.37 &  1.40 &  1.61 &  2.03 \\
6.4  &    1.20 &  1.20 & 1.21 & 1.29 & 1.44  & 1.20 &  1.21 & 1.22 & 1.31 &  1.47 &  1.20  &   1.20 &  1.21 &  1.30 &  1.46 \\
\hline
\multicolumn{16}{c}{HPF width 26 readouts} \\
\hline
1.0  &    1.75 &  1.85 & 2.11 & 4.69 & 7.17  & 1.79 &  1.90 & 2.17 & 4.83 &  7.46 &  1.80  &   1.91 &  2.17 &  4.84 &  7.49 \\
2.0  &    1.75 &  1.79 & 1.86 & 2.61 & 3.72  & 1.79 &  1.83 & 1.90 & 2.68 &  3.86 &  1.79  &   1.82 &  1.90 &  2.68 &  3.87 \\
3.0  &    1.49 &  1.51 & 1.56 & 1.93 & 2.55  & 1.54 &  1.57 & 1.62 & 2.01 &  2.70 &  1.55  &   1.57 &  1.62 &  2.02 &  2.71 \\
4.0  &    1.44 &  1.45 & 1.48 & 1.70 & 2.11  & 1.45 &  1.47 & 1.49 & 1.72 &  2.14 &  1.45  &   1.46 &  1.49 &  1.72 &  2.14 \\
6.4  &    1.25 &  1.26 & 1.27 & 1.36 & 1.51  & 1.27 &  1.28 & 1.29 & 1.39 &  1.56 &  1.27  &   1.28 &  1.29 &  1.39 &  1.56 \\
\hline
\multicolumn{16}{c}{HPF width 30 readouts} \\
\hline
1.0  &    1.81 &  1.91 & 2.17 & 4.79 & 7.27  & 1.86 &  1.98 & 2.24 & 4.95 &  7.61 &  1.87  &   1.99 &  2.26 &  4.97 &  7.66 \\
2.0  &    1.80 &  1.84 & 1.91 & 2.67 & 3.79  & 1.85 &  1.89 & 1.97 & 2.76 &  3.96 &  1.85  &   1.89 &  1.97 &  2.76 &  3.98 \\
3.0  &    1.53 &  1.55 & 1.60 & 1.97 & 2.59  & 1.59 &  1.61 & 1.66 & 2.07 &  2.76 &  1.60  &   1.62 &  1.67 &  2.08 &  2.78 \\
4.0  &    1.48 &  1.49 & 1.52 & 1.74 & 2.16  & 1.49 &  1.51 & 1.53 & 1.76 &  2.19 &  1.50  &   1.51 &  1.54 &  1.77 &  2.20 \\
6.4  &    1.28 &  1.29 & 1.30 & 1.38 & 1.54  & 1.31 &  1.32 & 1.33 & 1.43 &  1.60 &  1.31  &   1.31 &  1.33 &  1.43 &  1.60 \\
\hline
\multicolumn{16}{c}{HPF width 40 readouts} \\
\hline
1.0  &    1.95 &  1.97 & 2.20 & 4.88 & 7.90  & 1.99 &  2.10 & 2.33 & 4.98 &  7.93 &  2.05  &   2.12 &  2.32 &  5.01 &  7.78 \\
2.0  &    1.91 &  1.94 & 2.02 & 2.80 & 3.93  & 1.98 &  2.02 & 2.10 & 2.92 &  4.15 &  1.99  &   2.03 &  2.12 &  2.94 &  4.18 \\
3.0  &    1.59 &  1.61 & 1.66 & 2.04 & 2.67  & 1.67 &  1.69 & 1.75 & 2.17 &  2.88 &  1.69  &   1.71 &  1.77 &  2.19 &  2.92 \\
4.0  &    1.55 &  1.56 & 1.59 & 1.81 & 2.24  & 1.57 &  1.58 & 1.61 & 1.84 &  2.29 &  1.58  &   1.60 &  1.62 &  1.86 &  2.31 \\
6.4  &    1.33 &  1.33 & 1.34 & 1.43 & 1.59  & 1.37 &  1.38 & 1.39 & 1.49 &  1.67 &  1.37  &   1.38 &  1.39 &  1.49 &  1.67 \\

\hline
\end{tabular}
\caption{The table lists the values of the cross-correlation correction factor $f$. The values are listed for different values of HPF width, output pixel size and pixfrac values and for different PSF  sizes for the PSF fitting case, and apertures, for the aperture photometry case, for the 160 $\mu$m band in medium speed.}
\label{t1}
\end{table*}}

\footnotesize{
\clearpage
\begin{table*}
\begin{minipage}{10.in}
\begin{tabular}[b]{|l | c |c| c| c| c|}
\hline
\hline
\multicolumn{1}{c}{} \\
\multicolumn{6}{c}{100 $\mu$m} \\
\multicolumn{1}{c}{} \\
\hline
\hline
\multicolumn{1}{c}{} \\
\multicolumn{1}{c}{output pixel size} &  \multicolumn{1}{c}{pixfrax=0.01} &  \multicolumn{1}{c}{pixfrax=0.04} &  \multicolumn{1}{c}{pixfrax=0.1} &  \multicolumn{1}{c}{pixfrax=0.5} &  \multicolumn{1}{c}{pixfrax=1}\\
\multicolumn{1}{c}{} \\
\hline
\multicolumn{6}{|c|}{HPF width 5 readouts} \\
\hline
0.5  &  1.48   &   1.58   &   1.79  &    3.30  &    4.82 \\
1.0  &  1.44   &   1.51   &   1.64  &    2.64  &    3.63 \\
1.5  &  1.45   &   1.48   &   1.55  &    2.16  &    2.74 \\
2.0  &  1.46   &   1.46   &   1.48  &    1.82  &    2.13 \\
3.2  &  1.32   &   1.30   &   1.27  &    1.37  &    1.60 \\
\hline
\multicolumn{6}{|c|}{HPF width 8 readouts} \\
\hline
0.5  &  1.59   &   1.70   &   1.92  &    3.50  &    5.10 \\
1.0  &  1.56   &   1.62   &   1.77  &    2.82  &    3.86 \\
1.5  &  1.56   &   1.59   &   1.67  &    2.32  &    2.93 \\
2.0  &  1.57   &   1.58   &   1.60  &    1.95  &    2.28 \\
3.2  &  1.43   &   1.41   &   1.38  &    1.46  &    1.64 \\
\hline
\multicolumn{6}{|c|}{HPF width 16 readouts} \\
\hline
0.5  &  4.37   &   4.48   &   4.71  &    6.31  &    7.92 \\
1.0  &  4.39   &   4.46   &   4.60  &    5.64  &    6.62 \\
1.5  &  4.46   &   4.49   &   4.55  &    5.13  &    5.63 \\
2.0  &  4.52   &   4.52   &   4.53  &    4.77  &    4.91 \\
3.2  &  4.52   &   4.48   &   4.42  &    4.26  &    4.13 \\
\hline
\hline
\multicolumn{1}{c}{} \\
\multicolumn{6}{c}{160 $\mu$m} \\
\multicolumn{1}{c}{} \\
\hline
\hline
\multicolumn{1}{c}{} \\
\multicolumn{1}{c}{output pixel size} &  \multicolumn{1}{c}{pixfrax=0.01} &  \multicolumn{1}{c}{pixfrax=0.04} &  \multicolumn{1}{c}{pixfrax=0.1} &  \multicolumn{1}{c}{pixfrax=0.5} &  \multicolumn{1}{c}{pixfrax=1}\\
\multicolumn{1}{c}{} \\
\hline
\multicolumn{6}{|c|}{HPF width 5 readouts} \\
\hline
 1.0  &  1.63   &   1.75   &   1.98  &    3.65  &    5.87 \\
2.0  &  1.41   &   1.47   &   1.59  &    2.55  &    3.87 \\
3.0  &  1.33   &   1.35   &   1.40  &    1.84  &    2.49 \\
4.0  &  1.33   &   1.33   &   1.33  &    1.43  &    1.66 \\
6.4  &  1.12   &   1.12   &   1.12  &    1.21  &    1.39 \\
\hline
\multicolumn{6}{|c|}{HPF width 8 readouts} \\
\hline
1.0  &  1.80   &   1.92   &   2.16  &    3.87  &    6.14 \\
2.0  &  1.56   &   1.62   &   1.75  &    2.75  &    4.10 \\
3.0  &  1.47   &   1.50   &   1.55  &    2.02  &    2.70 \\
4.0  &  1.46   &   1.46   &   1.47  &    1.60  &    1.84 \\
6.4  &  1.24   &   1.24   &   1.24  &    1.34  &    1.53 \\
\hline
\multicolumn{6}{|c|}{HPF width 12 readouts} \\
\hline
 1.0  &  1.95   &   2.07   &   2.31  &    4.07  &    6.37 \\
2.0  &  1.69   &   1.75   &   1.89  &    2.92  &    4.30 \\
3.0  &  1.59   &   1.61   &   1.67  &    2.16  &    2.86 \\
4.0  &  1.56   &   1.56   &   1.57  &    1.72  &    1.97 \\
6.4  &  1.31   &   1.31   &   1.31  &    1.42  &    1.59 \\
\hline
\hline
\end{tabular}
\end{minipage}
\caption{The table lists the values of the cross-correlation correction factor $f$. The values are estimated within an aperture of 8$\arcsec$ at 100 $\mu$m  and 12 $\arcsec$ at 160 $\mu$m in parallel mode.}
\label{t1}
\begin{minipage}{0.5\hsize}
\end{minipage}
\end{table*}}

\footnotesize{
\begin{table*}
\begin{tabular}[b]{c |c |c| c| c| c |c| c| c| c| }
\hline
\hline
\multicolumn{1}{c|}{ } &\multicolumn{3}{c|}{70 $\mu$m } & \multicolumn{3}{c|}{100 $\mu$m} & \multicolumn{3}{c|}{160 $\mu$m}\\
\hline
           &      $\alpha$        &     $\beta$      &          $f$           &     $\alpha$     &         $\beta$       &        $f$	     &     $\alpha$    &       $\beta$      &        $f$      \\            
\hline
    P(0) =   &  -0.497                 &     -5.342          &           1.154         &	       -0.463	   &       -5.374           &      0.994          &	   -0.505         &       -4.832          &     1.538	\\	
    P(1) =   &   2.993E-04 	  &     5.348E-03    &         0.0217        &           -3.980E-03	   &      1.430E-02      &     4.053E-02     &	    8.239E-03	 &    -2.956E-03      &      4.363E-02	\\	
    P(2) =   &  -1.215E-05 	  &   -9.653E-05    &          -3.011E-04     &	    1.805E-04	   &   -4.527E-04      &   -1.070E-03     &	   -2.481E-04	 &     1.936E-04      &    -7.810E-04\\	
    P(3) =   &   1.679E-07 	  &    7.521E-07    &          1.214E-06     &	   -2.555E-06	   &    5.353E-06      &   1.056E-05        &	    8.066E-06	 &     5.019E-05      &     6.336E-06\\	
    P(4) =   &  -1.478E-02         &        1.352         &         -0.267          &	    -7.125E-03    &        1.364          &     -0.216           &	   -5.190E-02	  &    0.589 	      &    -0.552	\\
    P(5) =   &  -9.573E-05 	  &   -1.873E-04    &         -5.679E-04    &	     1.060E-03	   &    -3.263E-03      &   -2.478E-03     &	   -1.918E-03	   &  -1.195E-02         &    -2.085E-03	\\	
    P(6) =   &   1.488E-06 	  &   -8.969E-06    &         -3.416E-05     &	   -7.913E-06	   &    2.412E-05      &   3.463E-05         &	   -7.211E-05	   &  -7.523E04       &     2.198E-06\\	
    P(7) =   &   8.441E-03 	  &      -0.559        &           0.197         &	      2.292E-02	   &      -0.554           &      0.197            &	    3.972E-02	   &   -4.724E-02         &     0.155	\\	
    P(8) =   &   3.609E-06 	  &    2.367E-04    &        2.454E-04     &	   -2.084E-04	   &    6.528E-04      &  -2.750E-05       &	    5.019E-04	  &    5.337E-03      &      1.153E-04\\	
    P(9) =   &  -1.327E-03 	  &      0.085          &           -0.0403     &	    -4.633E-03	   &      8.250E-02      &    -4.207E-02     &	   -4.968E-03	  &   -6.778E-03      &     -1.375E-02	\\	
    P(10) =  &   2.140E-02 	   &     12.627         &             4.193       &	        0.678        &         10.129         &        4.459            &	    0.410	          &    7.677              &      5.708	\\	
    P(11) =  &   1.536E-05	  &    -8.255E-04   &          0.045         &	      1.198E-03   &     -3.090E-03     &      2.314E-02     &	   -3.590E-03	  &   -3.144E-02	 &     2.202E-02	\\
    P(12) =  &  -6.051E-07	  &     7.682E-06   &        -1.106E-04    &	    -1.894E-05   &     5.626E-05       &   -1.280E-04       &	   -2.742E-04	  &   -2.204E-03      &     -1.861E-04\\	
    P(13) =  &   9.150E-03	  &        0.846        &           -3.419       &	       -0.165         &        0.959          &       -3.401         &	   -7.413E-02	  &    0.733	      &     -2.467	\\	
    P(14) =  &  -7.047E-06	  &     8.808E-05   &         -7.320E-03    &	     3.674E-04      &    -8.039E-04     &    -4.975E-03    &	    2.318E-03	  &    1.794E-02&     -1.451E-03\\	
    P(15) =  &  -2.027E-03	  &       -0.133       &           0.536         &	       1.330E-02   &       -0.132          &       0.545         &	   -1.083E-03	  &   -7.636E-02	 &      0.239	\\
    P(16) =  &  -8.156E-02	  &        -29.741   &         1.466            &	        -1.189       &        -25.049        &        1.135         &	   -8.873E-02	  &  -18.897	      &      1.011	\\	
    P(17) =  &   4.760E-06	  &     8.721E-04   &         -1.830E-02    &	     5.7761E-07   &     -1.894E-03     &    -1.375E-03    &	    6.611E-03	  &    7.993E-02	      &     -2.507E-03	\\
    P(18) =  &   1.795E-04	  &       -0.121       &           0.475          &	       7.570E-02   &       -0.184           &       0.385         &	    2.154E-02	          &   -0.431	      &     -0.027	\\	
    P(19) =  &   5.410E-02	  &         18.070     &           -1.489        &	        0.623        &         15.536           &       -1.305       &	   -0.188	           &    11.218	      &     -0.413	\\	

\hline
\end{tabular}
\caption{List of best fit parameters of the polynomial functions estimated to retieve the best value of slope and intercept of the error map-coverage map relation and the best value of the cross correlation correction factor in any band and speed.}
\label{t1}
\end{table*}}

\end{document}